\documentclass[seceq]{ptptex}
\usepackage{amsmath,graphicx,bm,wrapft,rotfloat}
\renewcommand{\Re}{\mathop{\rm Re}}
\renewcommand{\Im}{\mathop{\rm Im}}
\newcommand{\br}{\bm{r}}
\newcommand{\bp}{\bm{p}}
\newcommand{\bq}{\bm{q}}
\newcommand{\bM}{\bm{M}}
\newcommand{\bI}{\bm{I}}
\newcommand{\bTheta}{\bm{\Theta}}
\newcommand{\arccosh}{\mathop{\rm arccosh}}
\newcommand{\arcsinh}{\mathop{\rm arcsinh}}
\newcommand{\erf}{\mathop{\rm erf}}
\newcommand{\sign}{\mathop{\rm sign}}
\newcommand{\Ai}{\mathop{\rm Ai}}
\newcommand{\Gi}{\mathop{\rm Gi}}
\newcommand{\rE}{{\mathrm{E}}}
\newcommand{\rF}{{\mathrm{F}}}
\newcommand{\rPi}{\mathchar"05}
\newcommand{\eps}{\varepsilon}

\newcommand{\gsim}{\mathop{\vcenter{%
 \offinterlineskip\hbox{$>$}\hbox{$\sim$}}}}
\newcommand{\gl}{\mathop{\vcenter{%
 \offinterlineskip\hbox{$>$}\hbox{$<$}}}}
\newcommand{\pp}[2]{\frac{\partial #1}{\partial #2}}
\newcommand{\ppp}[3]{\frac{\partial^2 #1}{\partial #2\partial #3}}
\newcommand{\tI}{\tilde{I}}
\newcommand{\tK}{\tilde{K}}
\newcommand{\sfig}[2]{%
 \centerline{\includegraphics[width=#1\textwidth]{#2}}}

\newcommand{\dfig}[4]{%
 \noindent
 \begin{minipage}{.48\textwidth}
 \centerline{\includegraphics[width=#1\textwidth,bb=#2]{#3}}
 \end{minipage}
 \hfill
 \begin{minipage}{.48\textwidth}
 \centerline{\includegraphics[width=#1\textwidth,bb=#2]{#4}}
 \end{minipage}}
\newcommand{\JOP}[1]{\JL{J.\ of\ Phys.,#1}}
\newcommand{\brcase}[2]{\mathchoice%
 {\genfrac\{\}{0pt}0{#1}{#2}}
 {\genfrac\{\}{0pt}1{#1}{#2}}
 {\genfrac\{\}{0pt}2{#1}{#2}}
 {\genfrac\{\}{0pt}3{#1}{#2}}}
\title{
Symmetry Breaking and Bifurcations \\
in the Periodic Orbit Theory: II \\
--- {\it Spheroidal Cavity} ---}
\author{%
Alexander G. {\sc Magner},$^{a,b,c}$
Ken-ichiro {\sc Arita},$^d$
Sergey N. {\sc Fedotkin}$^{b,c}$ \\
and Kenichi {\sc Matsuyanagi}$^e$}
\inst{
$^a$Research Center for Nuclear Physics, Osaka University,
Osaka 567-0047, Japan \\
$^b$Institute for Nuclear Research, 03680 Prospekt Nauki 47,
Kiev--28, Ukraine \\
$^c$Institute for Theoretical Physics, University of
Regensburg, D-93040 Regensburg, Germany \\
$^d$Department of Physics, Nagoya Institute of Technology,
Nagoya~466-8555, Japan \\
$^e$Department of Physics, Graduate School of Science, Kyoto
University, Kyoto~606-8502, Japan} \abst{ We derive a
semiclassical trace formula for the level density of the
three-dimensional spheroidal cavity.  To overcome the divergences
and discontinuities occurring at bifurcation points and in the
spherical limit, the trace integrals over the action-angle
variables are performed using an improved stationary phase
method.  The resulting semiclassical level density oscillations
and shell energies are in good agreement with quantum-mechanical
results.  We find that the births of three-dimensional orbits
through the bifurcations of planar orbits in the equatorial plane
lead to considerable enhancement of shell effect for
superdeformed shapes.}

\begin{document}
\maketitle

\section{Introduction}

The periodic orbit theory (POT)
\cite{gutzpr,gutz,bablo,strumag,bt76,bt77,creagh,strusem,mfimbrk,book}
is a nice tool for studying the correspondence between classical and
quantum mechanics and, in particular, the interplay of deterministic
chaos and quantum-mechanical behavior.  But also for systems with
integrable or mixed classical non-linear dynamics, the POT leads to a
deeper understanding of the origin of shell structure in finite
fermion systems such as nuclei,\cite{strusem,frisk,ak95,fispot}
metallic clusters,\cite{nish,erice,frau98} and mesoscopic
semiconductors.\cite{qdot,nak99,chan,nak00,nak01}  Bifurcations of
periodic orbits may play significant roles, e.g., in connection with
the superdeformations of atomic
nuclei,\cite{strusem,mfimbrk,ak95,sak98,ask98,mfamb} and were recently
shown to affect the quantum oscillations observed in the
magneto-conductance of mesoscopic devices.\cite{chan,nak00} This
phenomenon is observed for some control parameters (shapes, magnetic
field etc.) of the potential well, for which the orbits bifurcate and
new type of periodic orbits emerge from the original ones.
Examples can be found, e.g., in elliptic billiard and spheroidal
cavity.\cite{strusem,mfimbrk,ak95,ask98,mfamb,nish91,nish92,ask97,mfammsb}
In elliptic billiard, short diametric orbits with repetitions
bifurcate at certain values of deformation parameter,
and new orbits with hyperbolic caustics (butterfly-shaped orbit etc.)
emerge from them.  In spheroidal cavity, periodic orbits in the
equatorial plane bifurcate, and new three-dimensional orbits emerge.

The semiclassical trace formulae connect the quantum-mechanical
density of states with a sum over the periodic orbits of the
classical system.\cite{gutzpr,gutz,bablo,strumag}  In these
formulae, divergences arise at critical points where bifurcations
of periodic orbits occur or where symmetry breaking (or restoring)
transitions take places.  At these points, the standard stationary
phase method (SSPM),%
\footnote{
In this paper SSPM denotes the standard stationary phase method and
its extension to continuous
symmetries.\cite{bablo,strumag,bt76,creagh}}
used in the semiclassical evaluation of the trace integrals, breaks
down.  Various ways of avoiding these divergences have been
studied,\cite{bablo,bt76,creagh97} some of them employing uniform
approximations.%
\cite{creagh97,tgu,sie97,richens,MeierBrackCreagh,bremen,sie96,ssun,hhun}
Here we employ an improved stationary-phase method (ISPM) for the
evaluation of the trace integrals in the phase-space representation,
which we have derived for the elliptic billiard\cite{mfammsb} and very
recently for the spheroidal cavity.\cite{mfamb}

The singularities of the SSPM near the bifurcation points are due
to the peculiarities of its asymptotic expansions. In the
ISPM,\cite{mfamb,mfammsb}
the catastrophe integrals\cite{chester,fed:jvmp} are evaluated
more exactly within the finite integration limits in the
phase-space trace formula
\cite{bablo,bt76,mfimbrk,mfamb,mfammsb,bruno},
and one can overcome the singularity problem due to bifurcations,
which occur when the stationary points lie near the ends of the
integration region in the action-angle variables.  We can also
take into account the stationary points outside the classically
accessible region (``ghost orbits'').\cite{bt76}  This method is
particularly useful for integrable systems where integration
limits are easily obtained. This theory has been developed in
Ref.~\citen{mfammsb} for the case of the bifurcations through
which periodic orbit families with maximal degeneracy emerge from
the orbits with smaller degeneracy.
The essential difference between our method presented in this
paper and that with the uniform approximation of
Refs.~\citen{richens} and \citen{sie96} is that we improve the
calculation of the {\it angle} part of the phase-space trace
integral for the orbits with smaller degeneracies. Taking the
elliptic billiard as an example, we have applied the ISPM to the
integration over the angle variable for short diametric orbits,
and derived the improved trace formula which is continuous
through all bifurcation points including the circular limit and
the separatrix. We have then shown that significant enhancements
of the shell effect in level densities and shell structure
energies occur at deformations near the bifurcation points. Away
from the bifurcation points, our result reduces to the extended
Gutzwiller trace formula,\cite{strumag,strusem,mfimbrk,book} and
for the leading-order families of periodic orbits, it is
identical to that of Berry and Tabor\cite{bt76}.

The major purpose of this paper is to extend our semiclassical ISPM to
the three-dimensional (3D) spheroidal cavity,\cite{mfamb} which may be
taken as a simple (highly idealized) model for a heavy deformed
nucleus\cite{strusem,frisk} or a deformed metallic
cluster,\cite{nish,erice} and to specify the role of periodic orbit
bifurcations in the shell structure responsible for superdeformations.
Although the spheroidal cavity is integrable, it exhibits all the
difficulties mentioned above (i.e., symmetry breaking and
bifurcations) and therefore gives rise to an exemplary case study of a
non-trivial 3D system.  We apply the ISPM for the bifurcating orbits
and succeed in reproducing the superdeformed shell structure by the
POT, hereby observing a considerable enhancement of the shell effect
near the bifurcation points.

\section{Classical mechanics for the spheroidal cavity }
\label{sec:sphercav}

The semiclassical trace formulas for the oscillating part of the level
density for the spheroidal cavity are determined by the characteristic
properties of the classical periodic
families.\cite{strusem,mfimbrk,sak98,ask98,mfamb,nish91,nish92,ask97}
This section is an outlook of the definitions and solutions of the
classical mechanics for the spheroidal cavity in line of
Refs.~\citen{strusem,mfimbrk,ask98} and \citen{ask97}.  They will be
used for the semiclassical derivations of the trace formulas improved
at the bifurcation points.  We shall pay special attention to the
3D periodic orbits which emerge through
bifurcations and play
important roles as the semiclassical origin of superdeformed shell
structure.\cite{strusem,ask98,ask97}

\subsection{General periodic-orbit formalism}
\label{sec:genform}

We characterize the spheroid by its ratio of semi-axes $\eta =b/a$
keeping its volume fixed, and consider the prolate case with $\eta >
1$, where the major axis coincides with the symmetry axis.  We first
transform the Cartesian coordinates $(x,y,z)$ into the usual
cylindrical coordinates $(\rho,z,\varphi)$, where
$\rho=\sqrt{x^2+y^2}$, which are expressed in terms of the spheroidal
coordinates $(u,v,\varphi)$
\begin{equation}
\rho = \zeta\:\cos u\:\sinh v, \quad
   z = \zeta\:\sin u\:\cosh v, \quad
\zeta= \sqrt{b^2-a^2}
\label{ellcoor}
\end{equation}
with
\begin{equation}
-\frac{\pi}{2} \leq u \leq \frac{\pi}{2}, \quad
0 \leq v < \infty, \quad
0 \leq \varphi \leq 2\pi.
\label{ellcoorrange}
\end{equation}
The values of $\pm\zeta$ define the positions of the foci of the
spheroid lying on the $z$-axis.  Taking into account the volume
conservation condition $a^2b=R^3$, one has $b=R\eta^{2/3}$ and
$a=R\eta^{-1/3}$.  As is well known, the Hamilton-Jacobi equations
separate in the coordinates $(u,v,\varphi)$ for the spheroidal cavity.

In the Hamilton-Jacobi formalism, classical dynamics is determined by the
partial actions.  In the spheroidal coordinates these are given by
\begin{subequations}
\label{actionuv}
\begin{eqnarray}
I_u &=&\frac{p\zeta}{\pi}\int_{-u_c}^{u_c}
  du \sqrt{\sigma_1-\sin^2 u - \frac{\sigma_2}{\cos^2 u}},
  \label{actionu} \\
I_v &=&\frac{p\zeta}{\pi}\int_{v_c}^{v_b}
  dv \sqrt{\cosh^2 v-\sigma_1- \frac{\sigma_2}{\sinh^2 v}},
  \label{actionv} \\
I_\varphi &=&|l_z|=p\zeta \sqrt{\sigma_2},
  \label{actionphi}
\end{eqnarray}
\end{subequations}
where $l_z$ is the projection of the angular momentum onto the
symmetry axis, and $p=\sqrt{2m\eps}$, $m$ is the particle mass.  In
Eqs.~(\ref{actionuv}) we introduced new ``action'' variables $\sigma_1$
and $\sigma_2$ related to the turning points $-u_c$, $u_c$ and $v_c$,
$v_b$
along the trajectory in the $(u,v)$ coordinates; $u=u_c$ and
$v=v_c$ are the (hyperbolic and elliptic) caustic surfaces,
\begin{subequations}
\label{caustvcuc}
\begin{equation}
\cosh v_c=\left\{\frac12 (1+\sigma_1) + \left[\frac14 (1-\sigma_1)^2+
\sigma_2\right]^{1/2}\right\}^{1/2},
\end{equation}
\begin{equation}
\sin u_c=\left\{\frac12 (1+\sigma_1) - \left[\frac14 (1-\sigma_1)^2+
\sigma_2\right]^{1/2}\right\}^{1/2},
\end{equation}
\end{subequations}
and $v=v_b$ is the spheroid boundary, given by
$\cosh v_b=\eta/\sqrt{\eta^2-1}$.
The condition that the kinetic energy must be positive determine the
limits for the variables $\sigma_1$ and $\sigma_2$,
\begin{eqnarray}
\sigma_1^- &=& \sigma_2 \leq \sigma_1 \leq \frac{\eta^2}{\eta^2-1}
 -\sigma_2 \left(\eta^2-1\right)=\sigma_1^+, \nonumber\\
\sigma_2^- &=& 0 \leq \sigma_2 \leq \frac{1}{\eta^2-1}=\sigma_2^+.
\label{sigmalim}
\end{eqnarray}
These inequalities together with $2\pi$ intervals for the
corresponding angle variables determine the tori of the classically
accessible motion with the boundaries $\sigma_1^\pm(\sigma_2)$
and $\sigma_2^\pm$.

According to Eqs.~(\ref{actionuv}), the particle energy $\eps$ is a
function of only the action variables $I_u$, $I_v$ and $I_{\varphi}$,
$\eps=H\left(I_u,I_v,I_\varphi\right)$ due to integrability of the
system under consideration.  These relations define the partial
frequencies $\omega_u$, $\omega_v$ and $\omega_\varphi$ through
$\omega_j=\partial H/\partial I_j$.  The periodicity conditions for
the classical trajectories are significantly simplified in terms of
the partial frequencies $\omega_j$.  Introducing the new variables
$\kappa$ and $\theta$,
\begin{equation}
\kappa=\frac{\sin u_c}{\cosh v_c},\quad
\theta=\arcsin\left(\frac{\cosh v_c}{\cosh v_b}\right),
\label{kappatheta}
\end{equation}
along with the energy $\eps$ instead of the partial actions $I_u$,
$I_v$ and $I_{\varphi}$ (or $\sigma_1$ and $\sigma_2$), they read
\begin{subequations}
\label{omegauvf}
\begin{eqnarray}
\frac{\omega_u}{\omega_v} &\equiv& \frac12
 \left[1-\frac{\rF(\theta,\kappa)}{\rF(\kappa)}\right]
 =\frac{n_u}{n_v}, \\
\frac{\omega_\varphi}{\omega_u} &\equiv& \frac{2}{\pi}
 \left[\left(1-\left(\frac{\kappa}{\bar{\kappa}}\right)^2\right)
 \left(1-\bar{\kappa}^2\right)\right]^{1/2}
 \biggl\{\rPi\left(\left(\frac{\kappa}{\bar{\kappa}}\right)^2,
 \kappa\right)-\rF(\kappa) \nonumber\\
&&+ \left[\rPi\left({\bar \kappa}^2,\kappa\right)
 -\rPi\left(\theta,\bar{\kappa}^2,\kappa\right)\right]\big{/}
 \left[1-\frac{\rF(\theta,\kappa)}{\rF(\kappa)}\right]\biggr\}
 =\frac{n_\varphi}{n_u}.
\end{eqnarray}
\end{subequations}
Here, $n_u$, $n_v$ and $n_{\varphi}$ are co-prime integers,
$n_u=1,2,\cdots;\: n_v \ge 2 n_u;\: n_v \ge 2n_\varphi,\:
n_\varphi=1,2,\cdots$, and
$\bar{\kappa}=\sqrt{\eta^2-1}/(\eta\sin\theta)$.  $\rF$ and $\rPi$
are elliptic integrals of 1st and 3rd kinds
(see Appendix~\ref{app:curv} for their definitions).  The
periodicity condition (\ref{omegauvf}) relates
$\kappa(\sigma_1,\sigma_2)$ and $\theta(\sigma_1,\sigma_2)$ for a
given periodic orbit $\beta$ to the integers $n_u$, $n_v$ and
$n_\varphi$, which together with the number of repetitions $M$
define this orbit; i.e., $\beta=M(n_v,n_\varphi,n_u)$.

\subsection{Three-dimensional periodic orbits}

The 3D periodic orbits (3DPO) $M(n_v,n_\varphi,n_u)$
form two-parameter (${\cal K}=2$) families for a given energy $\eps$
since the number ${\cal K}$ of free continuous parameters specifying
an orbit with fixed energy and the same action is
two.~\cite{strusem,mfimbrk}
The condition for 3DPO is the existence of real roots
$(\kappa,\theta)$ of Eq.~(\ref{omegauvf}).  They appear at the
deformation $\eta=\eta_{\rm bif}$ given by
\begin{equation}
\eta_{\rm bif}=\frac{\sin(\pi n_\varphi/n_v)}{\sin(\pi n_u/n_v)},
\quad (n_u=1,2,\cdots, ~ n_v\geq 2n_\varphi+1, ~ n_\varphi=2,3,\cdots)
\label{etamin}
\end{equation}
where $\kappa=0$ and $\theta=\pi (1-{2n_u}/n_v)/2$,
and exist for larger deformation $\eta>\eta_{\rm bif}$.
These roots determine the caustics (the spheroid $v=v_c$ and the
hyperboloids $u=\pm u_c$) of the periodic
orbit $M(n_v,n_\varphi,n_u)$ through Eq.~(\ref{kappatheta}).  These
caustics are confocal to the boundary of the spheroid $v=v_b$.

\begin{figure}[t]
\sfig{1}{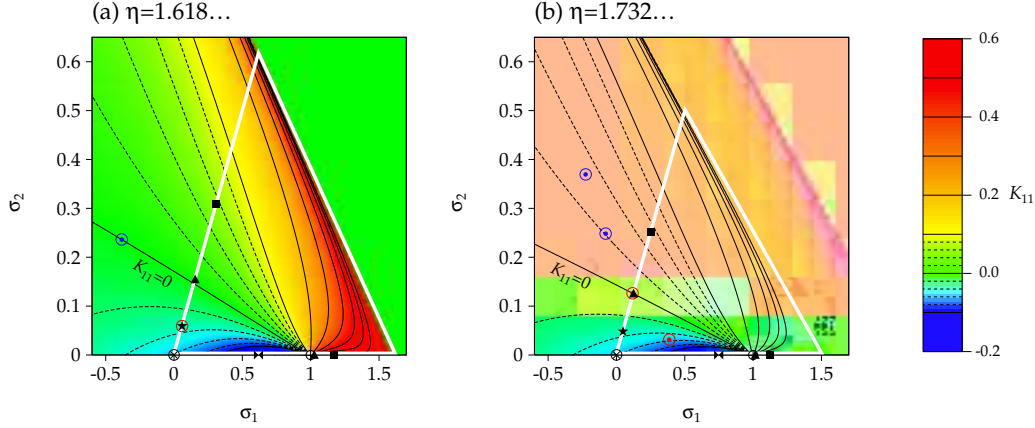}
\caption{\label{fig:curvature}
The triangle of the classically accessible region determined by
Eq.~(\ref{sigmalim}) is indicated by white lines in the
$(\sigma_1,\sigma_2)$ plane at bifurcation deformations
(a) $\eta=1.618\ldots$ and (b) $\eta=\sqrt{3}$.
The red and blue dots with circles indicate
the 3DPO stationary points inside (really existing 3DPO) and
outside (``ghost'' 3DPO) of this triangle region, respectively. Several
examples of the stationary points are indicated: on the $\sigma_2=0$~side,
the short 2DPO (elliptic triangle, square, and hyperbolic ``butterfly'');
on the $\sigma_2=\sigma_1$ side, the short EQPO (triangle, square, star,
and diameter (black crossed circle)).
The long diameter (separatrix) is located at ($\sigma_1=1,\sigma_2=0$).
The color and the contour curves indicate (in unit of $p\zeta/\pi$) the
curvature $K_{11}$ defined by Eq.~(\ref{Kn}).}
\end{figure}

Figure~\ref{fig:curvature} shows the stationary points corresponding
to the 3DPO for two bifurcation points $\eta_{\rm bif}$ given by
(\ref{etamin}).  The physical tori region (\ref{sigmalim}) in the
variables $\sigma_i$ is the triangle.  At $\eta_{\rm bif}=1.618\ldots$
(Fig.~\ref{fig:curvature}a), the stationary point for the 3DPO
$(5,2,1)$ coincides with that for the star-shaped $(5,2)$ orbit in the
equatorial plane (discussed below) lying on the boundary with
$\sigma_2=\sigma_1$, and moves toward inside of the physical tori
\begin{wrapfigure}[13]{r}{66mm}
\sfig{.3}{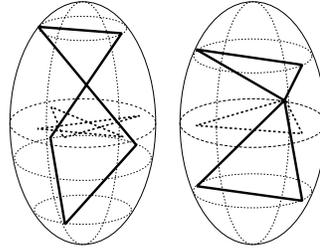}
\caption{\label{fig:orbit_3d} Short 3D periodic orbits (5,2,1)
and (6,2,1) bifurcated from the equatorial plane orbits (5,2) and
2(3,1), respectively.  Their projections on the equatorial plane
are also represented by thick-dashed lines.}
\end{wrapfigure}
region for larger deformations.  At $\eta_{\rm bif}=\sqrt{3}$
(Fig.~\ref{fig:curvature}b), the stationary point for the 3DPO
$(6,2,1)$ lies at the boundary side and coincides with that for
triangular orbits in the equatorial plane. At these bifurcation
deformations, the lengths of the 3DPO $(5,2,1)$ and $(6,2,1)$ coincide
with those of the star $(5,2)$ and the doubly repeated triangle
$2(3,1)$, respectively.  Figure~\ref{fig:orbit_3d} illustrates some
short 3DPO, and their projections onto the equatorial plane which
remind us of the parent equatorial orbits.

\subsection{Orbits in the meridian plane}

Equation~(\ref{omegauvf}) have partial solutions for
$\kappa(\sigma_1,\sigma_2)$ and $\theta(\sigma_1,\sigma_2)$ which
correspond to the separate families of orbits, i.e.
two-dimensional periodic orbits (2DPO), in the meridian planes
(containing the symmetry axis $z$) and in the equatorial plane.
First, we consider the special solutions of Eq.~(\ref{omegauvf})
corresponding to the two-parametric (${\cal K}=2$) 2DPO families
in the meridian plane.\cite{strusem,mfimbrk}
For these orbits, $\sigma_2=0$ and $\sigma_1$ is in the regions
\begin{equation}
0 < \sigma_1 < 1, \qquad 1 < \sigma_1 < \frac{\eta^2}{\eta^2-1},
\label{sigmaparlim}
\end{equation}
for the hyperbolic 2DPO (with hyperbolic caustics $u=\pm u_c$) and
the elliptic 2DPO (with elliptic caustics $v=v_c$), respectively.
The periodicity condition~(\ref{omegauvf}b) becomes the
identity $\omega_{\varphi}/\omega_u \equiv 1$~ ($n_{\varphi}=1$,
$n_u=1$), and $\theta$ is fixed by
\begin{equation}
\theta=\theta_h
 = \arcsin\left(\frac{\sqrt{\eta^2-1}}{\eta}\right), \qquad
\theta=\theta_e
 = \arcsin\left(\frac{\sqrt{\eta^2-1}}{\kappa\eta}\right),
\label{thetaeh}
\end{equation}
for the hyperbolic and elliptic 2DPO, respectively.
For $\kappa$, we only have the
condition Eq.~(\ref{omegauvf}a).  This $\kappa$ determines
$\sigma_1$ and thus $I_u$ and $I_v$ ($I_{\varphi}=0$ since
$\sigma_2=0$) through
\begin{equation}
\kappa=\kappa_h=\sqrt{\sigma_1},\qquad
\kappa=\kappa_e=\frac{1}{\sqrt{\sigma_1}}, \label{kappaeh}
\end{equation}
for the hyperbolic and elliptic orbits, respectively.

\begin{figure}[t]
\sfig{.6}{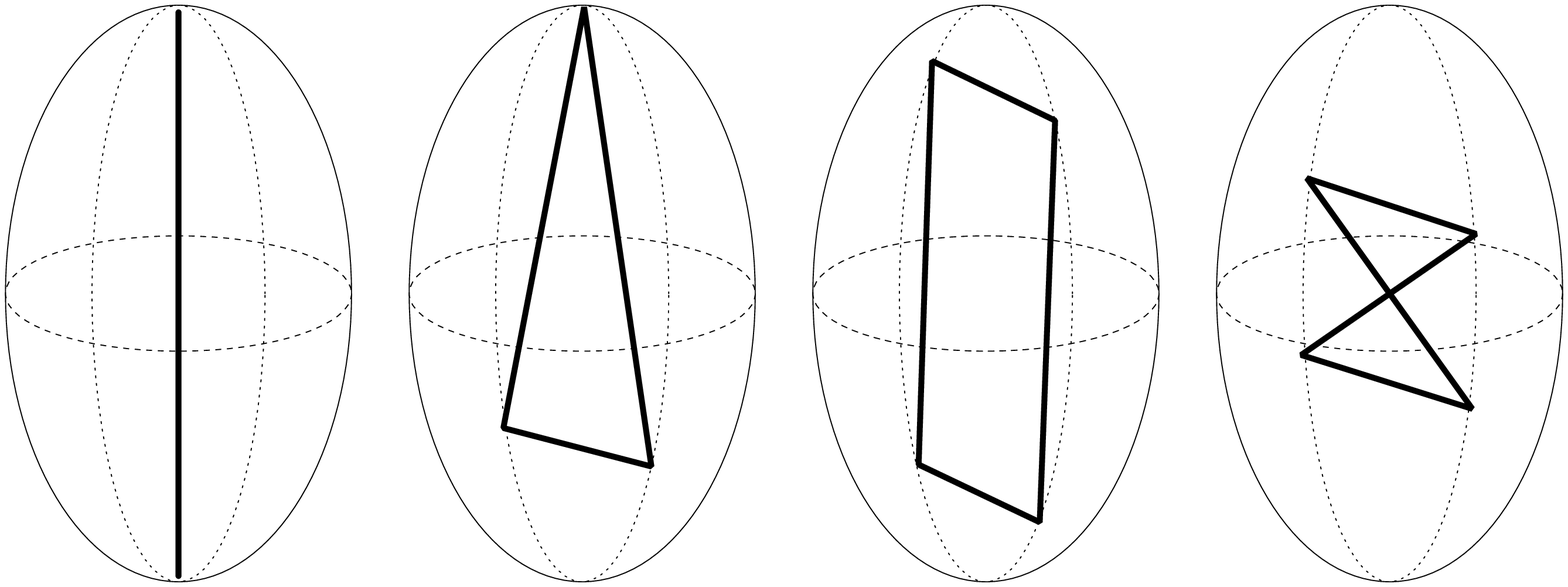}
\caption{\label{fig:orbit_meridian} Some short meridian-plane
orbits in the prolate spheroidal cavity: From the left to the
right; the isolated long diameter $(2,1,1)$, the elliptic
triangular $(3,1,1)$, the elliptic rhomboidal $(4,1,1)$, the
hyperbolic ``butterfly'' $(4,2,1)$.}
\end{figure}

Some examples of the hyperbolic and elliptic orbits lying along
the triangular boundary side $\sigma_2=0$ are indicated in
Fig.~\ref{fig:curvature}; see also their geometrical illustrations
in Fig.~\ref{fig:orbit_meridian}.  The hyperbolic and elliptic
tori parts are separated by the separatrix point
($\sigma_1=1,\sigma_2=0$) related to the long diameter (see
below).  Another end point of the hyperbolic tori coincides with
the stationary point ($\sigma_1=\sigma_2=0$) for the diametric
orbit in the equatorial plane.  We can think of these hyperbolic
and elliptic orbits as being periodic in the plane
$\varphi=\mbox{const}$ and we call them ``meridian-plane periodic
orbits''.

For the elliptic case, the solution $\kappa$ of
Eq.~(\ref{omegauvf}) with $\theta=\theta_e(\kappa)$ exists for any
$n_u=1,2,\cdots$ and $n_v \geq 2 n_u+1, (n_\varphi=n_u)$ at any
deformation $\eta >1$.  Examples are the triangles
$(n_v=3,n_\varphi=1,n_u=1)$, the rhomboids $(4,1,1)$ and the
star-shaped orbits $(5,2,2)$ as one-parameter families in the
meridian plane.  The root $\kappa$ found from
Eq.~(\ref{omegauvf}) gives the elliptic caustics with $u_c=\pi/2$
in Eq.~(\ref{kappatheta}) and the semi-axes
$a_c=\zeta\sqrt{1-\kappa^2}/ \kappa$ and $b_c=\zeta/\kappa$.

For the hyperbolic case, the solutions $\kappa$ can be found for
$n_u=1,2,3,\cdots$ and even $n_v$ ($n_v \geq 2(n_u+1)$).  In
Fig.~\ref{fig:curvature} the ``butterfly'' orbit $(4,2,1)$ is
indicated as an example.  The families of these orbits appear for
$\eta > \eta_{\rm bif}$ with
\begin{equation}
\eta_{\rm bif} = \left[\sin\left(\frac{\pi n_u}{n_v}\right)
 \right]^{-1}.
\label{etaminhyp}
\end{equation}
This is the deformation where the diametric orbits $M(2,1)$ with $M
\geq 2$ in the equatorial plane bifurcate and the hyperbolic orbits
emerge from them.  Their hyperbolic caustics are expressed in terms of
the root $\kappa$ of Eq.~(\ref{omegauvf}) and Eq.~(\ref{kappatheta})
with $v_c=0$.  The parameters $a_c$ and $b_c$ of these caustics are
given by $a_c=\zeta\sqrt{1-\kappa^2}$ and $b_c=\zeta\kappa$.

\subsection{Orbits in the equatorial plane}

\begin{figure}[t]
\sfig{.6}{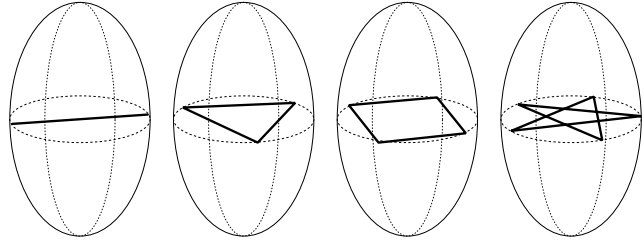}
\caption{\label{fig:orbit_equatorial} Some short equatorial-plane
orbits: From the left to the right, the short diameter $(2,1)$,
the triangular $(3,1)$, the rhomboidal $(4,1)$, and the
star-shaped $(5,2)$.}
\end{figure}

In the equatorial plane with $z=0$, the separate families of
regular polygons and diameters are the same as in a circular
billiard\cite{bablo} of radius $a$.  The restriction $z=0$
decreases the number $\cal K$ to ${\cal K} = 1$.  This single
parameter corresponds to the angle of rotation of the polygons and
the diameters about the symmetry axis $z$.
Figure~\ref{fig:orbit_equatorial} illustrates the most important
(shortest) equatorial-plane periodic orbits (EQPO); the diameters
$M(n_v=2,n_\varphi=1)$, triangles $M(3,1)$, squares $M(4,1)$ and
star-shaped orbits $M(5,2)$.  They satisfy, from inequalities
(\ref{sigmalim}),
\begin{equation}
\sigma_1=\sigma_2,\qquad
0 \leq \sigma_2 \leq \frac{1}{\eta^2-1}.
\label{sigmaeqlim}
\end{equation}
Therefore their stationary points lie along the $\sigma_2=\sigma_1$
side in the triangle, as indicated in Fig.~\ref{fig:curvature}.

The caustic parameters $u_c$ and $v_c$ for these families are defined by
$u_c=0$ and $v_c=\arcsinh[a \cos(\pi n_\varphi/n_v)/\zeta]$.
The solutions of Eq.~(\ref{omegauvf}) for these orbits are
$\kappa=0$ and $\theta=\arcsin\sqrt{1-\sin^2(\pi n_\varphi/n_v)/\eta^2}$.

\subsection{Diametric orbits along the symmetry axis}

In the spheroidal cavity, there is also a diametric orbit along
the $z$-axis (see Fig.~\ref{fig:orbit_meridian}).
It is isolated (${\cal K}=0$), since we have two
additional restrictions $x=0$ and $y=0$ decreasing ${\cal K}$ by
one unit more than in the previous case.  The solution of
Eq.~(\ref{omegauvf}) for this orbit is $\kappa=1$ and
$\theta=\arcsin(\sqrt{\eta^2-1} / \eta)$.  Its stationary point
coincides with the separatrix values ($\sigma_1=1, \sigma_2=0$)
corresponding formally to the caustic parameters ($u_c=\pi/2,~
v_c=0$); see the circle point with the vertical diameter in
Fig.~\ref{fig:curvature}.  (In Fig.~\ref{fig:curvature}b, it is
very close to the stationary point for the elliptic orbits (3,1,1)
in the meridian plane, which lies slightly on the right along the
$\sigma_2=0$ side.)

\subsection{Bifurcations}

At the deformations $\eta_{\rm bif}$ given by Eq.~(\ref{etamin}),
the EQPO $M(n_v,n_\varphi)$ bifurcate and the 3DPO or the
hyperbolic 2DPO $M(n_v,n_\varphi,n_u)$ emerge. We encounter the
breaking-of-symmetry problem at these bifurcation points since
the degeneracy (symmetry) parameter ${\cal K}$ changes there, for
instance, from ${\cal K}=1$ of the EQPO to ${\cal K}=2$ of the
3DPO through them.  Before the bifurcations ($\eta<\eta_{\rm
bif}$), the stationary points $\sigma_i$ of the 3DPO and the
hyperbolic 2DPO are situated outside of the triangular tori
region (\ref{sigmalim}), and give rise to complex
$(\kappa,\theta)$ and complex caustics.  Such formal orbits are
called ``complex'' or ``ghost'' orbits.\cite{bt76}  They cross
the $\sigma_1=\sigma_2$ boundary through the stationary point of
the EQPO at bifurcations ($\eta=\eta_{\rm bif}$) and then moves
into the triangular tori region for larger $\eta$. In
Fig.~\ref{fig:curvature} are also indicated the stationary points
for the 3DPO lying outside the physical tori region ((6,2,1) in
Fig.~\ref{fig:curvature}a, (7,2,1) and (8,2,1) in
Fig.~\ref{fig:curvature}b). The equatorial diameters $M(2,1)$
correspond to the limiting case, $\sigma_1=\sigma_2=0$.  They
bifurcate into themselves (${\cal K}=1$) and the hyperbolic 2DPO
$(2M,M,1)$ in the meridian plane (${\cal K}=2$) at the
deformations given by Eq.~(\ref{etaminhyp}).

The spherical limit is a special bifurcation point.  Namely,
the planar regular polygons (${\cal K}=3$) and diameters (${\cal
K}=2$) bifurcate into the meridian 2DPO (${\cal K}=2$), EQPO
(${\cal K}=1$) and the isolated long diameter (${\cal K}=0$) for
deformations $\eta>1$.  Another special point is the separatrix
($\sigma_1=1, \sigma_2=0$) related to the long diameter.  Near this
point the complicated 3DPO, elliptic and hyperbolic 2DPO having
large values of ($n_u, n_v$) and $n_u/n_v$ close to $1/2$ appear.
Similar bifurcations of the 3DPO, EQPO and elliptic 2DPO appear
near other boundary values of $\sigma_i$ in the triangular tori on
its ``creeping'' side $\sigma_1=\sigma_1^+(\sigma_2)$ where some
kinds of 3D ``creeping'' orbits having large values of $n_v$ but
with finite and generally different $n_u$ and $n_\varphi$ appear.
This is in analogy with the ``creeping'' singularities discussed
for the elliptic orbits in the elliptic billiard\cite{mfammsb}
near the maximum value of $\sigma_1$, $\sigma_1^{\rm(cr)} =\cosh^2
v_b=\eta^2 / (\eta^2-1)$, according to Eq.~(\ref{knfzb1}) at the
right vertex in the ``meridian-plane orbit'' side $\sigma_2=0$.
The 3DPO with large number of the corners $n_v$ and finite
$n_u=n_\varphi$ approach the ``creeping'' elliptic orbits in the
meridian plane.  Another vertex corresponds to the creeping EQPO
having large values of $n_v$ and $n_\varphi$ but for
$n_v/n_\varphi \rightarrow 1/2$.

The bifurcation point related to an appearance of ``creeping''
orbits cannot be reached for any finite deformation.  However,
even for finite deformations like superdeformed shapes, the
solutions for $\sigma_1$ and $\sigma_2$ (related to the roots
$\kappa$ and $\theta$ of the periodic-orbit conditions
(\ref{omegauvf})) can be close to the ``creeping'' values of
$\sigma_1$ and $\sigma_2$ (related to their boundary values of
(\ref{sigmalim})).  In such cases we have to take into account such
bifurcations in the trace formulas for the level density.  The
bifurcations of the 3D and 2D diameter-like orbits with $n_v/n_u$
close to $1/2$ near the separatrix are rather long, however, so
that they are not important for the shell effects discussed below.

\section{Trace formulas for the prolate spheroid}

\subsection{Phase-space trace formula in action-angle variables}

The level density $g(\eps)$ is obtained from the Green function
$G(\br,\br';\eps)$ by taking the imaginary part of its trace:
\begin{equation}
g(\eps) = \sum_n \delta\left(\eps-\eps_n\right)
 = -\frac{1}{\pi}\Im\int d\br'' \int d\br'
G(\br',\br'';\eps)~ \delta (\br''-\br'),
\label{trace}
\end{equation}
where $\eps_n$ is the single-particle energy.  Following
Ref.~\citen{mfammsb}, we apply now the Gutzwiller's trajectory
expansion for the Green function
$G(\br,\br',\eps)$.\cite{gutzpr,gutz,book}  After simple
transformations,\cite{mfammsb} we obtain the phase-space trace formula
in the action-angle variables $(\bI,\bTheta)$,
\begin{eqnarray}
g_{\rm scl}(\eps)&=&\frac{1}{(2\pi\hbar)^3} \Re\sum_\alpha
\int d\bTheta'' \int d\bI'
\delta\left(\eps - H\left(\bI',\bTheta'\right)\right)
\nonumber\\
&& \times \exp\left[\frac{i}{\hbar}
\left(S_\alpha\left(\bI',\bI'',t_\alpha\right)+
\left(\bI''-\bI'\right)\cdot \bTheta''\right)
- i \frac{\pi}{2} \nu_\alpha\right],
\label{pstraceactang}
\end{eqnarray}
where the sum is taken over all classical trajectories $\alpha$,
$\bI=\{I_u,I_v,I_\varphi\}$ are the actions for the spheroidal
cavity, $\bTheta=\{\Theta_u,\Theta_v,\Theta_\varphi\}$ the
conjugate angles, and $\nu_\alpha$ phases related with the Maslov
indices.\cite{fed:jvmp,masl,fed:spm,masl:fed}  The
phase-space trace formula (\ref{pstraceactang}) is especially useful
for integrable systems since the Hamiltonian $H$ does
not depend on the angle variables $\bTheta$ in this case, i.e.,
$H=H(\bI)$.  The action
\begin{equation}
S_\alpha\left(\bI',\bI'',t_\alpha\right)
=-\int_{\bI'}^{\bI''}d\bI\cdot
\bTheta(\bI)
\label{actioni}
\end{equation}
is related to the standard definition
\begin{equation}
S_\alpha\left(\bTheta', \bTheta'',\eps\right)=
\int_{\bTheta'}^{\bTheta''} d\bTheta\cdot
\bI(\bTheta)
\label{actionr}
\end{equation}
by the Legendre transformation
\begin{equation}
S_\alpha(\bTheta',\bTheta'',\eps)+
\bI''\cdot(\bTheta'-\bTheta'')=
S_\alpha(\bI',\bI'',t_\alpha)+\bTheta''\cdot(\bI''-\bI'),
\label{legentrans}
\end{equation}
$t_\alpha$ being the time for classical motion along
the trajectory $\alpha$.  The phase $\nu_\alpha$ will be
specified below.

\subsection{Stationary phase method and classical degeneracy}
\label{sec:spmdeg}

It should be emphasized that {\em even for integrable systems} the
trace integral (\ref{pstraceactang}) is more general than the
Poisson-sum trace formula which is the starting point of
Refs.~\citen{bt76,richens,sie96} for the semiclassical derivations
of the level density.  These two trace formulas become identical
when the phase of the exponent does not depend on the angle
variables $\bTheta$.  In this case, the integral over angles in
(\ref{pstraceactang}) gives simply $(2 \pi)^n$ where $n$ is the
dimension of the system ($n=3$ for the spheroidal cavity), and the
stationary condition for all angle variables are identities in the
$2\pi$ interval.  This is true for the {\em most degenerate}
classical orbits like the elliptic and hyperbolic 2DPO in the
meridian plane and the 3DPO with ${\cal K}=n-1=2$.  On the other hand,
for orbits with smaller degeneracy like the EQPO (${\cal K}=1$) and
the isolated long diameter (${\cal K}=0$), the exponent phase
strongly depends on angles and possesses a definite stationary
point.  Therefore, we have to integrate over such angles by the
ISPM in the same way as for the bifurcations of the isolated
diameters in the elliptic billiard.\cite{mfammsb}

\subsection{Stationary phase conditions}
\label{sec:statcondactang}

Due to the energy conserving $\delta$-function, we can exactly
take the integral over $I_v'$ in Eq.~(\ref{pstraceactang}) and
result in
\begin{eqnarray}
g_{\rm scl}(\eps) &=&
\frac{1}{(2\pi\hbar)^3}\Re\sum_\alpha
\int d\Theta_u'' \int d\Theta_v''
\int d\Theta_\varphi'' \int dI_u' \int dI_\varphi'
\frac{1}{|\omega_v'|}\nonumber\\
&&\times \exp\left[\frac{i}{\hbar}
\left(S_\alpha\left(\bI',\bI'',t_\alpha\right)+
\left(\bI''-\bI'\right)\cdot\bTheta''\right)
- i \frac{\pi}{2} \nu_\alpha\right].
\label{pstraceactang1}
\end{eqnarray}
The integration limits for $I_u$ and $I_\varphi$
are determined by their relations
to the variables ($\sigma_1$, $\sigma_2$) and by
the boundaries given by Eq.~(\ref{sigmalim}).
One of the trajectory $\alpha_0$ in the sum (\ref{pstraceactang1})
is the special one
which corresponds to the smooth level density $g_{\rm TF}$ of the
Thomas-Fermi model.\cite{book}
For all other trajectories, we first
write down the stationary phase condition for the action
variables $I_u'$ and $I_\varphi'$~:
\begin{subequations}
\label{statcondi}
\begin{equation}
\left(\pp{S_\alpha(\bI',\bI'',t_\alpha)}{I_u'}\right)^*
   - \Theta_u''
\equiv \Theta_u' - \Theta_u''
= 2\pi M_u,
\end{equation}
\begin{equation}
\left(\pp{S_\alpha(\bI',\bI'',t_\alpha)}{I_\varphi'}\right)^*
   - \Theta_\varphi''
\equiv \Theta_\varphi' - \Theta_\varphi''
= 2\pi M_\varphi,
\end{equation}
\end{subequations}
where $\bM=(M_u,M_v,M_\varphi)=M(n_u,n_v,n_\varphi)$ and $M$ are
integer numbers which indicate the numbers of revolution along
the primitive periodic orbit $\beta$.  The superscript asterisk
means that we take the quantities at the stationary point
$I_u'=I_u^*$ and $I_\varphi'=I_\varphi^*$.  We next use the
Legendre transformation (\ref{legentrans}).  Then, the stationary
phase conditions with respect to angles $(\Theta_u, \Theta_v,
\Theta_\varphi)$ are given by
\begin{equation}
\left(\pp{S_\alpha(\bTheta',\bTheta'',\eps)}{\bTheta''}
    + \pp{S_\alpha(\bTheta',\bTheta'',\eps)}{\bTheta'}\right)^*
\equiv \bI'' - \bI' = 0.
\label{statcondt}
\end{equation}

In the following derivations we have to decide whether the stationary
phase conditions (all or partially) given by Eqs.~(\ref{statcondi})
and (\ref{statcondt}) are identities or
equations for the specific stationary points.  For this purpose
we have to  calculate separately
the contributions from the most degenerate 3DPO,
the 2DPO families in the meridian plane (${\cal K}=2$) and those
from orbits with smaller degeneracy like EQPO (${\cal K}=1$) and
the isolated long diameter (${\cal K}=0$).
The latter two kinds of orbit are different from the former two
kinds in the above mentioned decision concerning the integration over
angles $\bTheta$.

\subsection{Three-dimensional orbits and meridian-plane orbits}

The most degenerate 3DPO and the meridian-plane (elliptic and
hyperbolic) 2DPO having the same values of action occupy some 3D
finite areas between the corresponding caustic surfaces specified
above.  In this case the stationary phase conditions
(\ref{statcondt}) for the integration over all angle variables
$\Theta_u$, $\Theta_v$ and $\Theta_\varphi$ are identities.  The
integrand does not depend on the angle variables and the result of
the integration is $(2\pi)^3$.  Since Eq.~(\ref{statcondt}) is
identically satisfied (the action does not depend on the angles
like the Hamiltonian $H(\bI)$) we have the conservation of the
action variables, $I_u'=I_u''=I_u$ and
$I_\varphi'=I_\varphi''=I_\varphi$, along the classical trajectory
$\alpha$.  The integrals over all $\Theta$ in
Eq.~(\ref{pstraceactang}) yield  $(2\pi)^3$ and we are left with
the Poisson-sum trace formula,\cite{bt76,book}
\begin{eqnarray}
g_{\rm scl}(\eps)
&=&\frac{1}{\hbar^3}\Re\sum_{\bM}\int d\bI\:
 \delta \left(\eps-H(\bI)\right)\:
 \exp\left[\frac{2\pi i}{\hbar}\bM\cdot\bI
 - i\frac{\pi}{2}\nu_{\bM}\right] \nonumber\\
&=&\frac{1}{\hbar^3}\Re\sum_{\bM}\int dI_u \int dI_\varphi\:
 \frac{1}{|\omega_v|}\:
 \exp\left[\frac{2\pi i}{\hbar}\bM\cdot\bI - i\frac{\pi}{2}\nu_{\bM}
 \right].
\label{poissonsum}
\end{eqnarray}
It is convenient to transform the integration variables
($I_u, I_\varphi$) to ($\sigma_1,\sigma_2$) defined by
Eq.~(\ref{actionuv}),
\begin{eqnarray}
g_{\rm scl}(\eps)=\frac{1}{\hbar^3}\Re\sum_{\bM} p\zeta
\int_{\sigma_2^-}^{\sigma_2^+} \frac{d\sigma_2}{2\sqrt{\sigma_2}}
\int_{\sigma_1^-}^{\sigma_1^+} d\sigma_1\:
\pp{I_u}{\sigma_1}\:\frac{1}{|\omega_v|}\:
\exp\left[\frac{2\pi i}{\hbar}\,\bM\cdot\bI-i\frac{\pi}{2}\nu_{\bM}
\right].\nonumber\\
\label{Poisvarsigma}
\end{eqnarray}
The integration limits are much simplified when written in terms
of $\sigma_i^\pm$ ($i=1,2$) and form the triangular region shown
in Figs.~1.  We then integrate over $\sigma_i$
expanding the exponent phase about the stationary point
$\sigma_i=\sigma_i^*$,
\begin{eqnarray}
2\pi\left(\bM\cdot\bI\right) &\equiv&
S_\alpha\left(\bI,\bI'',t_\alpha\right)+
\left(\bI''-\bI\right)\cdot\bTheta'' \nonumber\\
&=& S_\beta(\eps)
+\frac12 \sum_{i,j} J_{ij}^\beta(\sigma_i-\sigma_i^*)(\sigma_j-\sigma_j^*)
+\cdots,
\label{actionexpu}
\end{eqnarray}
where $S_\beta(\eps)$ is the action along the periodic orbit $\beta$,
\begin{equation}
S_\beta(\eps)=2\pi M \left[n_u I_u^* +
n_v I_v\left(\eps,I_u^*,I_\varphi^*\right)+
n_\varphi I_\varphi^*\right],
\label{actionpo}
\end{equation}
and $I_v(\eps,I_u,I_\varphi)$ the solution of the energy conservation
equation $\eps=H(I_u,I_v,I_\varphi)$ with respect to $I_v$.
Here the single prime index is omitted for simplicity.
The $J_{ij}^\beta$
is the Jacobian stability factor with respect to $\sigma_i$
along the energy surface,
\begin{equation}
J_{ij}^\beta=\left(\ppp{S_\alpha}{\sigma_i}{\sigma_j}
\right)_{\sigma_i=\sigma_i^*}=2\pi M n_v K_{ij}^\beta,
\label{jacobupar}
\end{equation}
and $K_{ij}^\beta$ is the $(2\times 2)$ curvature matrix of the energy
surface
taken at the stationary point $\sigma_i=\sigma_i^*$ (at the
periodic orbit $\beta$),
\begin{equation}
K_{ij}^\beta = \ppp{I_v}{\sigma_i}{\sigma_j}
+\frac{\omega_u}{\omega_v}\ppp{I_u}{\sigma_i}{\sigma_j}
+\frac{\omega_\varphi}{\omega_v}
\ppp{I_\varphi}{\sigma_i}{\sigma_j}. \label{Kn}
\end{equation}
See Appendix~\ref{app:curv} for the explicit expressions of these
curvatures.  As we shall see below, the off-diagonal curvature
$K_{12}$ is non-zero for variables $\sigma_i$.

Then we use the ISPM, where we keep exact finite limits for
integration over $\sigma_i$, and we finally obtain
\begin{equation}
\delta g_{\brcase{\rm 3D}{\rm 2D}}^{(2)}(\eps)=
\frac{1}{\eps_0}\Re\sum_\beta A_\beta^{(2)}\,
\exp\left(ikL_\beta-i \frac{\pi}{2}\nu_\beta\right),
\label{deltag3d2d}
\end{equation}
where $\eps_0=\hbar^2/2mR^2$ ($R^3=a^2b$ due to the volume
conservation condition).  The sum runs over all two-parameter
families of the 3DPO or the meridian-plane (elliptic and
hyperbolic) 2DPO,
$A^{(2)}_\beta$ is the amplitude for a 3DPO or a 2DPO,%
\footnote{
The expression (\ref{amp3d2d}) is valid also for the 2DPO
($\sigma_2^*=0$), since the product $\sigma_2 K_{22}$ is finite for
any $\sigma_2$ (see Appendix~\ref{app:curv}).}
\begin{eqnarray}
A_{\brcase{\rm 3D}{\rm 2D}}^{(2)}&=& \frac{1}{4\,\pi}\,
\frac{L_\beta \zeta}{(M n_v R)^2 \sqrt{\sigma_2^* \left|\det
K^\beta\right|}}\, \left[\pp{I_u}{\sigma_1}
\right]_{\sigma_i=\sigma_i^*} \erf\left({\cal Z}_1^-,{\cal
Z}_1^+\right)
\erf\left({\cal Z}_2^-,{\cal Z}_2^+\right). \nonumber\\
\label{amp3d2d}
\end{eqnarray}
Here, $L_\beta$ represents ``length'' of the periodic
orbit $\beta$
\begin{eqnarray}
L_\beta &=& \frac{2 \pi M n_v p}{m \omega_v} \nonumber\\
 &=& 2 M n_v  b \sin\theta \left[\rE(\theta,\kappa)-
\frac{\rF(\theta,\kappa)}{\rF\left(\kappa\right)}\,
\rE\left(\kappa\right)+
\cot\theta \sqrt{1-\kappa^2\sin^2\theta}\right],
\label{length}
\end{eqnarray}
where $\theta$ and $\kappa$ are defined by the roots of
periodic-orbit equations (\ref{omegauvf}) ($S_\beta=pL_\beta$ for
cavities).  This ``length'' taken at the stationary points
$\sigma_i^*$~ (the real positive roots of Eq.~(\ref{omegauvf})
through Eqs.~(\ref{caustvcuc}) and (\ref{kappatheta})) inside the
finite integration range (\ref{sigmalim}) represents the true
length of the corresponding periodic orbit $\beta$.  For other
stationary points, the ``length'' is nothing else than the
function (\ref{length}) continued analytically outside the tori
determined by (\ref{sigmalim}).  We introduced it formally instead
of $\omega_v $ by the equation $\omega_v=2 \pi p M n_v/m L_\beta $
for convenience.  In Eq.~(\ref{amp3d2d}) we introduced also the
generalized error function $\erf\left({\cal Z}^{-},{\cal
Z}^{+}\right)$ of the two complex arguments ${\cal Z}^{-}$ and
${\cal Z}^{+}$,
\begin{equation}
\erf\left(z^-,z^+\right)=\frac{2}{\sqrt{\pi}}
\int_{z^-}^{z^+} dz\,e^{-z^2}=\erf(z^+)-\erf(z^-),
\label{errorf}
\end{equation}
with $\erf(z)$ being the simple error function.\cite{abramov}
The arguments of these error functions are given by
\begin{subequations}\label{argerroru}
\begin{eqnarray}
{\cal Z}_1^{\beta\,\pm}
 &=& \sqrt{-i\pi Mn_v K^\beta_{11}/\hbar}
 \left(\sigma_1^\pm(\sigma_2^*)-\sigma_1^*\right), \\
{\cal Z}_2^{\beta\,\pm}
 &=& \sqrt{-i\pi Mn_v (\det K^\beta/K^\beta_{11})/\hbar}
 \left(\sigma_2^\pm-\sigma_2^*\right),
\end{eqnarray}
\end{subequations}
in terms of the finite limits $\sigma_i^\pm$ given by
(\ref{sigmalim}), and taken at the stationary point
$\sigma_2=\sigma_2^*$.  We note that, for the 3DPO $M(3t,t,1)$
with $t=2,3,...$, the curvature $K^\beta_{11}$ is zero at any
deformation.  For such orbits, one should use
\begin{subequations}
\label{argerroru2}
\begin{eqnarray}
{\cal Z}_1^{\beta\,\pm}
&=&\sqrt{-i \pi M n_v(\det K^\beta/K^\beta_{22})/\hbar}\,
\left(\sigma_1^\pm(\sigma_2^*)-\sigma_1^*\right), \\
{\cal Z}_2^{\beta\,\pm}
&=&\sqrt{-i \pi M n_v K^\beta_{22}/\hbar}\,
\left[\sigma_2^\pm - \sigma_2^* + \frac{K_{12}^\beta}{K_{22}^\beta}
\left(\sigma_1^\pm(\sigma_2^*)-\sigma_1^*\right)\right],
\end{eqnarray}
\end{subequations}
in place of (\ref{argerroru}).  The latter limits (\ref{argerroru2})
are derived by transforming the integration variable $\sigma_2$ to
$\sigma_2-(K_{12}/K_{22})(\sigma_1-\sigma_1^*)$.

Let us consider the stationary points $\sigma_i^*$ far from the
bifurcation points.  This means that they are located far from the
integration limits.  Accordingly, one can transform the generalized
error functions to the complex Fresnel functions with the real limits
and then extend the upper limit to $\infty$ and the lower one to
$-\infty$.  In this way we obtain asymptotically the Berry-Tabor
result of the standard POT,\cite{bt76} which is identical to the
extended Gutzwiller's result\cite{mfimbrk} for the most degenerate (3D
and meridian-plane) orbit families,
\begin{eqnarray}
A_{\brcase{\rm 3D}{\rm 2D}}^{(2)}(\eps)
=\frac{1}{\pi}\,
\frac{L_\beta \zeta}{(M n_v R)^2\sqrt{\sigma_2^* |\det K^\beta|}}\,
\left[\pp{I_u}{\sigma_1}\right]_{\sigma_i=\sigma_i^*}.
\label{amp3d2das}
\end{eqnarray}

The constant part of the phase $\nu_\beta$ in Eq.~(\ref{deltag3d2d}),
which is independent of $\eta$ and $\eps$, can be found by making use
of the above asymptotic expression and applying the Maslov-Fedoryuk
theory.\cite{fed:jvmp,masl,fed:spm,masl:fed} This theory relates the
Maslov index $\mu_\beta$ with the number of turning and caustic points
for the orbit family $\beta$.  For the 3DPO, the total asymptotic
phase $\nu_\beta$ is given by
\begin{eqnarray}
\nu_{\rm 3D} = \mu_{\rm 3D} - \frac12 \epsilon_{\rm 3D}
+2(M n_u-1), \quad
\mu_{\rm 3D} = M(3n_v + 2n_u).
\label{maslovphase3d}
\end{eqnarray}
Here $\mu_\beta$ denotes the Maslov index, the numbers of caustic and
turning points traversed by the orbit.  $\epsilon_\beta$ represents
the difference of the numbers of positive and negative eigenvalues of
curvature $K^\beta$.%
\footnote{
Since the dimension of $K^\beta$ is 2, $\epsilon_\beta$ is written as
$\epsilon_\beta=\sign(K_1^\beta)+\sign(K_2^\beta)$, where $K_i^\beta$
is the $i$-th eigenvalue of $K^\beta$.  It is also calculated by
$\epsilon_\beta=\sign(K_{11}^\beta)+\sign(\det K^\beta/K_{11}^\beta)$
for $K_{11}^\beta\ne 0$, and
$\epsilon_\beta=\sign(K_{22}^\beta)+\sign(\det K^\beta/K_{22}^\beta)$
for $K_{22}^\beta\ne 0$.  Here, $\sign(x)=\pm 1$ for $x\gl 0$ and 0
for $x=0$.}
For the hyperbolic and elliptic meridian 2DPO, one obtains
\begin{equation}
\nu_{\rm 2DH}=\mu_{\rm 2DH} - \frac12 \epsilon_{\rm 2DH} +
2\left(M n_u-1\right), \quad \mu_{\rm 2DH}=2 M
\left(n_v+n_u\right) \label{maslovphase2dh}
\end{equation}
and
\begin{equation}
\nu_{\rm 2DE}=\mu_{\rm 2DE} - \frac12 \epsilon_{\rm 2DE}
+ 2\left(M n_u- 1\right), \quad
\mu_{\rm 2DE}=3 M n_v
\label{maslovphase2de}
\end{equation}
respectively.  Note that
the total phase includes the argument of the complex amplitude
(\ref{amp3d2d}), and depends on both deformation and energy.

Near the bifurcation deformations, the stationary points $\sigma_i^*$
are close to the boundary of the finite area (\ref{sigmalim}).  In
such cases the asymptotics of the error functions are not good
approximations, and we have to carry out the integration over
$\sigma_i$ in the calculation of the error functions in
Eq.~(\ref{amp3d2d}) exactly within the finite limits.  One should also
note that the contributions from ``ghost'' periodic orbits are
important near the bifurcation points.  It makes the trace formula
continuous as function of $\eta$ at all bifurcations.

When the stationary phase points $\sigma_i^*$ are close to other
boundaries of the tori, one has also to take the integrals with the
finite limits; for instance, near the triangular side
$\sigma_1=\sigma_1^+(\sigma_2)$ where we have the ``creeping'' points
for the 3DPO inside the tori (\ref{sigmalim}) and the meridian
elliptic 2DPO near the end point ($\sigma_1=\sigma_1^+$, $\sigma_2=0$)
with a large number of the vertices $n_v \rightarrow
\infty$.  Another sample of such special bifurcation point is the
separatrix ($\sigma_1=1, \sigma_2=0$) where 3DPO and hyperbolic
2DPO have a finite limit $n_u/n_v \rightarrow 1/2$ for $n_v
\rightarrow \infty$ and $n_u \rightarrow \infty$.  In this case the
curvature $K_{11}$ becomes infinite and the amplitude
(\ref{amp3d2d}) approaches zero.  Thus, to improve the trace
formula near the bifurcations, we have to evaluate the generalized
error integral $\erf({\cal Z}_i^{\beta\,-},{\cal Z}_i^{\beta\,+})$
(or corresponding complex Fresnel functions\cite{abramov}) in
Eq.~(\ref{amp3d2d}) within the finite limits ${\cal
Z}_i^{\beta\,\pm}$ given by Eqs.~(\ref{argerroru}) or
(\ref{argerroru2}).

For the spheroidal cavity we have another bifurcation at the
spherical limit where the ``azimuthal'' Jacobian $J_{22}^\beta$
and $J_{12}^\beta$ (\ref{jacobupar}) ($\sigma_2 \propto
I_\varphi^2$) vanishes.\cite{mfimbrk} This is the reason for the
divergence of the standard POT result (\ref{amp3d2das}) in the
spherical limit.  Our improved trace formula (\ref{amp3d2d}) is
finite in the spherical limit, because the ``azimuthal''
generalized error function $\erf({\cal Z}_2^{\beta\,-}, {\cal
Z}_2^{\beta\,+})$ is proportional to $\sqrt{J_{22}^\beta}$ in
this limit and thus this ``azimuthal'' Jacobian is exactly
canceled by that coming from the denominator of
Eq.~(\ref{amp3d2d}).  Thus, as shown in Ref.~\citen{mfimbrk}, the
elliptic 2DPO term (${\cal K}$=2) in the level density approach
the spherical Balian-Bloch result for the most degenerate planar
orbits with larger degeneracy (${\cal K}=3$),
\begin{eqnarray}
\delta g_{\rm sph}^{(3)}(\eps) &=&
\frac{\sqrt{kR}}{\eps_0}\sum_{t \geq 1,\,q>2t}
\sin\left(\frac{2 \pi t}{q}\right)
\sqrt{\frac{\sin(2\pi t/q)}{q\pi}} \nonumber\\
&& \times\sin{\left[2kR q\sin\left(\frac{\pi t}{q}\right)-
\frac{3\pi}{2}q-(t-1)\pi-\frac{\pi}{4}\right]},
\label{sph3}
\end{eqnarray}
where $t=M n_u$ and $q=M n_v$. Note that Eq.~(\ref{sph3}) can be
derived directly from the phase-space trace formula
(\ref{pstraceactang}) or from the Poisson-sum trace formula, both
rewritten in terms of the spherical action-angle variables.

\subsection{Equatorial-plane orbits}

We cannot apply the Poisson-sum trace formula (\ref{poissonsum})
for equatorial-plane orbits, because,
although the stationary-phase conditions for $\Theta_\varphi''$
and $\Theta_v''$  in Eq.~(\ref{statcondt}) are identities,
it is not the case for the angle variable $\Theta_u''$.
We thus apply the ISPM for the integration over $\Theta_u''$.

Coming back to Eq.~(\ref{pstraceactang1}) we transform the
phase-space trace formula to new ``parallel'' ($\Theta_v'';I_v'$)
and ``perpendicular'' ($\Theta_u'',\Theta_\varphi'';I_u',I_\varphi'$)
variables as explained in Appendix~\ref{app:trace_eq}
for more general (integrable and non-integrable) systems.
We then make the integration over the
($I_u',I_\varphi'$) variables in terms of the ISPM by transforming
them into the $\sigma_i$ variables.
Next, we consider the
integration over the {\em angle} variable $\Theta_u''$ by the ISPM,
as there is the {\em isolated} stationary point
$\Theta_u^*=0$ or integer multiple of $2\pi$.
We expand the exponent phase in a power series of $\Theta_u''$
about $\Theta_u^*=0$,
\begin{eqnarray}
S_\alpha\left(\bI,\bI'',t_\alpha\right)+
\left(\bI''-\bI\right)\cdot \bTheta'' \nonumber\\
&& \hspace{-12em}= pL_{\rm EQ}
+\frac12 \sum_{ij}J_{ij}^{\rm EQ}(\sigma_i-\sigma_i^*)(\sigma_j-\sigma_j^*)
+\frac12 J_\perp^{\rm EQ}~
\left(\Theta_u''\right)^2+ \cdots,
\label{actionexptu}
\end{eqnarray}
where the stationary point $\sigma_1^*=\sigma_2^*\equiv\sigma^*$ is
given by
\begin{equation}
\sigma^*=
\left(\frac{I_\varphi^*}{p \zeta}\right)^2
=\frac{a^2 \cos^2\phi}{\zeta^2}
=\frac{\cos^2\phi}{\eta^2-1}, \qquad
I_\varphi^*=p\,a\cos\phi.
\label{sigmaperpstar}
\end{equation}
$L_{\rm EQ}$ is the length of the equatorial polygon with $n_v$
vertices and $M$ rotations, and is given by
\begin{equation}
L_{\rm EQ}=2 M n_v R \sin \phi, \quad \phi=\pi n_\varphi/n_v.
\end{equation}
In this way one finally obtains the amplitudes $A^{(1)}_{\rm EQ}$
for EQPO,
\begin{equation}
\delta g_{\rm EQ}^{(1)}(\eps)=\frac{1}{\eps_0}\Re\sum_{\rm EQ}
A^{(1)}_{\rm EQ}\exp\left\{i\left(kL_{\rm EQ}-\frac{\pi}{2}
\nu_{\rm EQ}\right)\right\},
\label{pstraceEQ}
\end{equation}
\begin{eqnarray}
A_{\rm EQ}^{(1)}= \sqrt{\frac{\sin^3\phi}{\pi Mn_v k R \eta
F_z^{\rm EQ}}}\: \erf\left({\cal Z}_{1}^-,{\cal Z}_{1}^+\right)\,
\erf\left({\cal Z}_2^-,{\cal Z}_2^+\right)\, \erf\left({\cal
Z}_3^-,{\cal Z}_3^+\right), \label{ampEQ}
\end{eqnarray}
see Appendix~\ref{app:trace_eq} for a detailed derivation.  Here,
${\cal Z}_i^\pm$ are the limits given by
Eqs.~(\ref{argerroru}) or (\ref{argerroru2}) for $i=1,2$, and
${\cal Z}_3^-=0, {\cal Z}_3^+={\cal Z}_\perp^+$ from
Eq.~(\ref{limitsEQ3}).
The latter is related to the finite limits $0 \leq \Theta_u \leq
\pi /2$ for the angle $\Theta_u$ in the trace integration in
Eq.~(\ref{pstraceactang1}), taking into account explicitly the
factor 4 due to the time-reversal and spatial symmetries.

For the total asymptotic phase $\nu_{\rm EQ}$, one finds
\begin{equation}
\nu_{\rm EQ}=\mu_{\rm EQ}+\frac12, \quad
\mu_{\rm EQ}=3M n_v,
\label{masloveq}
\end{equation}
where $\mu_{\rm EQ}$ is the Maslov index.  We calculated this
phase using the Maslov-Fedoryuk theory\cite{masl:fed} at a point
asymptotically far from the bifurcations.  Note that the total
phase is defined as the sum of the asymptotic phase $\nu_{\rm
EQ}$ and the argument of the amplitude $A_{\rm EQ}$,
Eq.~(\ref{ampEQ}), so that it depend on $kR$ and $\eta$ through
the complex arguments of the product of the error functions.  In
the derivations of Eq.~(\ref{ampEQ}) we have taken into account
the off-diagonal curvature as in the previous subsection, but
much smaller corrections due to the mixed derivatives of the
action $S_\alpha$ with respect to $\Theta_u''$ and $\sigma_i$ are
neglected, taking $\sigma_i=\sigma_i^*$ in
Eq.~(\ref{actionexptu}).

The bifurcation points are associated with zeros of the stability
factor $F_z^{\rm EQ}$ and given by
\begin{equation}
\eta_{\rm bif}=\frac{\sin\phi}{\sin\left(n\phi /M\right)},
\qquad n=1,2,\cdots, M.
\label{etabif}
\end{equation}
The bifurcation points most important
for the superdeformed shell structure are listed in Table~\ref{table:bifpts}.
\begin{table}[t]
\caption{\label{table:bifpts}
Bifurcation points of some short periodic orbits.}
\begin{center}
\begin{tabular}{c@{\quad}l|c@{\quad}l}
\hline\hline
periodic orbit & ~$\eta_{\rm bif}$\qquad &
periodic orbit & ~$\eta_{\rm bif}$\qquad \\
\hline
(4,2,1) & $\sqrt{2}$ & (6,3,1) & 2        \\
(5,2,1) & 1.618...   & (7,3,1) & 2.247... \\
(6,2,1) & $\sqrt{3}$ & (8,3,1) & 2.414... \\
(7,2,1) & 1.802...   & (9,3,1) & 2.532... \\
(8,2,1) & 1.848...   &         &          \\
\hline
\end{tabular}
\end{center}
\end{table}

When the stationary points are located inside the finite
integration region far from the ends, we transform the error
functions in Eq.~(\ref{ampEQ}) into the Fresnel ones and extend
their arguments to $\pm\infty$, except for the case when the lower
limit is exactly zero.  According to the definitions of the limit,
Eqs.~(\ref{argerroru}) and (\ref{limitsEQ3}), for ${\cal
Z}_i^\pm$, we have asymptotically ${\cal Z}_i^+ \rightarrow
+\infty ~(i=1,2,3)$, ${\cal Z}_1^-={\cal Z}_3^- \rightarrow 0$,
${\cal Z}_2^- \rightarrow 0$ for diameters and ${\cal Z}_2^-
\rightarrow -\infty$ for the other EQPO.  Finally, we arrive at the
standard Balian-Bloch formula\cite{bablo} for the
amplitude $A^{(1)}_{\rm EQ}$,
\begin{eqnarray}
A^{(1)}_{\rm EQ}=\frac{f_{\rm EQ}}{\sqrt{\pi kR \eta}}\,
\sqrt{\frac{\sin^3\phi}{M n_v F_z^{\rm EQ}}}
\label{ampgeqbb}
\end{eqnarray}
where $f_{\rm EQ}=1$ for the diameters and 2 for the other EQPO
($\erf({\cal Z}_2^-,{\cal Z}_2^+) \rightarrow f_{\rm EQ}$ in this
limit).

As seen from Eq.~(\ref{ampgeqbb}), there is a divergence at the
bifurcation points where $F_z^{\rm EQ} \rightarrow 0$.  We
emphasize that our ISPM trace formula (\ref{pstraceEQ}) has no
such divergences.  Indeed, the stability factor $F_z^{\rm EQ}$
responsible for this divergence is canceled by $F_z^{\rm EQ}$ from
the upper limit ${\cal Z}_3^+$, Eq.~(\ref{limitsEQ3}), of the last
error function in Eq.~(\ref{ampEQ}), ${\cal Z}_3^+ \propto
\sqrt{F_z^{\rm EQ}}$~, and we obtain the finite result at the
bifurcation point:
\begin{eqnarray}
A_{\rm EQ}^{(1)}=
\frac{\eta^{1/3}\sin\phi\sqrt{\eta^2-\sin^2\phi}}{\sqrt{2i(\eta^2-1)}M n_v}\:
\erf\left({\cal Z}_1^-,{\cal Z}_1^+\right)\,
\erf\left({\cal Z}_2^-,{\cal Z}_2^+\right).
\label{ampeqbif}
\end{eqnarray}

It is very important to note that there is a local enhancement of the
amplitude (\ref{ampeqbif}) by a factor of order $\sqrt{kR}$~%
\footnote{
The parameter of our semiclassical expansion is in practice
$\sqrt{kL_\beta}\left(\propto\sqrt{kR}\right)$.  It is actually large
for 3D orbits ($L_\beta\sim 10R$) associated with superdeformed shell
structures in nuclei.}
near the bifurcation point.  This enhancement is associated with a
change of the degeneracy parameter ${\cal K}$ by one unit locally near
the bifurcation point and results from exactly carrying out one
integration more than in the SSPM case.  In general, any change of the
degeneracy parameter ${\cal K}$ by $\Delta {\cal K}$ is accompanied by
the amplitude enhancement by $(kR)^{\Delta {\cal K}/2}$ because of the
$\Delta {\cal K}$ extra exact integrations.  These enhancement
mechanism of the amplitude obtained in the ISPM is quite general and
is independent of the specific choice of the potential shapes.

We mention that a more general trace formula which can be applied
also to non-integrable but axially symmetric systems
can be derived from the phase-space trace formula (see
Appendix~\ref{app:trace_eq}).

The contribution of the equatorial diameters in
Eq.~(\ref{pstraceEQ}) for deformations far from bifurcation points
reduces to the Balian-Bloch result for the spherical
diameters (${\cal K}=2$),
\begin{equation}
\delta g_{\rm sph}^{(2)}(\eps) = -\frac{1}{\eps_0}
\sum_M \frac{1}{2 \pi M}\,\sin(4MkR).
\label{sph2}
\end{equation}
The amplitude for the planar polygons in the equatorial plane
vanish in the spherical limit (see Appendix~\ref{app:trace_eq}).
Note that the contributions of the
planar polygons in the spherical cavity, Eq.~(\ref{sph3}),
are obtained as the limit of $A_{\rm 2D}^{(2)}$, Eq.~(\ref{amp3d2d}),
for elliptic orbits in the meridian plane.\cite{mfimbrk}

\subsection{Long diametric orbits and separatrics}

As mentioned in \S\ref{sec:sphercav}, the curvatures
$K_{ij}^\beta$ become infinite near the separatrix
($\sigma_1=1,\sigma_2=0$), see Appendix \ref{app:separatrix}.  This
separatrix corresponds to the isolated long diameters (${\cal
K}=0$) along the symmetry axis.  Thus, for the derivation of their
contributions to the trace formula, the expansion up to the second
order in action-angle variables considered above fails like for
the turning and caustic points in the usual phase space
coordinates.  However, we can apply the Maslov-Fedoryuk
theory \cite{fed:jvmp,masl,fed:spm,masl:fed} in a similar way as
the calculation of the Maslov indices associated with the turning
and caustic points but with the use of the action-angle variables
in place of the usual phase space variables.  This is similar to
the derivation of the long diametric term in the elliptic
billiard.\cite{mfammsb}

Starting from the phase space trace formula (\ref{pstraceactang1})
we note that the spheroidal separatrix problem differs from the
one for the elliptic billiard \cite{mfammsb} by the integrals over
the two azimuthal variables $\Theta_\varphi''$ and $I_\varphi'$
which are additional to the integrals over $\Theta_u''$ and
$I_u'$.  We expand the phase of exponent in
Eq.~(\ref{pstraceactang1}) with respect to the action $I_\varphi'$
and angle $\Theta_\varphi''$ about the stationary points
$I_\varphi^*=0$ and an arbitrary $\Theta_\varphi^*$ (for instance,
$\Theta_\varphi^*=0$), and take into account the {\it third} order
terms, in a similar way as for the variables $\Theta_u''$ and
$I_u'$ (see Appendix~\ref{app:separatrix}).  Note that we consider
here small deviations from the long diameters, and
$\Theta_\varphi^*$ determines the azimuthal angle of the final
point $\br''$ of this trajectory near the symmetry axis.

After the procedure explained in Appendix~\ref{app:separatrix}, we obtain
\begin{eqnarray}
\delta g^{(0)}_{\rm LD}(\eps) &=&
\frac{\pi b}{2\eps_0 R}\Re\sum_M \frac{1}{kR}\:
e^{ik L_{\rm LD} - i\frac{\pi}{2}\nu_{\rm LD}}
\prod_{j=1}^2 e^{\frac{2i}{3}\left[
 (w_j^\parallel)^{3/2}+(w_j^\perp)^{3/2}\right]}
\frac{\left(w_j^\parallel w_j^\perp\right)^{1/4}}
   {\sqrt{|c_{2,j}^\parallel c_{2,j}^\perp|}} \nonumber\\
&& \times [\Ai(-w_j^\parallel)+i\Gi(-w_j^\parallel)] \nonumber\\
&& \times [\Ai(-w_j^\perp,{\cal Z}_\perp^-,{\cal Z}_\perp^+)
   +i\Gi(-w_j^\perp,{\cal Z}_\perp^-,{\cal Z}_\perp^+)]
\label{deltagld}
\end{eqnarray}
(see Appendix~\ref{app:separatrix} for notations used here).

For finite deformations
and sufficiently large $kR$, i.e. for large
$p\zeta \propto kR \sqrt{\eta^2-1}$, near the separatrix
$\sigma_1 \rightarrow 1, \sigma_2 \rightarrow 0$,
the incomplete Airy functions in this equation can be approximated by
the complete ones.  Thus, Eq.~(\ref{deltagld}) reduces to
the standard Gutzwiller's
result for the isolated diameters,\cite{bablo,mfimbrk}
\begin{eqnarray}
\delta g^{(0)}_{\rm LD}(\eps)=
\frac{2 b}{\pi\eps_0 kR^2} \sum_M \frac{1}{|F_{xy}^{\rm LD}|}
\cos\left[k L_{\rm LD}(M) - \frac{\pi}{2} \nu_{\rm LD}\right],
\label{deltaglgutz}
\end{eqnarray}
with the length $L_{\rm LD}(M)=4Mb=4M\eta^{2/3}R$ and the
stability factor $F_{xy}^{\rm LD}$ for long diameters given by
Eq.~(\ref{gutzstabfactld}).

For the calculation of the asymptotic phase $\nu_{\rm LD}$
we use this asymptotic expression and calculate the Maslov indices
$\mu_{\rm LD}$ by the Maslov-Fedoryuk theory,\cite{masl:fed}
\begin{equation}
\nu_{\rm LD}=\mu_{\rm LD}+2, \quad \mu_{\rm LD}=4M.
\label{maslLD}
\end{equation}
The additional phases, dependent on deformation and energy,
come from the arguments of the complex exponents and Airy
functions of the complex amplitude.

In the spherical limit, both the upper and lower limits of
the incomplete Airy functions in Eq.~(\ref{deltagld}) approach
zero and the angle integration has the finite limit $\pi/2$ (see
Appendix~\ref{app:separatrix}).  With this, the other factors
ensure that the amplitude for long diameters becomes zero; namely,
the long diametric contribution to the level density vanishes in
the spherical limit.

\section{Level density, shell energy and averaging}
\label{sec:avdensesc}

\subsection{Total level density}
\label{sec:totlevdens}

The total semiclassical level density improved at the bifurcations
can be written as a sum
over all periodic orbit families in the spheroidal cavity considered in
the previous section,
\begin{equation}
\delta g_{\rm scl}(\eps)=
\delta g_{\rm 3D}^{(2)}(\eps)+
\delta g_{\rm 2D}^{(2)}(\eps)+
\delta g_{\rm EQ}^{(1)}(\eps)+
\delta g_{\rm LD}^{(0)}(\eps)=
\sum_\beta \delta g_{\rm scl}^{(\beta)}(\eps),
\label{tracetotal}
\end{equation}
where the first two terms represent the contributions from the
most degenerate (${\cal K}=2$) families of periodic orbits,
the 3DPO and the meridian-plane 2DPO, given by Eq.~(\ref{deltag3d2d}),
the third term the EQPO given by Eq.~(\ref{pstraceEQ}),
and the fourth term the long diameters given by Eq.~(\ref{deltagld}).

\subsection{Semiclassical shell energy}
\label{sec:enshellcorr}

The shell energy $\delta E$ can be expressed in terms of
the oscillating part $\delta g_{\rm scl}^{(\beta)}(\eps )$
of the semiclassical level density (\ref{tracetotal})
as\cite{strumag,mfimbrk,book}
\begin{equation}
\delta E = \sum_{\beta} \left(\frac{\hbar}{t_{\beta}}\right)^2
\delta g_{\rm scl}^{(\beta)}(\eps_\rF),\qquad
N=\int_0^{\eps_\rF} d\eps\:g(\eps).
\label{descl1}
\end{equation}
Here, $t_\beta$ denotes the period for a particle moving with the Fermi
energy $\eps_\rF$ along the periodic orbit $\beta$,
\begin{equation}
t_\beta = M T_\beta = \frac{2\pi M}{\Omega_{\beta}},
\label{tbeta}
\end{equation}
$T_\beta$ being the primitive period ($M=1$),
$M$ the number of repetitions,
and $\Omega_\beta$ the frequency.  The Fermi energy
$\eps_\rF$ is determined by the second equation of (\ref{descl1})
where $N$ is the particle number.

In the derivation of Eq.~(\ref{descl1}) we used an expansion of
the amplitudes $A_\beta(\eps)$
about the Fermi energy $\eps=\eps_\rF$.
Although $A_\beta(\eps)$
are oscillating functions of the energy $\eps$
(or $kR$), we can apply such an expansion,
because $A_\beta$ are much more smooth compared to the
oscillations coming from the exponent function of $kL_\beta$.
The latter oscillations are responsible for the shell structure,
while the oscillations of $A_\beta$
merely lead to slight modulations with much smaller frequencies.

Thus, the trace formula for $\delta E$ differs from that for
$\delta g$ only by a factor $({\hbar/t_\beta
})^2=(\hbar^2k_\rF/mL_\beta )^2$ near the Fermi surface, i.e.
longer orbits are additionally suppressed by a factor
$1/L_\beta^2$.  The semiclassical shell energy is therefore
determined by short periodic orbits.

\subsection{Average level density}
\label{sec:averdensity}

For the purpose of presentation of the level density improved at the
bifurcations we need to consider only an average level density, thus
also avoiding the convergence problems that usually arise when one is
interested in a full semiclassical quantization.

The average level density is obtained by folding the level density with
a Gaussian of width $\Gamma$:
\begin{equation}
g_\Gamma(\eps) =
   \frac{1}{\sqrt{\pi}\Gamma}\int_{-\infty}^{\infty} d\eps'\,
   g(\eps')\, e^{-(\frac{\eps-\eps'}
   {\Gamma})^2} \quad.
\label{folding}
\end{equation}
The choice of the Gaussian form of the averaging function is
immaterial and guided only by mathematical simplicity.

Applying now  the averaging procedure defined above
to the semiclassical level density (\ref{tracetotal}),
one obtains~\cite{bablo,mfimbrk}
\begin{equation}
\delta g_{\Gamma,{\rm scl}}(\eps ) =
\sum_\beta \delta g_{\rm scl}^{(\beta)}(\eps)\,
e^{-(\frac{\Gamma M T_\beta}{2\hbar})^2}=
\sum_\beta \delta g_{\rm scl}^{(\beta)}(\eps)\,
e^{-(\frac{\gamma L_\beta}{2R})^2}.
\label{dgsclgamma}
\end{equation}
The latter equation is written specifically for cavity problems in
terms of the orbit length $L_{\beta}$ (in units of a typical length
scale $R$) and a dimensionless parameter $\gamma$,
\begin{equation}
   \Gamma = 2\gamma\sqrt{\eps \eps_0}~,
                                                     \label{gammas}
\end{equation}
where $\gamma$
is a dimensionless quantity for averaging with respect to $kR$.
Thus, the averaging yields an exponential decrease of the amplitudes
with increasing $L_{\beta}$ and $\gamma $.
In Ref.~\citen{mfimbrk}, the $\gamma$ is chosen to be 0.6.
In this case, all longer orbits are strongly damped and only the
short periodic orbits contribute to the oscillating part of the level
density.  For the study of the bifurcation phenomena
in the superdeformed region,
we need a significantly smaller value of $\gamma$.

Finally, we can say that the higher the degeneracy of an orbit,
the larger the volume occupied by the orbit family in the phase
space, and the shorter its length,  the more important is its
contribution to the average density of states.

\section{Quantum Spheroidal Cavity}

\subsection{Oscillating level density}

We calculated the quantum spectrum by the spherical wave
decomposition method\cite{pal95} in which wave functions are
decomposed into the spherical waves as
\begin{equation}
\psi_m(\br)={\sum_l}' C_l\,j_l(kr)\,Y_{lm}(\Omega).
\label{eq:wf_swdm}
\end{equation}
Here, $m$ denotes the magnetic quantum number, and $\sum'$ means that
$l$ is summed over even(odd) numbers for positive(negative) parity
states.  $j_l$ and $Y_{lm}$ are the usual spherical Bessel function
and spherical harmonics, respectively.  The
expansion coefficients $C_l$'s are determined so that the wave
function (\ref{eq:wf_swdm}) satisfies the Dirichlet boundary
condition
\begin{equation}
\psi_m(r=R(\Omega))=0
\label{eq:dirichlet}
\end{equation}
or equivalently,
\begin{equation}
\int d\Omega Y^*_{lm}(\Omega)\psi_m(r=R(\Omega))=0,
\quad ^\forall l.
\label{eq:bdcond}
\end{equation}
By inserting (\ref{eq:wf_swdm}) into (\ref{eq:bdcond}), one obtains the
matrix equation
\begin{equation}
{\sum_{l'}}' B_{ll'}(k) C_{l'} = 0, \quad
B_{ll'}(k)=\int d\Omega Y^*_{lm}(\Omega)
 j_{l'}(kR(\Omega))Y_{l'm}(\Omega).
\end{equation}
Truncating the summation $l$ by sufficiently large number $l_c$, one
can obtain the energy eigenvalue $\eps_n=\hbar^2k_n^2/2m$ by
searching the roots
\begin{equation}
\det{B(k_n)}=0.
\end{equation}
\begin{figure}[t]
\sfig{0.6}{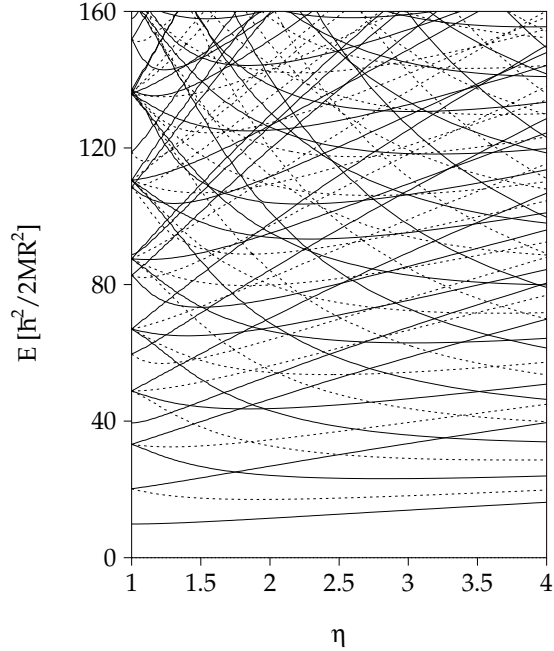}
\caption{\label{fig:level}
Single-particle spectrum for the spheroidal cavity as a function of
axis ratio $\eta$.  Solid and dashed curves represent the positive and
negative parity levels, respectively.}
\end{figure}
Figure~\ref{fig:level} shows the energy
level diagram for the prolate spheroidal cavity as functions of axis
ratio $\eta>1$.
\begin{figure}[t]
\sfig{0.6}{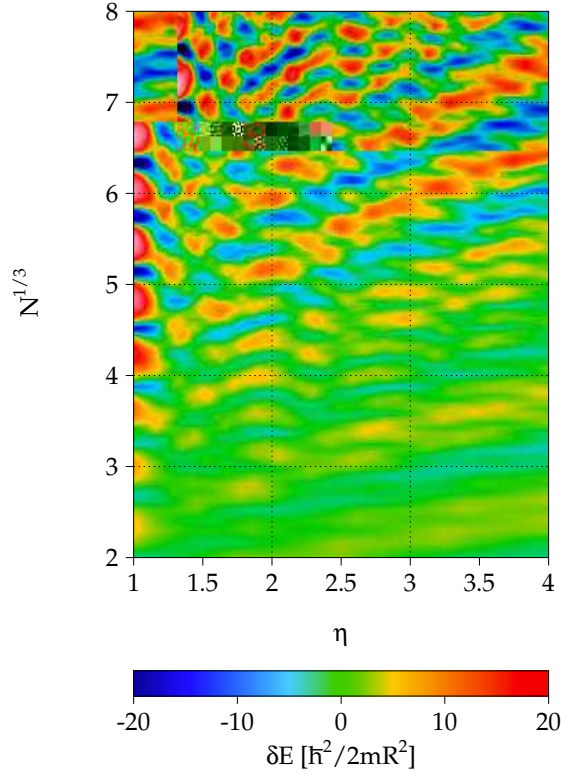}
\caption{\label{fig:energy}
Shell structure energy $\delta E$ as a function of $\eta$ and
$N^{1/3}$, where $N$ is the neutron (proton) number taking the
spin-degeneracy factor two into account.  Energies are counted in units of
$\hbar^2/2mR^2~ (\sim 30A^{-2/3}\mbox{MeV})$.}
\end{figure}
In Fig.~\ref{fig:energy}, we plot shell structure energy
\begin{equation}
\delta E(N,\eta) = \sum_{n=1}^N \eps_n(\eta) - \tilde{E}(N,\eta)
\end{equation}
as functions of $\eta$ and particle number $N$.  As well as the strong
shell effect at the spherical shape ($\eta=1$), one clearly see a
remarkable shell structure at the superdeformed shape ($\eta\sim 2$).

\begin{sidewaysfigure}[p]
\dfig{.8}{50 10 485 622}{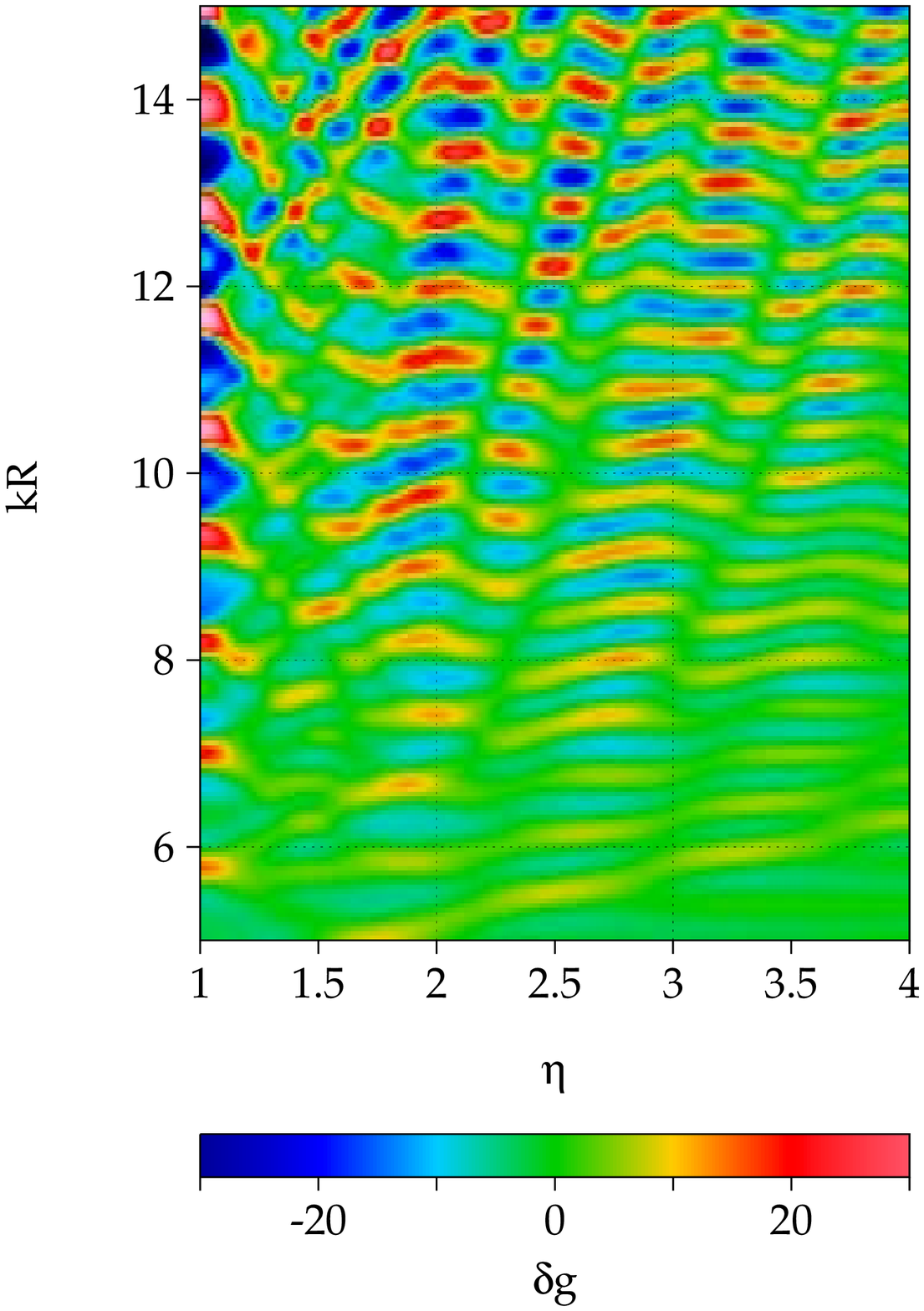}{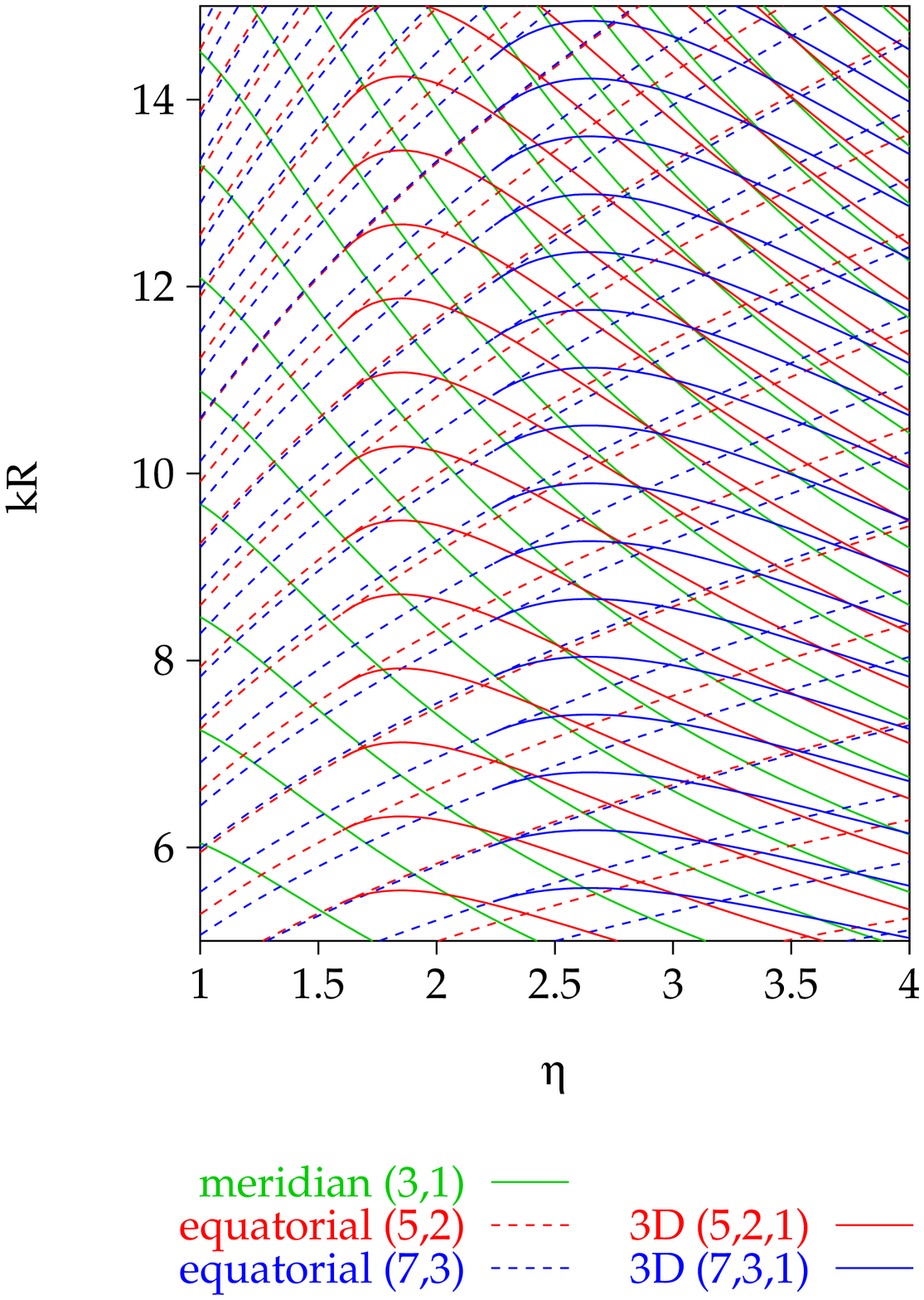}
\caption{\label{fig:density}
Oscillating part $\delta g$ of the single-particle level density
as a function of $\eta$ and $kR$ (left-hand side) and
stationary-action curves for several periodic orbits (right-hand
side).
Clear correspondence between the enhancement of the shell effect and
the periodic-orbit bifurcations is seen.}
\end{sidewaysfigure}

Next, we calculated the coarse-grained level density by the usual
Strutinsky smoothing procedure by taking wave number $k$ as
smoothing variable:
\begin{equation}
g_\gamma(k)=\frac{1}{\gamma}\int_0^\infty dk'R\,
f_M\left(\frac{kR-k'R}{\gamma}\right)g(k').
\end{equation}
As a smoothing function $f_M(x)$, we took a Gaussian with $M$-th order
curvature corrections
\begin{equation}
f_M(x)=\frac{1}{\sqrt{\pi}}e^{-x^2}L_{M/2}^{1/2}(x^2),
\end{equation}
where $L_n^{\alpha}(z)$ represents a Laguerre polynomial.
Eq.~(\ref{folding}) corresponds to the case of $M=0$.  In the
following, we took the order of curvature corrections $M=6$ and
smoothing width $\tilde{\gamma}=2.5$ with which we can nicely
satisfy the plateau condition.\cite{strut} The coarse-graining is
also performed by the same smoothing function but with smaller
$\gamma$.  We define the oscillating part of the level density by
subtracting the smooth part as
\begin{equation}
\delta g_\gamma(k)=g_\gamma(k)-g_{\tilde{\gamma}}(k).
\end{equation}
The left-hand side of Fig.~\ref{fig:density} shows $\delta
g_\gamma(k)$ with $\gamma=0.3$, as functions of $\eta$ and $kR$.
One will note that a remarkable shell structure emerges at $\eta \sim
2$, corresponding to the superdeformed shape.

Let us consider the mechanism of this strong shell effect.  If a single
orbit makes a dominant contribution to the periodic-orbit sum
\begin{eqnarray}
\delta g_{\rm scl}(\eps) = \sum_\beta
 a_\beta(k)\cos(kL_\beta-\pi\nu_\beta/2), \quad
 a_\beta(k)=A_\beta/\eps_0,
\label{posum}
\end{eqnarray}
the major oscillating pattern in $\delta g$ should be determined by
the phase factor of the dominant term.  In that case, the positions of
valley curves for $\delta g$ in the $(\eta, kR)$ plane are given by
\begin{equation}
kL_\beta-\pi\nu_\beta/2=(2n+1)\pi,
\qquad (n=0,1,2,\cdots).
\label{eq:sac}
\end{equation}
The right-hand side of Fig.~\ref{fig:density} shows the {\em stationary
action curves} (\ref{eq:sac}) for several periodic orbits.  Green
solid curves represent the triangular orbit in the meridian plane.  Other
longer meridian orbits also make the same behavior but with smaller
distances.
Red dashed curves represent the star-shaped orbit with five vertices
in the equatorial plane.  It causes bifurcates at $\eta=1.618\ldots$
and the 3D orbit (5,2,1)
appear (red solid curves).  The sequence ($n$,2,1) ($n=5,6,7,\cdots$)
make the same behavior shifting the bifurcation points a little bit to
larger $\eta$.  Comparing with the plot of quantum $\delta g$, one
notices a clear correspondence between the superdeformed shell
structure and the bifurcation of above star-shaped orbits.  One will
also note the correspondence between the bifurcations of the
equatorial-plane orbits ($n$,3) ($n=7,8,9,\cdots$) with the
hyperdeformed shell structure emerging at $\eta\simeq2.5$.  The
significant shell energy gain at the superdeformed shape obtained in
Fig.~\ref{fig:energy} is considered as a result of this strong shell
effect in the level density.

\subsection{Fourier analysis of level density}

Fourier analysis is a useful tool to investigate quantum-classical
correspondence in the level density.\cite{bablo}  Due to the simple
form of action integral $S_\beta=\hbar k L_\beta$, one can easily
Fourier transform the semiclassical level density $g_{\rm scl}(k)$
with respect to $k$.  Let us define the Fourier transform $F(L)$
by
\begin{equation}
F(L)=\int dk e^{-ikL}g(k).
\end{equation}
In actual numerical calculations, we multiply the integrand by
a Gaussian truncation function as
\begin{equation}
F_\Delta(L)=\frac{\Delta}{\sqrt{2\pi}}\int dk
 e^{-\frac12(k\Delta)^2} e^{-ikL} g(k).
\end{equation}
Inserting the semiclassical level density (\ref{posum}), the Fourier
transform is expressed as
\begin{equation}
F^{\rm scl}_\Delta(L)=\bar{F}_\Delta(L)+\frac12\sum_\beta
 e^{-i\pi\nu_\beta/2} a_\beta\left(i\pp{}{L}\right)
 \exp\left[-\frac12\left(\frac{L-L_\beta}{\Delta}
 \right)^2\right].
\label{eq:fourier_cl}
\end{equation}
This is a function which has peaks at the lengths of classical
periodic orbits $L=L_\beta$.  On the other hand, we can calculate
$F(L)$ by inserting the quantum mechanical level density
$g(k)=\sum_n \delta(k-k_n)$ as
\begin{equation}
F^{\rm qm}_\Delta(L)=\frac{\Delta}{\sqrt{2\pi}}\sum_n
 e^{-\frac12(k_n\Delta)^2} e^{-ik_n L}.
\label{eq:fourier_qm}
\end{equation}
It should present successive peaks at orbit lengths $L=L_\beta$.
Thus we can extract information on classical periodic
orbits from the quantum spectrum.
\begin{sidewaysfigure}[p]
\dfig{0.8}{50 10 485 622}{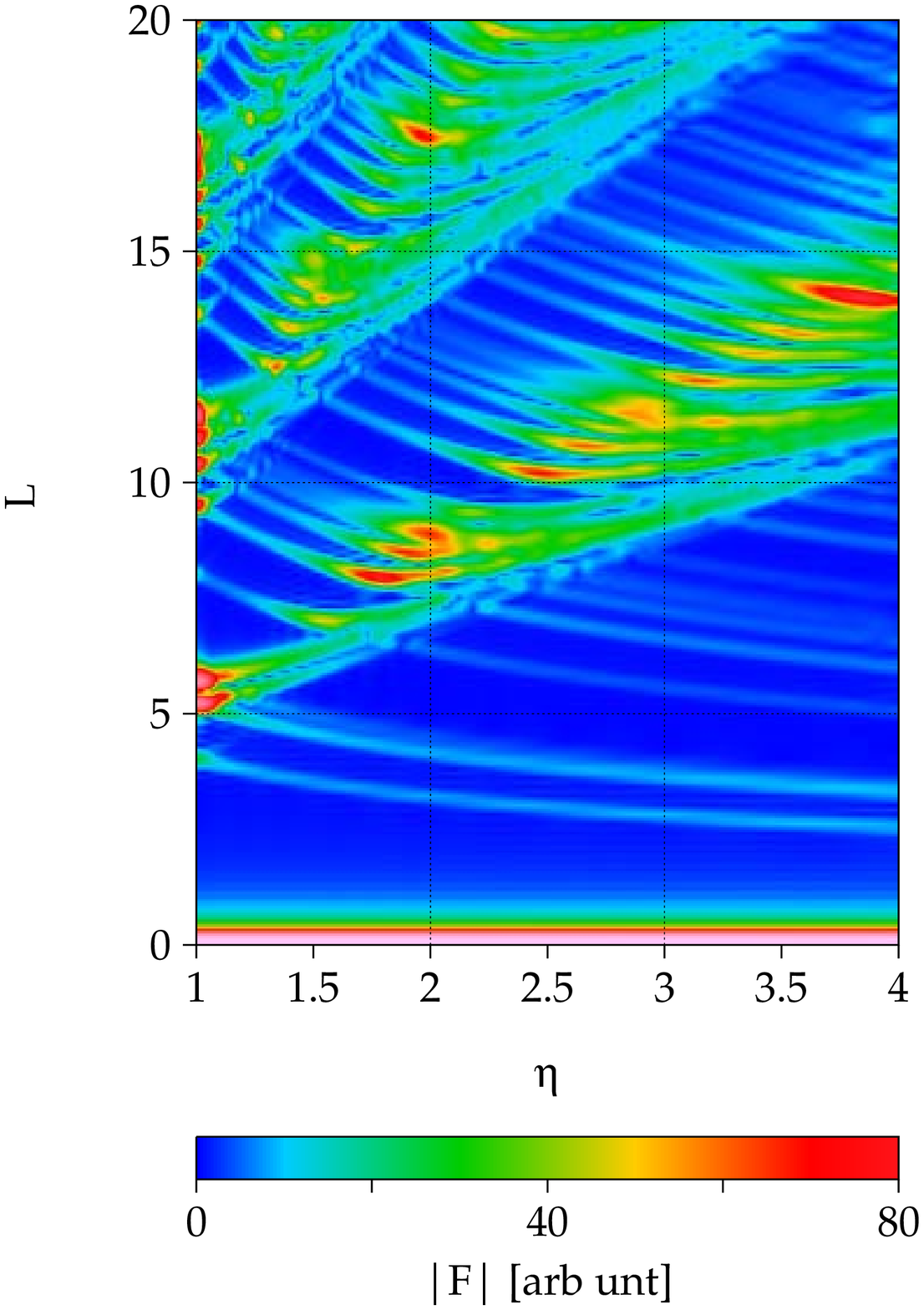}{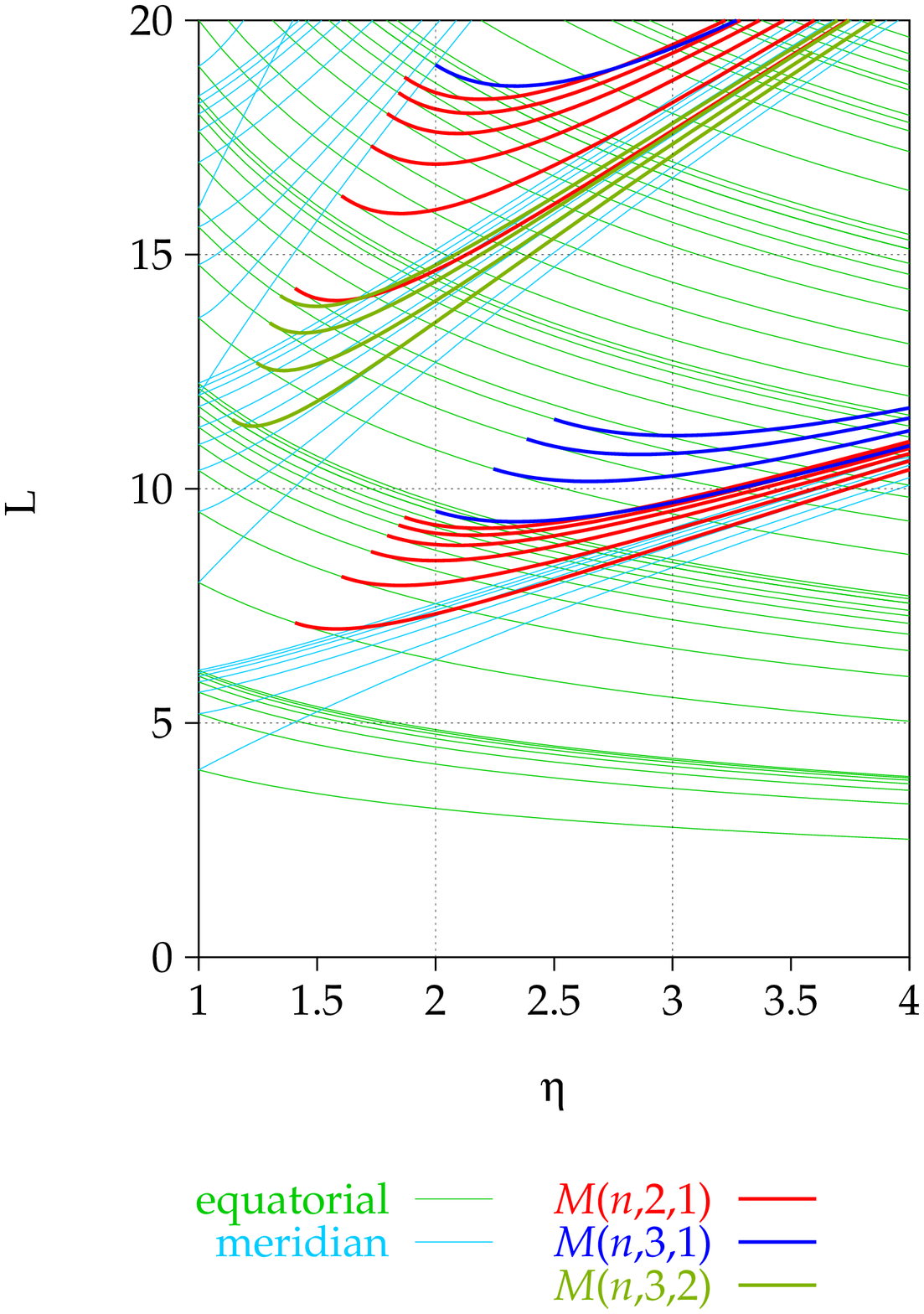}
\caption{\label{fig:fourier}
Fourier amplitude $|F(L)|$ of the single-particle level density
(left-hand side) and lengths of classical periodic orbits (right-hand
side).}
\end{sidewaysfigure}
In the left-hand side of
Fig.~\ref{fig:fourier}, we plot the Fourier transform
(\ref{eq:fourier_qm}) as a function of $L$ and $\eta$.  In the
right-hand side of Fig.~\ref{fig:fourier}, lengths of classical
periodic orbits $L_\beta(\eta)$ are shown.  Red curves represent the
orbits $M(n_v,2,1)$ $(n_v=4,5,6,\cdots)$.  We found strong Fourier
peaks at $\eta\simeq 2$ corresponding to the periodic orbits (5,2,1),
(6,2,1) and (7,2,1) just after the bifurcation points.  We also found
Fourier peaks at $\eta\simeq 2.5$ corresponding to the periodic orbits
(7,3,1) and (8,3,1) etc.  Thus, we can conclude that those periodic
orbit bifurcations play essential roles in the emergence of
superdeformed and hyperdeformed shell structures.

\subsection{Coarse-grained shell structure energy}

In order to prove that the shell structure at the superdeformed
shape is due to the bifurcated orbits, we calculated the
`coarse-grained' shell energy defined by
\begin{equation}
\delta\tilde{E}_\gamma(N)
=\int^{\tilde{k}_\rF(\gamma)}\eps(k)g_\gamma(k)dk
-\int^{\tilde{k}_\rF(\tilde{\gamma})}\eps(k)g_{\tilde{\gamma}}(k)dk,
\label{eq:dtilde_E}
\end{equation}
where the smoothed Fermi wave number $\tilde{k}_\rF$ in each term
is determined so that they satisfy the particle number condition
\begin{equation}
\int^{\tilde{k}_\rF(\gamma)}g_\gamma(k)dk
=\int^{\tilde{k}_\rF(\tilde{\gamma})}g_{\tilde{\gamma}}(k)dk
=N.
\label{eq:sfermi}
\end{equation}
By coarse-graining with width $\gamma$, a shell structure of
resolution $\Delta kR=\gamma$ is extracted.  Classical orbits relevant
for such a structure are those with lengths
\begin{equation}
L < L_{\rm max}=\frac{2\pi R}{\gamma}.
\end{equation}
Taking $\gamma=0.6$, contributions from periodic orbits with
$L\gsim 10R$ are smeared out.  Around the superdeformed shape,
bifurcated orbits have lengths $L\sim 10R$ and those contributions
are significantly weakened by smoothing with $\gamma=0.6$, and the
major oscillating pattern of $\delta E$ should disappear if those
bifurcated orbits are responsible for the superdeformed shell
effect.
\begin{figure}[t]
\sfig{0.6}{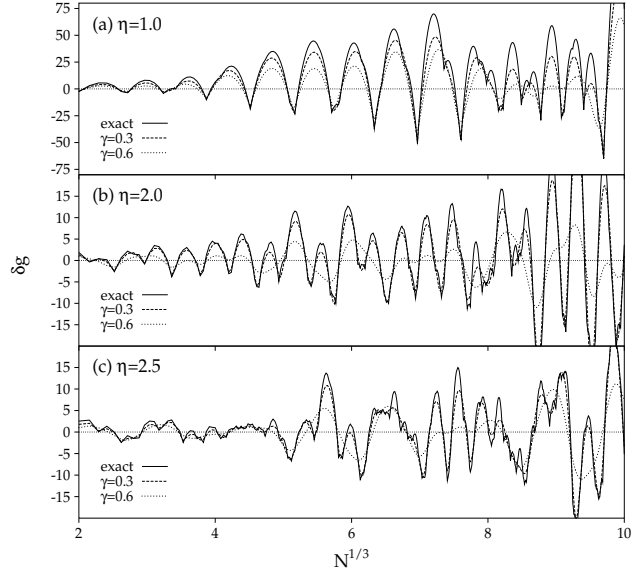}
\caption{\label{fig:scegam}
Shell structure energies plotted as functions of $N^{1/3}$.  Solid
curves represent the exact shell structure energies.  Dashed and
dotted curves represent those calculated by using the coarse-grained
level density $g_\gamma(k)$ with the smoothing width $\gamma=0.3$ and
$0.6$, respectively.}
\end{figure}
In Fig.~\ref{fig:scegam}, the coarse-grained shell
energies (\ref{eq:dtilde_E}) calculated for $\eta=1, 2, 3$, with
$\gamma=0.3$ and 0.6, are compared with the exact shell structure
energies.  In the upper panel, one observes that the spherical
shell structure survives after smoothing with $\gamma=0.6$,
indicating that the major structure is determined by orbits whose
lengths are sufficiently shorter than 10$R$.  On the other hand,
in the middle panel, one notices that the major oscillating
pattern of the superdeformed shell structure is considerably
broken after smoothing with $\gamma=0.6$.  The same argument is
valid also for $\eta=3$.  This strongly supports the significance
of bifurcated orbits for the superdeformed and hyperdeformed shell
structures.

\section{Enhancement of semiclassical amplitudes near the bifurcation points}

In this section, we present some results of the semiclassical ISPM
calculation, which clearly show enhancement phenomena of the
semiclassical amplitudes $|A_{\rm 3D}|$ and $|A_{\rm EQ}|$ near the
bifurcation points.

\begin{figure}[t]
\dfig{1}{88 45 554 340}{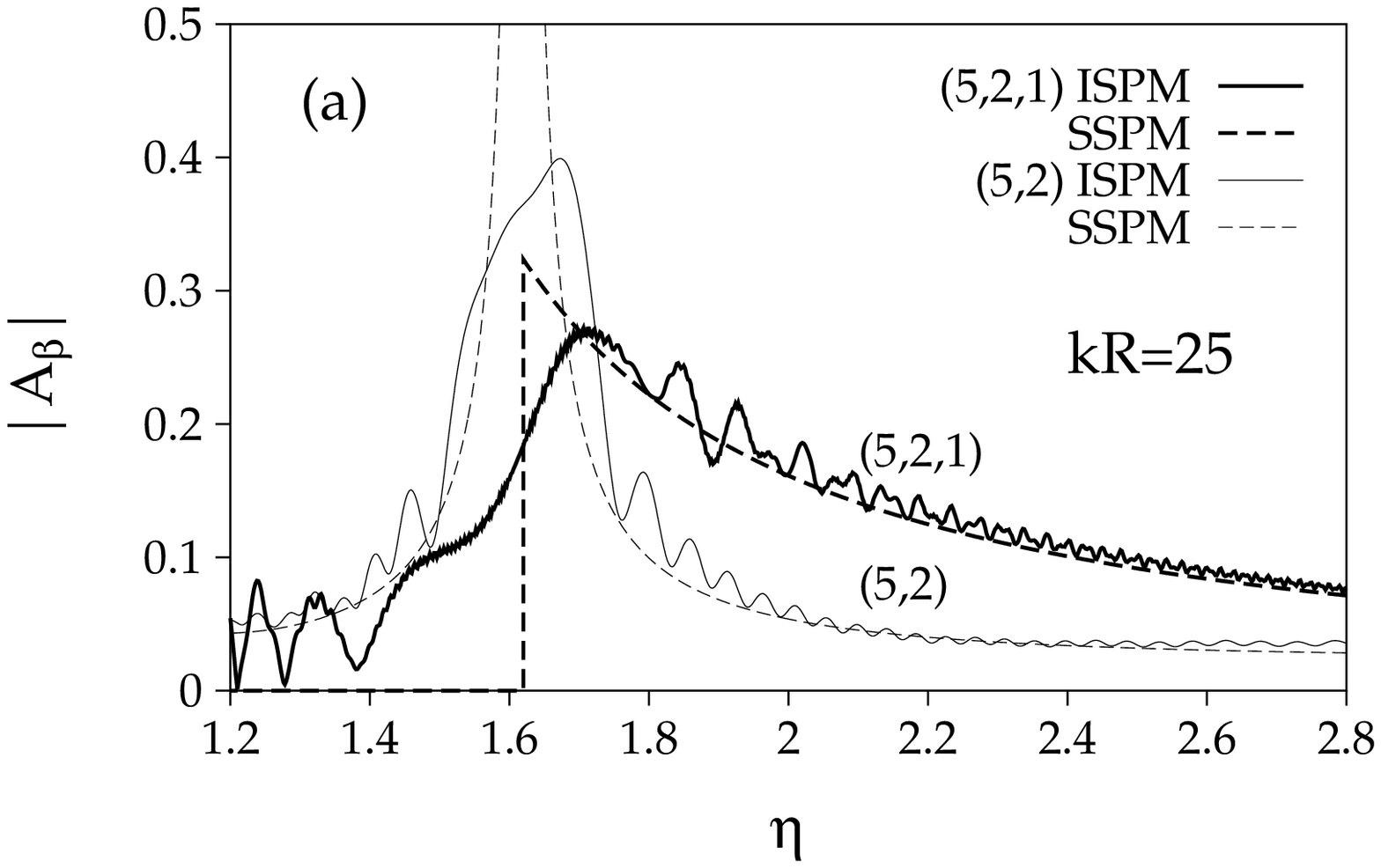}{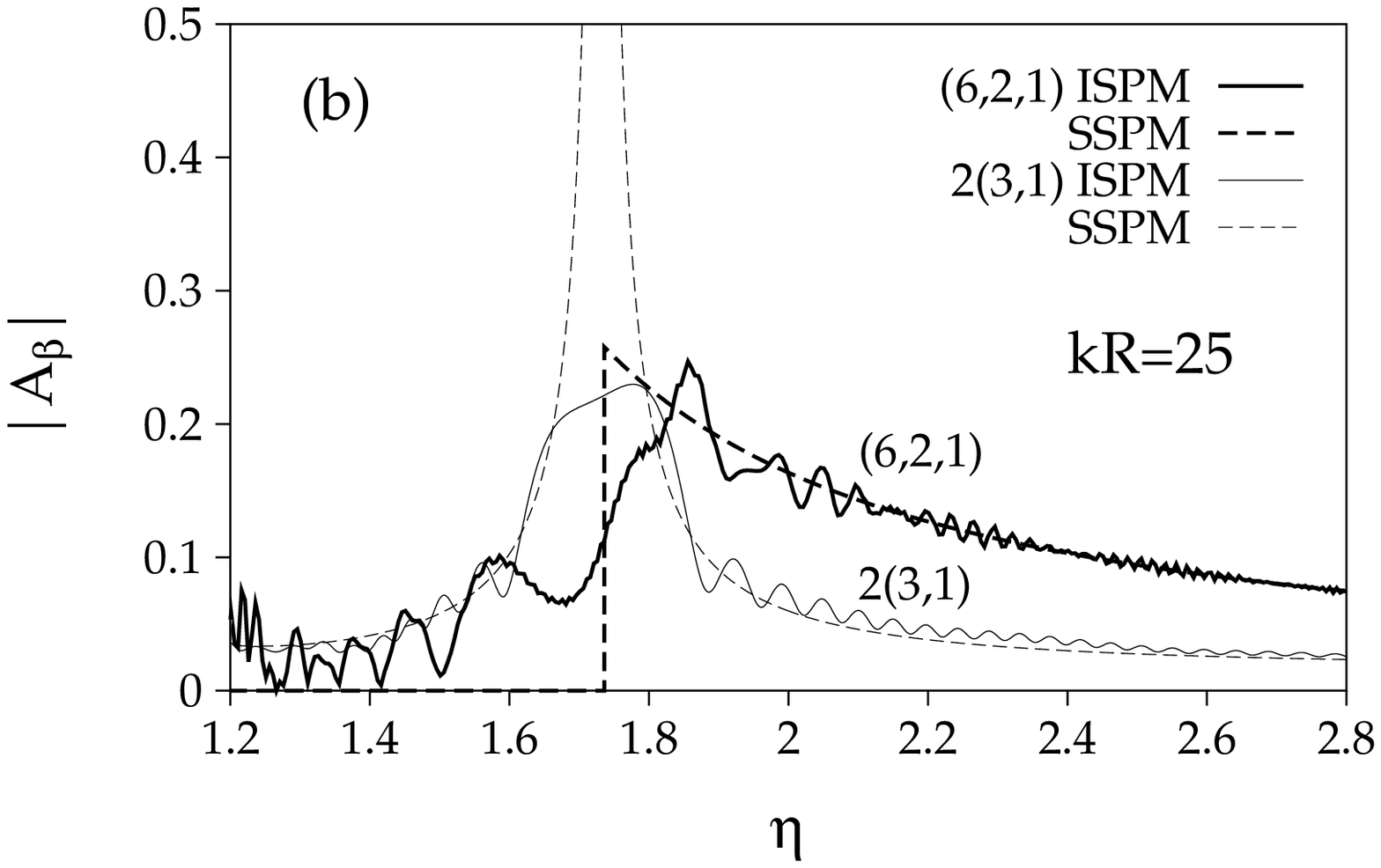}
\caption{\label{fig:sclamp} (a) Semiclassical amplitudes $|A_{\rm
3D}|$ for the 3DPO (5,2,1) and $|A_{\rm EQ}|$ for the EQPO (5,2),
calculated at $kR=25$ by the ISPM, are plotted as functions of the
deformation parameter $\eta$ by thick and thin solid curves,
respectively.  They are compared with the SSPM amplitudes
(dashed curves).
(b) The same as (a) but for the 3DPO (6,2,1) and the EQPO 2(3,1).}
\bigskip

\dfig{1}{88 45 554 340}{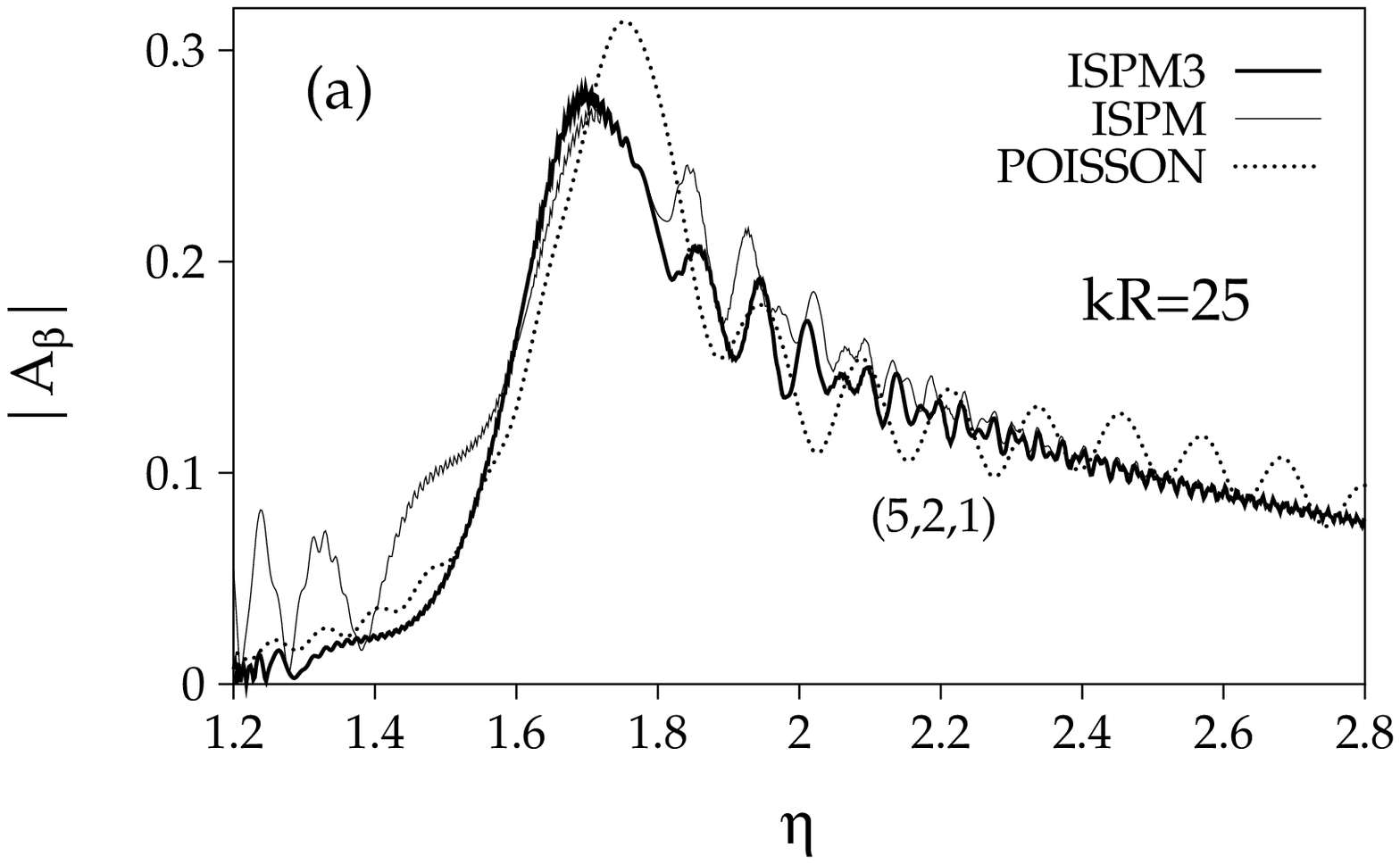}{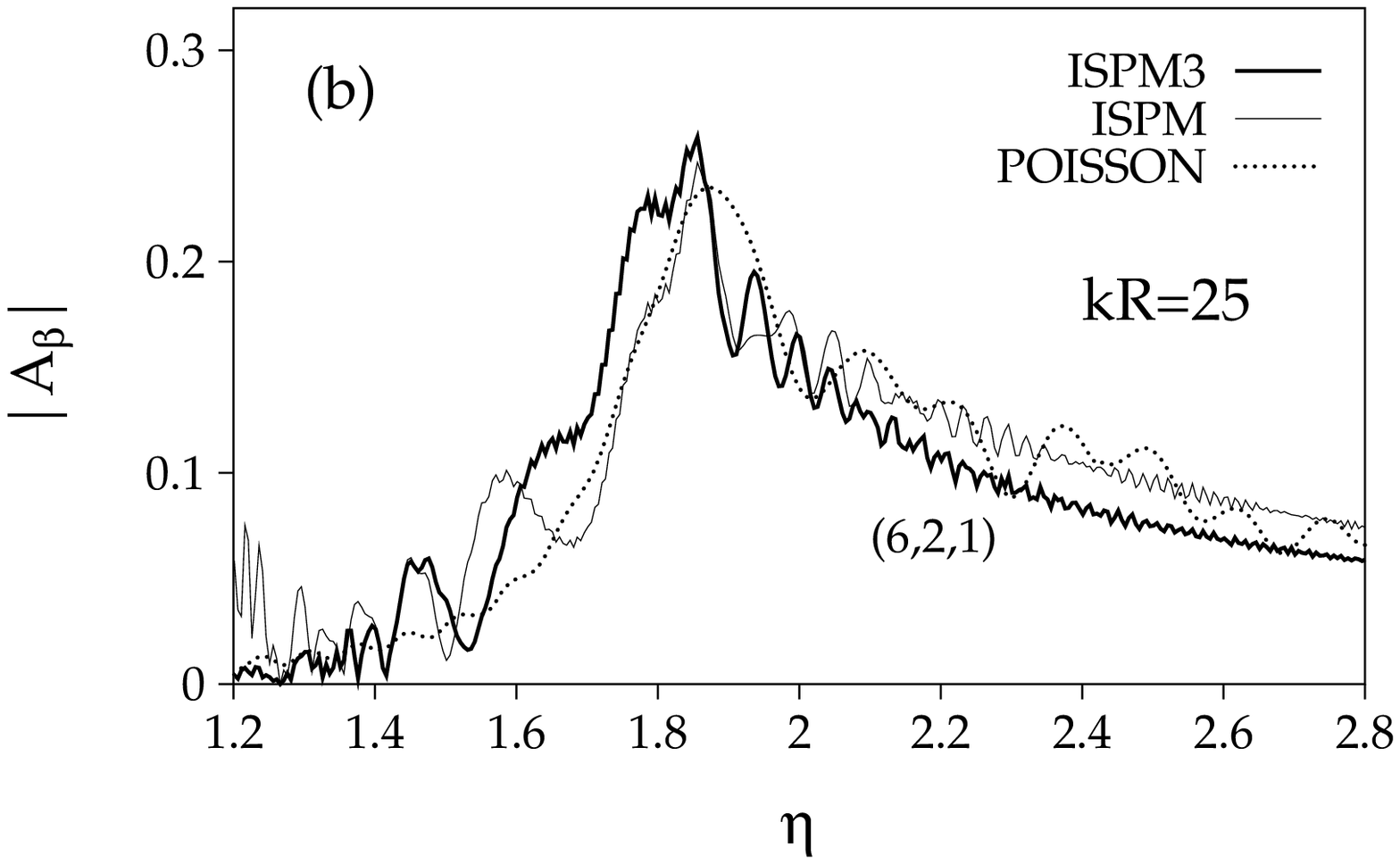}
\caption{\label{fig:sclamp_ispm3} (a) ISPM3 amplitudes for the 3DPO
(5,2,1), calculated at $kR=25$, are shown as function of deformation
$\eta$ by thick-solid curves.
For comparison, the ISPM amplitudes and the results of exact integration
in the Poisson-sum trace formula are plotted
with thin-solid and thick-dotted curves, respectively.
(b) The same as (a) but for the 3DPO (6,2,1).}
\end{figure}

Figure~\ref{fig:sclamp}a shows the modulus of the complex amplitude
$A_{\rm 3D}$ (Eq.~(\ref{amp3d2d})) for the 3DPO $(5,2,1)$ and $A_{\rm
EQ}$ (Eq.~(\ref{ampEQ})) for the EQPO $(5,2)$ as functions of the
deformation parameter $\eta$.  They are compared with those of the
SSPM.  The SSPM amplitude for the EQPO $(5,2)$ is divergent at the
bifurcation deformation $\eta_{\rm bif}=1.618\ldots$, while the ISPM
amplitude is finite and continuous through this bifurcation point with
a rather sharp maximum at this point.  This is due to a local change
of the symmetry parameter ${\cal K}$ from 1 to 2 at the bifurcation,
and the associated enhancement of the amplitude is of the order
$\sqrt{kR}$.  As seen from Fig.~\ref{fig:sclamp}a, the ISPM amplitude
for the $(5,2,1)$ is continuous through the bifurcation point and
exhibit a remarkable enhancement a little on the right of it.  It
approaches the SSPM amplitude given by Eq.~(\ref{amp3d2das}) away from
the bifurcation point.
The ISPM enhancement for the 3DPO is also of the order $\sqrt{kR}$,
because of the same change of the degeneracy parameter ${\cal K}$ from
1 to 2 as in the case of the bifurcating EQPO.
The same is true for the 3DPO (6,2,1) and the EQPO 2(3,1) as shown in
Fig.~\ref{fig:sclamp}b.

In Fig.~\ref{fig:sclamp_ispm3}, we consider corrections from the
3rd-order terms in the expansion of the action about the
stationary point. Here we incorporate the 3rd-order terms in the
$\sigma_1$ variable (ISPM3) which are expected to be important
for the 3DPO (6,2,1) whose curvature $K_{11}$ is identically zero
(see Appendix~\ref{app:3rd_order}).  We also show results of
exact integration in the Poisson-sum trace formula
(\ref{Poisvarsigma}) (marked POISSON).  One sees that the results
of the ISPM3 for the $(5,2,1)$ and $(6,2,1)$ orbits are in good
agreement with those of the ISPM in the most important regions
near the bifurcations and on the right-hand sides of them. It is
gratifying to see that the ISPM and the ISPM3 amplitudes $|A_{\rm
3D}|$ for $(5,2,1)$ and $(6,2,1)$ are also in good agreement with
the results of exact integration in the Poisson-sum trace
formula.  With the 3rd-order corrections, excessive ghost orbit
contributions in the ISPM (bumps in the ISPM amplitudes in the
left-hand side of the bifurcation point) are removed and better
agreement with the result of exact integration is obtained.
Except for that, the corrections due to the 3rd-order terms are
rather small, and good convergence is achieved up to the
second-order terms.

The amplitudes $|A_\beta|$ are slightly oscillating functions of $kR$.
Since the period of this oscillation is much larger than that of the
shell energy oscillation, one can use the expansion about the Fermi
energy $\eps_\rF$ (or $k_\rF R$) in the derivations of both the
semiclassical ISPM shell energy $\delta E_{\rm scl}$ and the
oscillating level density $\delta g_{\rm scl}$
(\ref{deltag3d2d}).
\begin{figure}[t]
\sfig{0.6}{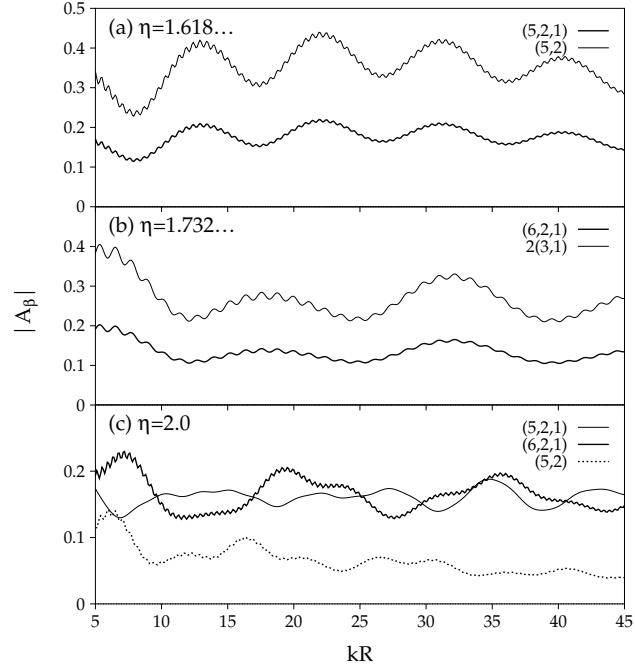}
\caption{\label{fig:sclamp3} (a) Semiclassical amplitudes $|A_{\rm
3D}|$ for the 3DPO (5,2,1) and $|A_{\rm EQ}|$ for the
EQPO (5,2) are plotted by bold and thin solid curves, respectively,
as functions of $kR$ at the bifurcation point $\eta=1.618\ldots$.
(b) The same as (a) but for the 3DPO (6,2,1) and the EQPO 2(3,1)
at the bifurcation point $\eta=1.732\ldots$.
(c) Semiclassical amplitudes $|A_{\rm 3D}|$ for (5,2,1),
(6,2,1) and $|A_{\rm EQ}|$ for (5,2) are plotted by thin-solid,
thick-solid and dotted curves, respectively, as functions of $kR$ at
$\eta=2.0$.}
\end{figure}
Figure~\ref{fig:sclamp3} shows the semiclassical
amplitudes $A_{\rm 3D}$ for the 3DPO (5,2,1) and $A_{\rm EQ}$ for the
EQPO $(5,2)$ as functions of $kR$ at $\eta=1.618\ldots$ (top panel)
and $\eta=2$ (bottom panel).  In this figure, the semiclassical
amplitudes $A_{\rm 3D}$ for the 3DPO (6,2,1) and $A_{\rm EQ}$ for the
EQPO 2(3,1) are also plotted as functions of $kR$ at the bifurcation
point $\eta=\sqrt{3}$ (middle panel).  We see that for $\eta=2$ the
amplitudes $|A_{\rm 3D}|$ for the 3DPO $(5,2,1)$ and $(6,2,1)$ become
much larger than the amplitude $|A_{\rm EQ}|$ for the EQPO.

\section{Comparison between quantum and semiclassical calculations}

In this section we present results of calculation of level
densities and shell energies with the use of the quantum
Strutinsky method and the semiclassical ISPM, and make comparisons
between the quantum and semiclassical calculations.  In the quantum
calculations, the averaging parameter $\gamma=0.3$ is used.

\begin{figure}[b]
\sfig{0.6}{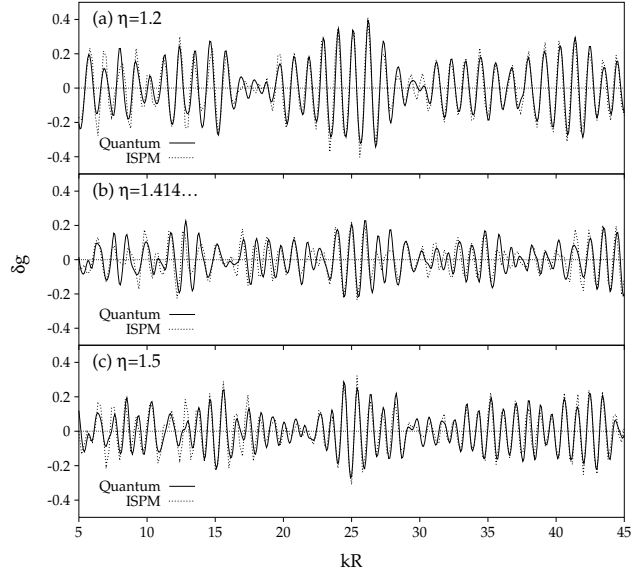}
\caption{\label{fig:scldens1} Oscillating level densities
evaluated by the semiclassical ISPM, and by the quantum
mechanics are shown by dotted and solid curves,
respectively, as functions of $kR$ for several deformations
$\eta$.}
\end{figure}

Figure~\ref{fig:scldens1} shows oscillating level densities $\delta g$
for relatively small deformations; QM and ISPM denote the $\delta g$
obtained by the quantum Strutinsky method and the semiclassical ISPM,
respectively.  For $\eta=1.2$ we obtain a good convergence of the
periodic orbit sum (\ref{tracetotal}) by taking into account the short
elliptic 2DPO with $n_v \leq 12$, $n_u=1$, the short EQPO with the
maximum vertex number $p_{\rm max}=M(n_v)_{\rm max}=14$, and the
maximum winding number $t_{\rm max}=M n_\varphi=1$
($M=1,n_\varphi=1$). The ISPM result is in good agreement with the
quantum result.  For the bifurcation point $\eta=\sqrt{2}$ of the
butterfly orbit $(4,2,1)$ and $\eta=1.5$ slightly on the right of it,
the convergence of the periodic-orbit sum is attained by taking into
account the contributions from the bifurcating orbits, $(4,2,1)$ and
the twice-repeated diameter $2(2,1)$ with $t_{\rm max}=2$, in addition
to the 2DPO and the EQPO considered in the $\eta=1.2$ case.

\begin{figure}[p]
\sfig{0.6}{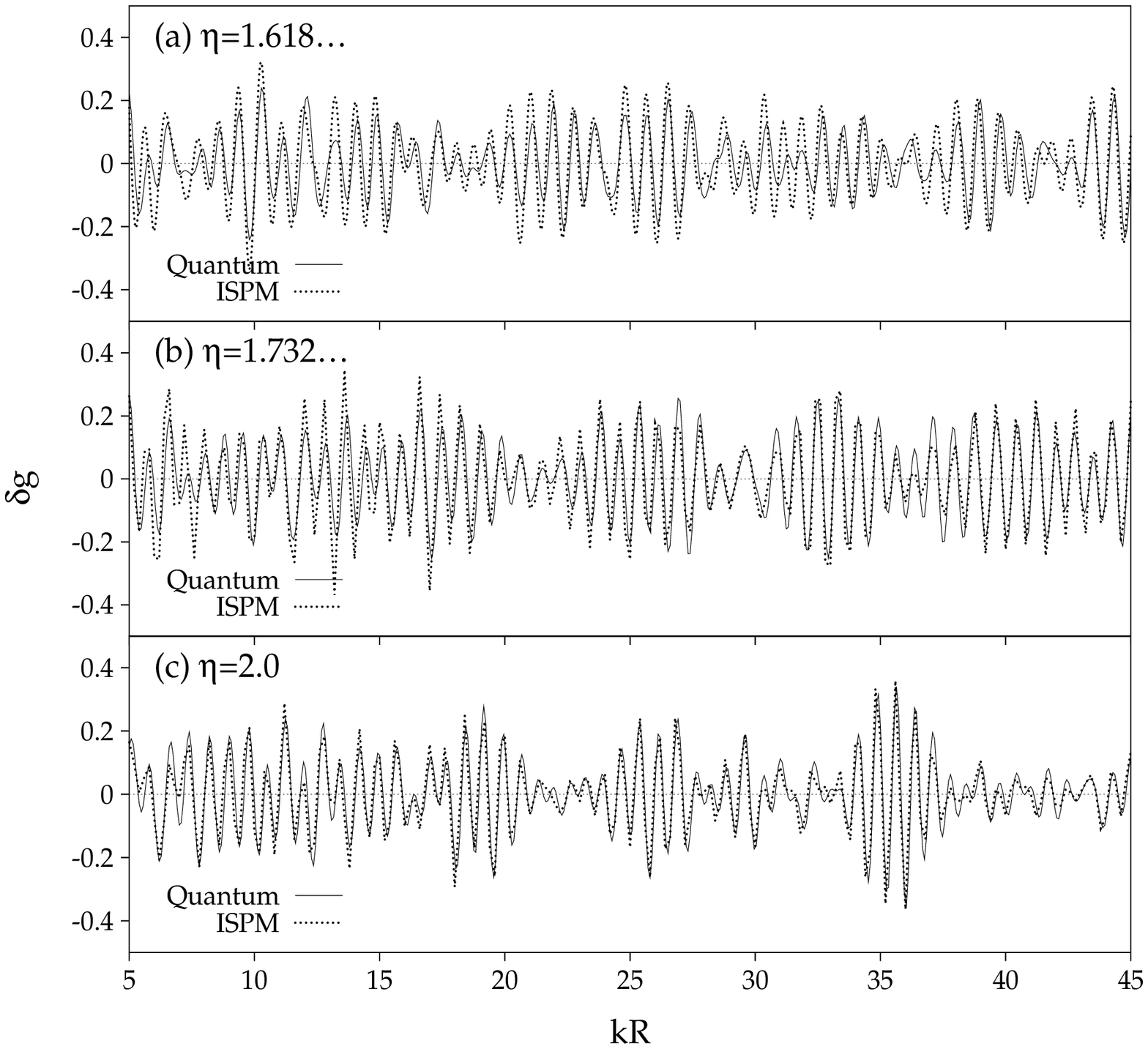}
\caption{\label{fig:scldens2} The same as
Fig.~\protect\ref{fig:scldens1} but for larger deformations.}
\bigskip

\sfig{0.6}{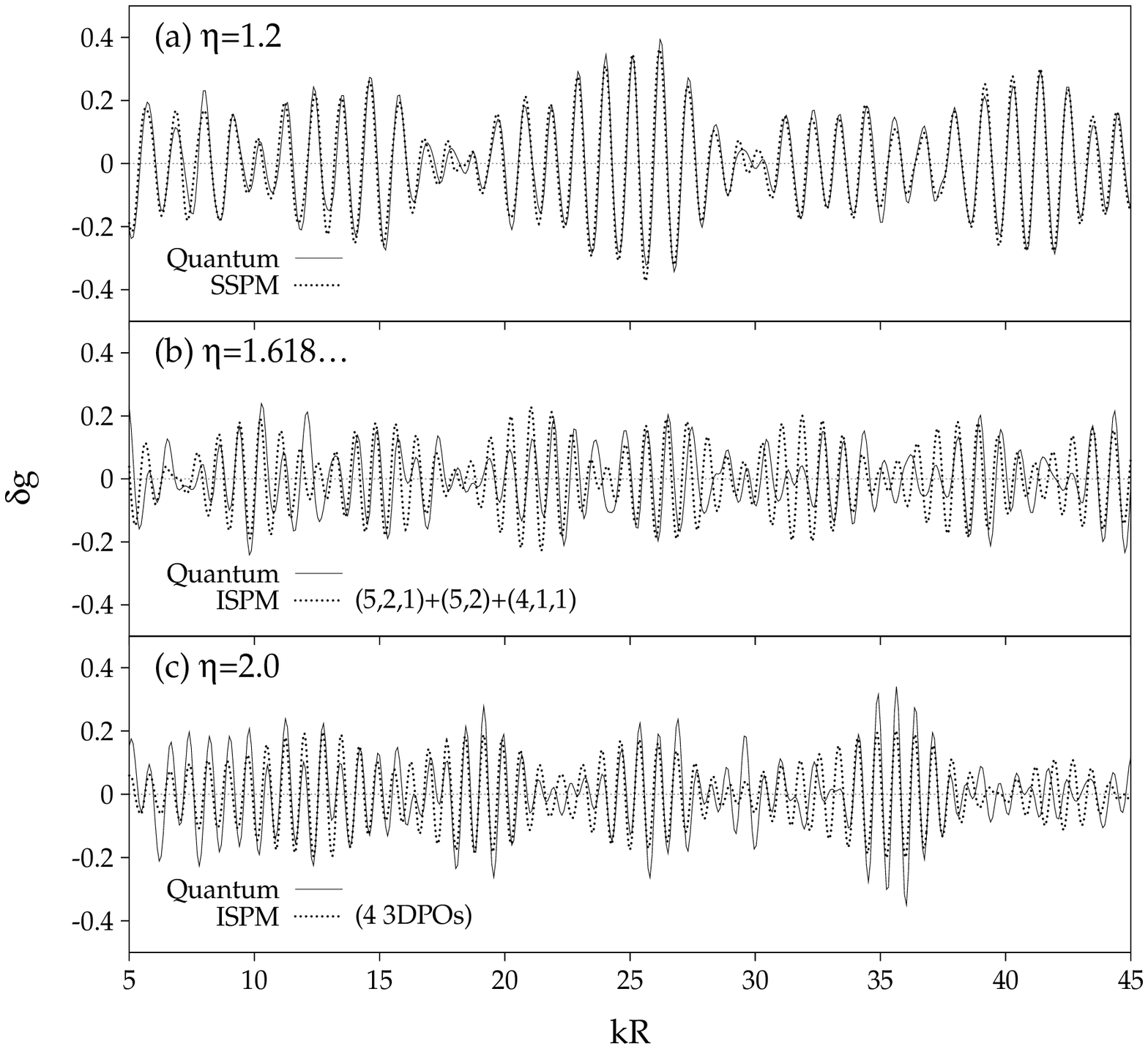}
\caption{\label{fig:scldens3} Comparison of the oscillating level
densities calculated by quantum mechanics (solid curves)
and those obtained by some specific semiclassical calculations
(dotted curves); (a) the top panel shows a comparison with
the SSPM result for $\eta=1.2$, (b) the middle panel shows the
ISPM result in which only the bifurcating 3DPO $(5,2,1)$, the EQPO
$(5,2)$, and the 2DPO butterfly $(4,2,1)$ are taken into account
for the POT sum in Eq.~(\ref{deltag3d2d}) for $\eta=1.618\ldots$,
(c) the bottom panel shows the ISPM result in which only the four
shortest 3DPO are taken into account for $\eta=2.0$.
}
\end{figure}

Figure~\ref{fig:scldens2} presents the oscillating level densities for
the bifurcation deformations; $\eta=1.618\ldots$ for the EQPO $(5,2)$,
$\eta=\sqrt{3}$ for the EQPO $2(3,1)$, and $\eta=2$ for the triply
repeated equatorial diameters $3(2,1)$.  It is interesting to compare
this figure with Fig.~\ref{fig:scldens3} where some results of
simplified semiclassical calculations are shown.
\begin{figure}[t]
\sfig{0.6}{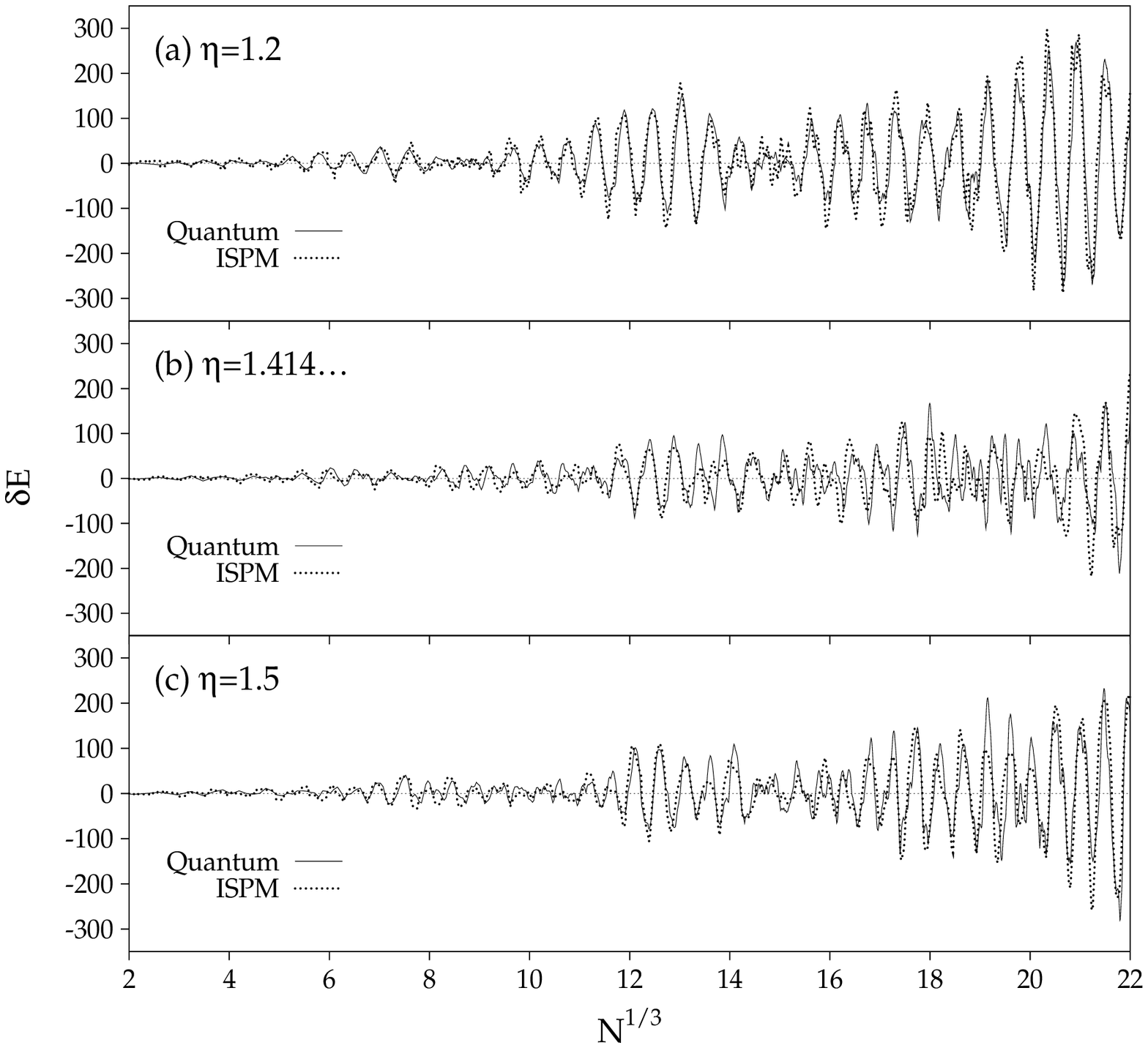}
\caption{\label{fig:sclene1} Semiclassical ISPM and quantum shell
energies (in unit of $\eps_0$) are plotted by dotted and
solid curves, respectively, as functions of $N^{1/3}$.}
\sfig{0.6}{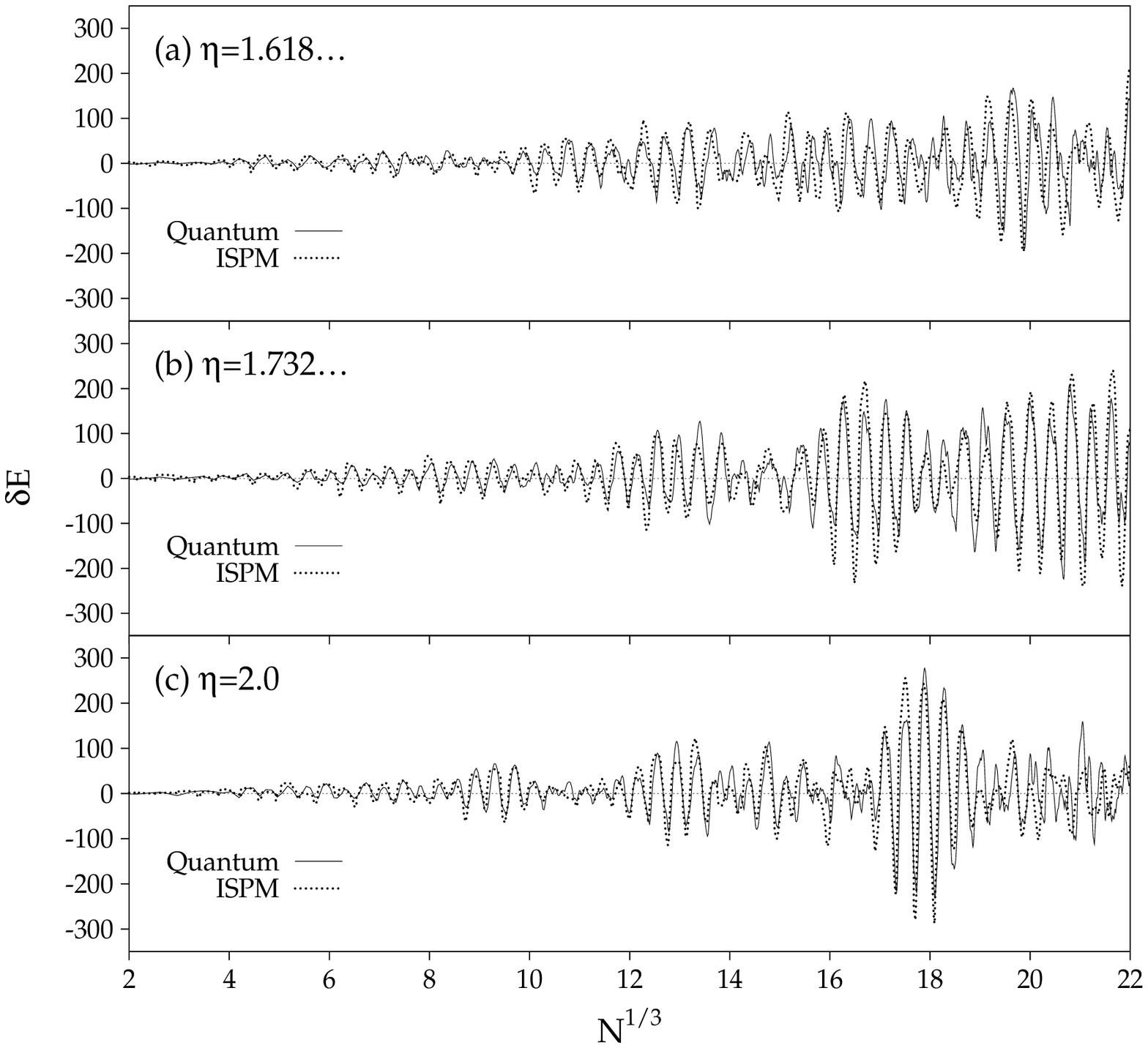}
\caption{\label{fig:sclene2} The same as
Fig.~\protect\ref{fig:sclene1} but for larger deformations.}
\end{figure}
In the top panel of
Fig.~\ref{fig:scldens3}, the SSPM is used instead of the ISPM.  We see
that the SSPM is a good approximation for $\eta=1.2$.  In the middle
and bottom panels, only bifurcating orbits are taken into account in
the periodic-orbit sum: only the 3DPO $(5,2,1)$, the EQPO $(5,2)$ and
the butterfly $(4,2,1)$ are accounted for in the middle panel, while
only the 3DPO $(5,2,1)$, $(6,2,1)$, $(7,2,1)$,$(8,2,1)$ in the bottom
panel.  By comparing with the corresponding ISPM results shown in
Fig.~\ref{fig:scldens3}, we see that, for $\eta=1.618\ldots$ and 2,
the major patterns of the oscillation are determined by these short
3DPO.

\begin{figure}[t]
\sfig{0.6}{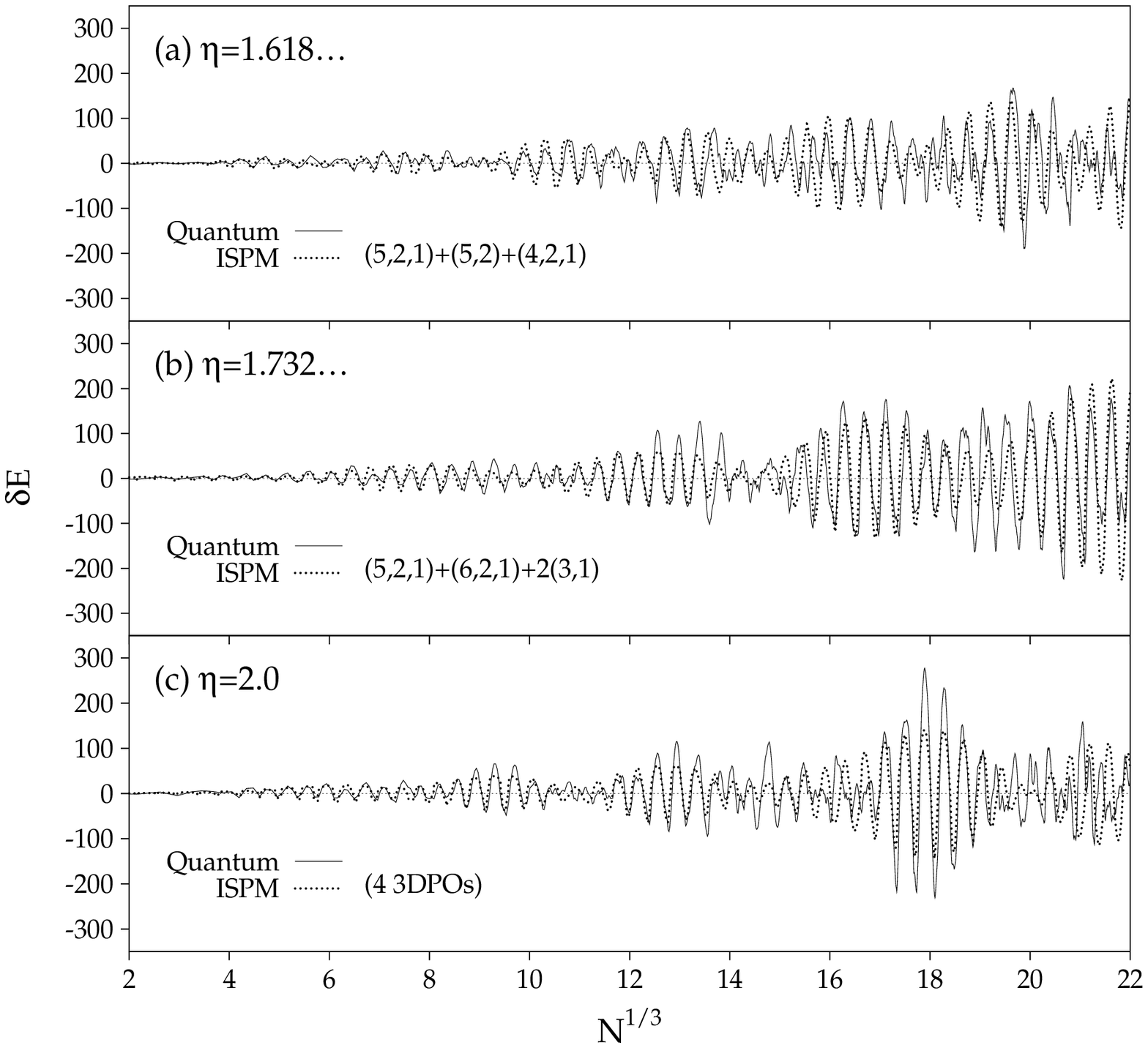}
\caption{\label{fig:sclene3} Comparison of quantum shell energies
(solid curves) with those obtained with specific
semiclassical calculations (dotted curves):
(a) the top panel shows the ISPM result for $\eta=1.618...$,
where only the bifurcating orbits (5,2,1), (5,2) and (4,2,1) are
taken into account,
(b) the middle panel shows for $\eta=1.732...$
the contributions of {\em only} the three orbits;
the 3DPO $(5,2,1)$ and $(6,2,1)$, and the EQPO $2(3,1)$,
(c) the bottom panel shows for $\eta=2.0$
the contributions of {\em only} the four shortest 3DPO to the ISPM sum.}
\end{figure}

Figures~\ref{fig:sclene1} and \ref{fig:sclene2} show the shell
energies which respectively correspond to the oscillating level
densities shown in Figs.~\ref{fig:scldens1} and
\ref{fig:scldens2}.  Again, we see good agreement between the
results of the semiclassical ISPM and the quantum calculations.
For $\eta=1.2$, a good convergence is attained by including only
the shortest elliptic 2DPO and EQPO, in the same way as for the
level density $\delta g$, see Ref.~\citen{mfimbrk}.  For
$\eta=\sqrt{2}$ and 1.5, properties of the ISPM shell energies
are similar to those considered for the elliptic billiard in
Ref.~\citen{mfammsb}.  Now, let us more closely examine the
bifurcation effects in the superdeformed region by comparing
Fig.~\ref{fig:sclene2} with Fig.~\ref{fig:sclene3}.
In the top
panel of Fig.~\ref{fig:sclene3}, we show the ISPM result for
$\eta=1.618\ldots$ in which only the bifurcating 3DPO $(5,2,1)$,
the short EQPO $(5,2)$ and the hyperbolic 2DPO $(4,2,1)$ are
taken into account. In the middle panel of this figure, we show
the ISPM shell energies at $\eta=1.732\ldots$, calculated by
taking into account only the 3DPO $(5,2,1)$, the bifurcating 3DPO
$(6,2,1)$ and the EQPO $2(3,1)$. These comparisons clearly
indicate that a few dominant periodic orbits determine the
property of quantum shell structure at those bifurcation
deformations. The bottom panel in this figure shows the
dominating contributions of only a few shortest 3DPO at
$\eta=2.0$. Evidently, the short 3DPO $(5,2,1)$, $(6,2,1)$,
$(7,2,1)$ and $(8,2,1)$ determine the major oscillating pattern
of the shell energy. Thus, we can say that they are responsible
for the formation of the shell structure at large deformations
around the superdeformed shape.  These results of calculation are
in good agreement with those obtained in Ref.~\citen{ask98} from
the analysis of the length spectra (Fourier transforms of the
quantum level densities).

\section{Conclusion}

We have obtained an analytical trace formula for the 3D spheroidal
cavity model, which is continuous through all critical deformations
where bifurcations of periodic orbits occur.  We find an enhancement of
the amplitudes $|A_\beta|$ at deformations $\eta\sim 1.6\mbox{--}2.0$
due to bifurcations of 3D orbits from the short 2D orbits in the
equatorial plane.  The reason of this enhancement is quite general and
independent of the specific potential shapes.  We believe that this is
an important mechanism which contributes to the stability of
superdeformed systems, also in the formation of the second minimum
related to the isometric states in nuclear fission.  Our semiclassical
analysis may therefore lead to a deeper understanding of shell
structure effects in superdeformed fermionic systems -- not only in
nuclei or metal clusters but also, e.g., in deformed semiconductor
quantum dots whose conductance and magnetic susceptibilities are
significantly modified by shell effects.

\section*{Acknowledgements}

A.G.M. gratefully acknowledges the financial support provided under
the COE Professorship Program by the Ministry of Education, Science,
Sports and Culture of Japan (Monbu-sho), giving him the opportunity to
work at the RCNP, and thanks Prof. H.~Toki for his warm hospitality
and fruitful discussions. We acknowledge valuable discussions with
Prof. M.~Brack.  Two of us (A.G.M. and S.N.F.)  thank also the
Regensburger Universit\"atsstiftung Hans Vielberth and Deutsche
Forschungsgemeinshaft (DFG) for the financial support.

\appendix
\section{Curvatures}
\label{app:curv}

\subsection{Three-dimensional orbits}
\label{app:3dpocurv}

The action is written as
\begin{equation}
S = 2 \pi M \left(n_v I_v + n_u I_u + n_\varphi I_\varphi \right),
\label{act1}
\end{equation}
where $I_u$, $I_v$ and $I_\varphi$ are the partial actions.
In a dimensionless form,
\begin{equation}
I_u=\frac{p\zeta}{\pi}\tI_u, \qquad
I_v=\frac{p\zeta}{\pi}\tI_v, \qquad
I_\varphi=\frac{p\zeta}{\pi}\tI_\varphi,
\label{dimensionless1}
\end{equation}
where
\begin{subequations}\label{iuvc1}
\begin{eqnarray}
\tI_u &=& 2\int_0^{z_-}
 \frac{dz}{1-z^2}\,\sqrt{(z^2-z_-^2)(z^2-z_+^2)}, \\
\tI_v &=& \int_{z_+}^{z_b}
 \frac{dz}{z^2-1}\,\sqrt{(z^2-z_-^2)(z^2-z_+^2)}, \\
\tI_\varphi &=& \pi\sqrt{\sigma_2}.
\end{eqnarray}
\end{subequations}
$z_\pm$ are related to the $\sigma_i$ variables by
\begin{equation}
z_+^2+z_-^2=\sigma_1+1,\qquad
z_+^2 z_-^2=\sigma_1-\sigma_2.
\label{zsigma1}
\end{equation}
In terms of the elliptic integrals, (\ref{iuvc1}) can be expressed as
\begin{subequations}\label{iuvce1}
\begin{eqnarray}
 \tI_u &=& \frac{2}{z_+}\,\left[(z_-^2-1)\rF(k)-\sigma_2 \rPi(z_-^2,k)
   +z_+^2 \rE(k)\right], \label{iuce1} \\
 \tI_v &=& \frac{1}{z_+}\,\left\{(z_+^2-z_-^2)
   \left[\rF(\varphi,k)-\rPi(\varphi,n,k)\right]-z_
   +^2 \rE(\varphi,k)\right\} + z_b \sin{\varphi}, \label{ivce1}
\end{eqnarray}
\end{subequations}
with
\begin{displaymath}
k = \frac{z_-}{z_+},\qquad
n = \frac{1-z_-^2}{1-z_+^2},\qquad
\varphi = \arcsin\sqrt{\frac{z_b^2-z_+^2}{z_b^2-z_-^2}},
\end{displaymath}
\begin{equation}
z_b = \cosh v_b = \frac{\eta}{\sqrt{\eta^2-1}}.
\label{knfzb1}
\end{equation}
Here, we used the standard definitions of the elliptic integrals
of the first and the third kind\footnote{%
The definitions of elliptic integrals (\ref{ellipticfuns})
are related with those in Ref.~\citen{abramov} as
$\rF(\theta,\kappa) \equiv \rF(\theta | \alpha)$ and
$\rPi(\theta,n,\kappa)=\rPi(n,\theta | \alpha)$~ ($\kappa=\sin\alpha$).}
\begin{subequations}
\label{ellipticfuns}
\begin{equation}
\rF(\varphi,k) = \int_0^\varphi\frac{dx}{\sqrt{1-k^2 \sin^2{x}}},
\label{elf1}
\end{equation}
\begin{equation}
\rE(\varphi,k) = \int_0^\varphi \sqrt{1-k^2 \sin^2{x}}\:dx,
\label{elef1}
\end{equation}
\begin{equation}
\rPi(\varphi,n,k) = \int_0^\varphi
\frac{dx}{(1-n \sin^2{x})\sqrt{1-k^2 \sin^2x}}.
\label{elpi1}
\end{equation}
\end{subequations}
and omit argument $\varphi=\pi/2$ for complete elliptic integrals.
The action (\ref{act1}) is written as
\begin{equation}
S=2p\zeta M\left(n_v\tI_v + n_u \tI_u + n_\varphi\tI_\varphi\right).
\label{actc1}
\end{equation}
The curvatures $K_{ij}$ of the energy
surface $\eps=H(\sigma_1,\sigma_2,\eps)$ are defined as
\begin{equation}
K_{ij}=\frac{p\zeta}{\pi}\tK_{ij}=
\ppp{I_v}{\sigma_i}{\sigma_j}+\frac{\omega_u}{\omega_v}\,
\ppp{I_u}{\sigma_i}{\sigma_j}+\frac{\omega_\varphi}{\omega_v}\,
\ppp{I_\varphi}{\sigma_i\sigma_j},
\label{Kntilde}
\end{equation}
and the frequency ratios in
Eq.~(\ref{Kntilde}) are given by
\begin{equation}
\frac{\omega_u}{\omega_v} \equiv
-\left(\pp{I_v}{I_u}\right)_{I_\varphi}=
-\frac{\partial \tI_v/\partial\sigma_1}
      {\partial \tI_u/\partial\sigma_1}\,,
\label{wuwv1}
\end{equation}
\begin{equation}
\frac{\omega_\varphi}{\omega_v} \equiv
-\left(\pp{I_v}{I_\varphi}\right)_{I_u}=
-\frac{2\sqrt{\sigma_2}}{\pi}\,
\left[\pp{\tI_v}{\sigma_2}+
\frac{\omega_u}{\omega_v}\pp{\tI_u}{\sigma_2}\right]\,.
\label{wfwv1}
\end{equation}
We used here the properties of Jacobians for the transformations
from the variables ($I_u,I_\varphi$) to ($\sigma_1,\sigma_2$).
For the first derivatives of the actions (\ref{iuvc1})
with respect to $\sigma_1$ and $\sigma_2$, one obtains
\begin{subequations}
\begin{equation}
\pp{\tI_u}{\sigma_1} = \frac{1}{z_+}\,\rF(k)\,,\qquad
\pp{\tI_v}{\sigma_1} = -\frac{1}{2z_+}\,\rF(\varphi,k)\,,
\label{diuvds1}
\end{equation}
\begin{equation}
\pp{\tI_u}{\sigma_2} = -\frac{1}{z_+}\,\rPi(z_-^2,k)\,,\qquad
\pp{\tI_v}{\sigma_2} = C_\rF\,\rF(\varphi,k)+C_\rPi\,\rPi(\varphi,n,k)\,,
\label{diuvds2}
\end{equation}
\end{subequations}
with
\begin{eqnarray}
C_\rF &=& \frac{z_+^2-1}{2z_+\sigma_2}
 = -\frac{1}{2z_+(z_-^2-1)}\,, \nonumber \\
C_\rPi &=& -\frac{z_+^2-z_-^2}{2z_+\sigma_2}
 = \frac{z_+^2-z_-^2}{2z_+(z_+^2-1)(z_-^2-1)}\,.
\end{eqnarray}
For the second derivatives of these actions, one obtains
\begin{subequations}
\begin{eqnarray}
\pp{^2 \tI_u}{\sigma_1^2}
&=& \frac{1}{2z_+^3}\,
   \left\{\frac{1}{k^2}\left[\rPi(k^2,k)-\rF(k)\right]
   \left(\pp{z_-^2}{\sigma_1}-k^2\pp{z_+^2}{\sigma_1}\right)
  -\pp{z_+^2}{\sigma_1}\rF(k)\right\}\,,
\label{d2iuds12} \\
\pp{^2\tI_v}{\sigma_1^2}
&=& -\frac{1}{4z_+^3}\,
   \biggl\{\frac{1}{k^2}
    \left[\rPi(\varphi,k^2,k)-\rF(\varphi,k)\right]
    \left(\pp{z_-^2}{\sigma_1}-k^2\pp{z_+^2}{\sigma_1}
   \right)\nonumber\\
& & \hspace{13.5em}
  -\pp{z_+^2}{\sigma_1}\,\rF(\varphi,k)
  +\frac{2z_+^2}{\Delta_\varphi}\,\pp{\varphi}{\sigma_1}\biggr\},
\label{d2ivds12} \\
\pp{^2 \tI_u}{\sigma_2^2}
&=& \frac{1}{2z_+^5 k_1^2}
   \left[\rPi(z_-^2,k)+2z_+^2\, \pp{\rPi(z_-^2,k)}{n}
    +\frac{1+k^2}{k}\,\pp{\rPi(z_-^2,k)}{k}
   \right],
\label{d2iuds22} \\
\frac{\partial^2 \tI_v}{\partial\sigma_2^2}
&=& \pp{C_\rF}{\sigma_2}\,\rF(\varphi,k)
  + C_\rF\left(\frac{1}{\Delta_\varphi} \pp{\varphi}{\sigma_2}
  + \pp{\rF(\varphi,k)}{k} \pp{k}{\sigma_2}\right)
  + \pp{C_\rPi}{\sigma_2}\,\rPi(\varphi,n,k)
\nonumber\\
& & + C_\rPi \left(\pp{\rPi(\varphi,n,k)}{\varphi} \pp{\varphi}{\sigma_2}
  + \pp{\rPi(\varphi,n,k)}{n} \pp{n}{\sigma_2}
  + \pp{\rPi(\varphi,n,k)}{k} \pp{k}{\sigma_2}\right),
\nonumber\\
\label{d2ivds22}
\end{eqnarray}
and
\begin{eqnarray}
\frac{\partial^2 \tI_u}{\partial \sigma_1 \partial \sigma_2}
&=& -\frac{1}{2z_+^5k_1^2}\,
   \left[\rF(k)+\frac{1+k^2}{k}\,\pp{\rF(k)}{k}\right],
\label{d2iuds1ds2} \\
\frac{\partial^2 \tI_v}{\partial \sigma_1 \partial \sigma_2}
&=& \frac{1}{4z_+^5k_1^2}
    \left[\rF\left(\varphi,k\right)
     -\frac{(\sigma_1+1-2z_b^2)\tan\theta}
           {\Delta_\varphi z_b^2 \Delta_\theta^2 k_1}
     +\frac{1+k^2}{k}\, \pp{\rF(\varphi,k)}{k}\right].
\nonumber\\
\label{d2ivds1ds2}
\end{eqnarray}
\end{subequations}
Here,
\begin{equation}
\Delta_x = \sqrt{1-k^2\sin^2 x},
\quad k_1 = \sqrt{1-k^2},
\quad \theta = \arcsin\left(\frac{z_+}{z_b}\right),
\end{equation}
and
\begin{equation}
\pp{z_\pm^2}{\sigma_1}
= \frac12\,\left[1 \pm \frac{\sigma_1-1}
  {\sqrt{(\sigma_1-1)^2+4\sigma_2}}\right]
= \frac12\,\left[1 \pm \frac{z_+^2 + z_-^2 -2}{z_+^2-z_-^2}\right]\,,
\label{dzpm2ds11}
\end{equation}
\begin{equation}
\pp{\varphi}{\sigma_1}
= \frac12\,
  \frac{\pp{z_-^2}{\sigma_1}(z_b^2-z_+^2)
       -\pp{z_+^2}{\sigma_1}(z_b^2-z_-^2)}
       {(z_b^2-z_-^2)\sqrt{(z_b^2-z_+^2)(z_+^2-z_-^2)}}\,,
\label{dfids11}
\end{equation}
\begin{equation}
\pp{k^2}{\sigma_2}
= -\frac{1+k^2}{z_+^4k_1^2}\,,
\end{equation}
\begin{eqnarray}
\pp{C_\rF}{\sigma_2}
&=& \frac{z_-^2-2z_+^2-1}{4z_+^3(1-z_-^2)^2(z_+^2-z_-^2)},
\nonumber\\
\pp{C_\rPi}{\sigma_2}
&=& -\frac{\sigma_2(3z_+^2+z_-^2)-2z_+^2(z_+^2-z_-^2)^2}
          {4z_+^3\sigma_2^2(z_+^2-z_-^2)}\,,
\label{dcfds2dcpids21}
\end{eqnarray}
\begin{equation}
\pp{\varphi}{\sigma_2}
= \frac{\left(2z_b^2-(\sigma_1+1)\right)\tan{\theta}}
       {2z_b^2z_+^4k_1^3\Delta_\theta^2},
\label{dfds2}
\end{equation}
\begin{equation}
\pp{n}{\sigma_2}
= \frac{\sigma_1-1}{(1-z_+^2)^2(z_+^2-z_-^2)},
\label{dnds2}
\end{equation}
\begin{equation}
\pp{z_\pm^2}{\sigma_2}=\pm\frac{1}{z_+^2-z_-^2}.
\label{dzpm2ds2}
\end{equation}
Derivatives of elliptic integrals are given by
\begin{equation}
\pp{\rF(\varphi,k)}{k}
= \frac1k \left[\rPi(\varphi,k^2,k)-\rF(\varphi,k)\right]\,,
\label{dFdk1}
\end{equation}
\begin{equation}
\pp{\rPi(\varphi,n,k)}{\varphi}
= \frac{1}{(1-n\sin^2{\varphi})\Delta_\varphi}\,,
\label{dFdkdpidf1}
\end{equation}
\begin{equation}
\pp{\rPi(\varphi,n,k)}{n}
= \frac1n \left[\rPi_{21}(\varphi,n,k)-\rPi(\varphi,n,k)\right],
\label{dpidn1}
\end{equation}
\begin{equation}
\pp{\rPi(\varphi,n,k)}{k}
= \frac1k \left[\rPi_{13}(\varphi,n,k)-\rPi(\varphi,n,k)\right],
\label{dpidk1}
\end{equation}
with
\begin{equation}
\rPi_{ij}(\varphi,n,k)
= \int_0^\varphi
  \frac{dx}{(1-n \sin^2{x})^i(1-k^2\sin^2{x})^{j/2}}\,.
\label{pij1}
\end{equation}

\subsection{Meridian-plane orbits}
\label{app:meridpocurv}

For the meridian-plane orbits for which $I_\varphi=0$ ($\sigma_2=0$),
the actions $I_u$ and $I_v$ defined by Eq.~(\ref{actionuv}) can be
simplified.
In the dimensionless form (\ref{dimensionless1}) one obtains for the
elliptic orbits
\begin{subequations}\label{ensurfe}
\begin{eqnarray}
\tI_u &=& 2\sqrt{\sigma}~\rE\left(\frac{1}{\sqrt{\sigma}}\right),\\
\tI_v &=& \sqrt{\sigma}\left[
  \rE\left(\theta_e,\frac{1}{\sqrt{\sigma}}\right)-
  \rE\left(\frac{1}{\sqrt{\sigma}}\right)\right]
  +\frac{\sqrt{\eta^2-\sigma (\eta^2-1)}}{\eta \sqrt{\eta^2-1}}
\end{eqnarray}
\end{subequations}
Here we used the identity~\cite{Byrd:Friedman}
\begin{equation}
\rPi(\varphi,k^2,k)
=\left[\rE(\varphi,k)-k^2 \sin{\varphi}\cos{\varphi}
 /\sqrt{1-k^2\sin^2{\varphi}}\right]/(1-k^2)\,.
\label{piidentity1}
\end{equation}
In this subsection, we omit the suffix ``1'' for the variable
$\sigma_1$ for shortness. For the hyperbolic orbits,
\begin{subequations}\label{ensurfh}
\begin{eqnarray}
\tI_u &=& 2\left[\rE(\sqrt{\sigma})-(1-\sigma)~\rF(\sqrt{\sigma})\right],\\
\tI_v &=& (1-\sigma)\left[\rF(\sqrt{\sigma})-\rF(\theta_h,\sqrt{\sigma})
   \right] +\rE(\theta_h,\sqrt{\sigma}) \nonumber\\
&& \hspace{33mm} -\rE(\sqrt{\sigma})
   +\frac{\sqrt{\eta^2-\sigma (\eta^2-1)}}{\eta \sqrt{\eta^2-1}}
\end{eqnarray}
\end{subequations}
Equations~(\ref{ensurfe}) and (\ref{ensurfh}) may be regarded as
parametric equations in terms of the parameter $\sigma$~
for the energy surface of the meridian-plane orbits,
$\eps(\tI_u,\tI_v,\tI_\varphi=0)$,
for its elliptic and hyperbolic parts, respectively.

The curvature $K_{11}$ of the energy curve for the meridian-plane
orbits can be obtained by differentiating Eqs.~(\ref{ensurfe}) and
(\ref{ensurfh}) implicitly through the parameter $\sigma$. In this way
one obtains Eq.~(\ref{jacobupar}) with the dimensionless derivatives for
the elliptic orbits
\begin{subequations}
\label{derivate}
\begin{equation}
\pp{\tI_u}{\sigma} = \frac{1}{\sqrt{\sigma}}
  ~\rF\left(\frac{1}{\sqrt{\sigma}}\right),
\end{equation}
\begin{equation}
\pp{^2\tI_u}{\sigma^2} = -\frac{1}{2\sqrt{\sigma^3}}
  ~\rPi\left(\frac{1}{\sigma},\frac{1}{\sqrt{\sigma}}\right),
\end{equation}
\begin{equation}
\pp{\tI_v}{\sigma} = -\frac{1}{2\sqrt{\sigma}}
  ~\left[\rF\left(\frac{1}{\sqrt{\sigma}}\right)
  -\rF\left(\theta_e, \frac{1}{\sqrt{\sigma}}\right)\right],
\end{equation}
\begin{equation}
\pp{^2 \tI_v}{\sigma^2} = \frac{1}{4\sqrt{\sigma^3}}
  ~\left[\rPi\left(\frac{1}{\sigma}, \frac{1}{\sqrt{\sigma}}\right)
  -\rPi\left(\theta_e, \frac{1}{\sigma}, \frac{1}{\sqrt{\sigma}}\right)
  +\frac{\eta \sqrt{\eta^2-1}}{\sqrt{1-(1-\sigma^{-1})\eta^2}}\right].
\end{equation}
\end{subequations}
For the hyperbolic orbits,
\begin{subequations}
\label{derivath}
\begin{equation}
\pp{\tI_u}{\sigma} = \rF(\sqrt{\sigma}),
\end{equation}
\begin{equation}
\pp{^2 \tI_u}{\sigma^2} = \frac{1}{2\sigma}
  ~\left[\rPi(\sigma,\sqrt{\sigma}) - \rF(\sqrt{\sigma})\right],
\end{equation}
\begin{equation}
\pp{\tI_v}{\sigma} = \frac12
  ~\left[\rF(\theta_h,\sqrt{\sigma}) - \rF(\sqrt{\sigma})\right],
\end{equation}
\begin{equation}
\pp{^2 \tI_v}{\sigma^2} = \frac{1}{4\sigma}
  ~\left[\rPi(\theta_h,\sigma,\sqrt{\sigma})
  -\rPi(\sigma,\sqrt{\sigma})z+\rF(\sqrt{\sigma})
  -\rF(\theta_h,\sqrt{\sigma})\right].
\end{equation}
\end{subequations}
Thus, for elliptic orbits
\begin{equation}
\tK_{11}=\frac{1}{4\sqrt{\sigma^3}}
\left[\frac{\rF(\theta_e,\kappa)}{\rF(\kappa)}\rPi(\kappa^2,\kappa)
 -\rPi(\theta_e,\kappa^2,\kappa)
 +\frac{\sqrt{\eta^2-\sigma (\eta^2-1)}}{\eta \sqrt{\eta^2-1}}\right],
\end{equation}
and for hyperbolic orbits
\begin{equation}
\tK_{11}=-\frac{1}{4\sigma}
\left[\frac{\rF(\theta_h,\kappa)}{\rF(\kappa)}\rPi(\kappa^2,\kappa)
 -\rPi(\theta_h,\kappa^2,\kappa)\right].
\end{equation}

\subsection{Equatorial-plane orbits}
\label{app:eqpocurv}

For the equatorial limit $\sigma_2=\sigma_1\equiv\sigma$ one obtains
from (\ref{zsigma1})
\begin{equation}
z_-^2=0,\qquad z_+^2=\sigma+1\,.
\label{zpmeq1}
\end{equation}
We thus obtain in this limit ($k \to 0$)
\begin{displaymath}
\pp{\tI_u}{\sigma_1} = \frac{\pi}{2\sqrt{\sigma+1}}, \qquad
\pp{\tI_v}{\sigma_1} = -\frac{\varphi_{\rm EQ}}{2\sqrt{\sigma+1}},
\end{displaymath}
\begin{equation}
\pp{z_\pm^2}{\sigma_1} =
\brcase{\sigma/(\sigma+1)}{1/(\sigma+1)}\,.
\label{diuvds1eq}
\end{equation}
and
\begin{eqnarray}
\frac{\partial^2 \tI_u}{\partial \sigma_1^2} &=&
 \frac{\pi (1-2\sigma)}{8(\sigma+1)^{5/2}}, \nonumber\\
\frac{\partial^2 \tI_v}{\partial \sigma_1^2} &=&
 \frac{1}{8(\sigma+1)^{5/2}} \Biggl\{(2\sigma-1) \varphi_{\rm EQ} +
 \frac12\sin{(2\varphi_{\rm EQ})}
\nonumber\\
&& - \frac{2\sqrt{\sigma+1}\left[z_b^2(1-\sigma)-(\sigma+1)
 \right]}{z_b^2 \sqrt{z_b^2-(\sigma+1)}}\Biggr\},
\nonumber\\
\varphi_{\rm EQ} &=& \arcsin\frac{\sqrt{z_b^2-(\sigma+1)}}{z_b}
\label{d2iuvds12eq1}
\end{eqnarray}
Substituting (\ref{diuvds1eq}) and (\ref{d2iuvds12eq1})
into (\ref{Kntilde}) one finally obtains the equatorial limit
\begin{subequations}\label{kijeq}
\begin{equation}
\tK_{11}^{\rm EQ}=
\frac{z_b^2(2\sigma-1)+(\sigma+1)}
     {8z_b^2(\sigma+1)^2\sqrt{z_b^2-(\sigma+1)}}.
\end{equation}
In the same way, one obtains
\begin{eqnarray}
\tK_{22}^{\rm EQ}&=&
\frac{z_b^2(2-\sigma)+\sigma(\sigma+1)}
     {8z_b^2\sigma(\sigma+1)^2\sqrt{z_b^2-(\sigma+1)}}, \\
\tK_{12}^{\rm EQ}&=&
\frac{3z_b^2-(\sigma+1)}
     {8z_b^2(\sigma+1)^2\sqrt{z_b^2-(\sigma+1)}},
\end{eqnarray}
\end{subequations}
The determinant of the curvature matrix for EQPO becomes
\begin{equation}
\det\tK^{\rm EQ}=-\frac{1}{32z_b^2\sigma(\sigma+1)^2}
\label{detkeq}
\end{equation}
which is negative for any orbit and for any deformation $\eta>1$.
It shows that bifurcations of EQPO's occur only through the zeros of
stability factor $F_z^{\rm EQ}$.

\section{Derivation of trace formula for the equatorial-plane orbits}
\label{app:trace_eq}

We start with the phase-space trace formula\cite{mfimbrk,mfammsb,sie97,bruno}
\begin{eqnarray}
\delta g_{\rm scl}(\eps)&=&
\Re\sum_\alpha \int
\frac{d\bq'' d\bp'}{(2 \pi \hbar)^3}
~\delta \left(\eps-H\left(\bq',\bp'\right) \right)
\left|{\cal J}\left(\bp_\perp'',\bp_\perp'\right)\right|^{1/2} \nonumber\\
&&\times \exp\left\{\frac{i}{\hbar}
\left[S_\alpha\left(\bp',\bp'',t_\alpha\right)+
\left(\bp''-\bp'\right)\cdot\bq''\right]
-i\frac{\pi}{2}\nu_\alpha\right\}, \label{pstrace}
\end{eqnarray}
where the sum runs over all trajectories $\alpha$,
$\bq=\bq_\alpha(t,\bq'',\bp')$
determined by the fixed initial momentum $\bp'$ and the final
coordinate $\bq''$, $H\left(\bq,\bp\right)$ is the classical
Hamiltonian, $\nu_{\alpha}$ the phase related to the Maslov index,
number of caustics and turning
points.\cite{fed:jvmp,masl,fed:spm,masl:fed}
The $S_\alpha\left(\bp',\bp'',t_\alpha\right)$
is the action in the mixed phase-space representation,
\begin{equation}
S_\alpha\left(\bp',\bp'',t_\alpha\right)
=-\int_{\bp'}^{\bp''}~ d\bp \cdot \bq\left(\bp\right),
\label{actionppt}
\end{equation}
related to the standard definition of the action
$S_\alpha\left(\bq',\bq'',\eps\right)$,
\begin{equation}
S_\alpha\left(\bq',\bq'',\eps\right)
=\int_{\bq'}^{\bq''}~ d\bq \cdot \bp\left(\bq\right)
\label{actionrre}
\end{equation}
by the Legendre transformations (the integration by parts),
\begin{equation}
S_\alpha\left(\bp',\bp'',t_\alpha\right)=
S_\alpha\left(\bq',\bq'',\eps\right)+
\left(\bp'-\bp''\right)\bq'',
\label{legtrans}
\end{equation}
$t_\alpha$ being the time for motion of the particle along the trajectory
$\alpha$. The ${\cal J}\left(\bp_\perp'',\bp_\perp'\right)$
in Eq.~(\ref{pstrace})
is the Jacobian for the transformation from $\bp_\perp''$
to $\bp_\perp'$. Here, we introduced the local system
of the phase-space coordinates
$\bq=\left\{q_\parallel,\bq_\perp\right\}$ and
$\bp=\left\{p_\parallel,\bp_\perp\right\}$
splitting the vectors into the parallel and perpendicular
components with respect to the trajectory $\alpha$.

For the equatorial-plane periodic orbits (EQPO)
one of the perpendicular components
$\bq_\perp$ and $\bp_\perp$ can be taken along the symmetry
axis $z$, say $z$ and $p_z$, keeping for other perpendicular components
the same suffix, $q_\perp$ and $p_\perp$. After the
transformation to this local phase-space coordinate system
and integration over the ``parallel'' momentum $p_\parallel=p=\sqrt{2m\eps}$
by using the $\delta$-function in Eq.~(\ref{pstrace}),
one obtains for the contribution from the EQPO (${\cal K}=1$)
\begin{eqnarray}
\delta g_{\rm EQ}^{(1)}(\eps)&=&
\frac{1}{(2\pi \hbar)^3}\Re\sum_\alpha \int
\frac{dq_\parallel''}{|\dot{q}_\parallel''|}
\int d q_\perp''d p_\perp'
\int d z'' d p_z'
\left|{\cal J}\left(\bp'',\bp'\right)\right|^{1/2}\nonumber\\
&& \times \exp\left\{\frac{i}{\hbar}
\left[S_\alpha \left(\bp',\bp'',t_\alpha \right)+
\left(\bp''-\bp'\right)\cdot \bq''\right]
-i\frac{\pi}{2}\nu_\alpha\right\},
\label{pstrace1}
\end{eqnarray}
where $\dot{q}_\parallel=\partial H /\partial p_\parallel=p/m$
is the velocity.
In the spheroidal action-angle variables, $q_\parallel=\Theta_v$,
$p_\parallel=I_v$, $\dot{q}_\parallel=\omega_v$,
$q_\perp=\Theta_\varphi=\varphi$, $p_\perp=I_\varphi$,
$z=\Theta_u$, $p_z=I_u$, and we have
\begin{eqnarray}
\delta g_{\rm EQ}^{(1)}(\eps)&=&
\frac{1}{(2\pi \hbar)^3}\Re\sum_\alpha \int
\frac{d\Theta_v''}{|\omega_v|}
\int d\Theta_\varphi''\, dI_\varphi'
\int d\Theta_u''\, dI_u'
\left|{\cal J}\left(I_\varphi''I_u'',I_\varphi'I_u'\right)\right|^{1/2}
\nonumber\\
&& \times
\exp\left\{\frac{i}{\hbar}
\left[S_\alpha \left(\bI',\bI'',t_\alpha \right)+
\left(\bI''-\bI'\right)\cdot\bm{\Theta}''\right]
-i\frac{\pi}{2}\nu_\alpha\right\}, \label{pstrace1b}
\end{eqnarray}
We now perform the integrations using the
expansion of the action $S_\alpha$ about the stationary points
\begin{eqnarray}
S_\alpha\left(\bI',\bI'',t_\alpha\right)
 +\left(\bI''-\bI'\right)\cdot\bm{\Theta}'' \nonumber\\
&& \hspace{-13em} = S_\beta(\eps)
 + \frac12 \sum_{ij}J_{ij}(\sigma_i-\sigma_i^*)(\sigma_j-\sigma_j^*)
 + \frac12 J_\perp\left(z-z^*\right)^2+\cdots
\end{eqnarray}
Here we omit the corrections associated with mixed derivatives of
type $\partial^2S/\partial \Theta\partial I$ for simplicity.
$J_\perp$ is the Jacobian corresponding to the second
variation of the action $S_\alpha$ with respect to the angle variable
$\Theta_u$,
\begin{eqnarray}
J_\perp^{\rm EQ}=
\left(\frac{\partial^2 S_\alpha}{\partial {\Theta_u'}^2}+
2\frac{\partial^2 S_\alpha}{\partial \Theta_u'\partial \Theta_u''}+
\frac{\partial^2 S_\alpha}{\partial {\Theta_u'}^2}\right)_{\rm EQ}
= \left(-\frac{\partial I_u'}{\partial\Theta_u'}-
2\frac{\partial I_u'}{\partial\Theta_u''}+
\frac{\partial I_u''}{\partial\Theta_u''}\right)_{\rm EQ}.
\nonumber\\
\label{jacobeqperp}
\end{eqnarray}
This quantity can be expressed in terms of the curvatures $K^{\rm EQ}$
and the Gutzwiller's stability factor $F_z^{\rm EQ}$,
\begin{eqnarray}
F_z^{\rm EQ}&=&-\left[
\left(-\frac{\partial I_u'}{\partial \Theta_u'}-
      2\frac{\partial I_u'}{\partial \Theta_u''}+
       \frac{\partial I_u''}{\partial \Theta_u''}\right)\Big/
       \frac{\partial I_u'}{\partial \Theta_u''}\right]_{\rm EQ}
\nonumber\\
&=&4\sin^2\left[\frac12 M n_v
 \arccos\left(1-2\eta^{-2}\sin^2\phi\right)\right],
\label{gutzstabfactz}
\end{eqnarray}
as
\begin{equation}
J_\perp^{\rm EQ}
=-\frac{F_z^{\rm EQ}}{(J_u-J_{u\varphi}^2/J_\varphi)^{\rm EQ}}
=-\frac{F_z^{\rm EQ}}{2\pi M n_v
(K_u-K_{u\varphi}^2/K_\varphi)^{\rm EQ}}
\label{jacobperprelations}
\end{equation}
In these equations we used simple identical Jacobian transformations
\begin{eqnarray}
\left(\pp{I_u'}{\Theta_u''}\right)_{I_\varphi'}^{-1}
&=& \pp{(\Theta_u'',I_\varphi')}{(I_u',I_\varphi')}
 =  \pp{\Theta_u''}{I_u'}-\pp{\Theta_u''}{I_\varphi'}\,
    \pp{I_\varphi'}{I_u'}
 =  J_u - J_{u\varphi}\frac{J_{u\varphi}}{J_\varphi} \nonumber
\end{eqnarray}
The curvature $K_u^{\rm EQ}$ is the quantity $K_u$ defined in (\ref{Jufi}),
evaluated at stationary point $\sigma_1=\sigma_2=\sigma^*$ given by
Eq.~(\ref{sigmaperpstar}), and so on.

The integrand of (\ref{pstrace1b}) does not depend on the angles
($\Theta_v$, $\Theta_\varphi$) and we obtain simply $(2\pi)^2$ for
the integration over these angle variables.  We transform
integration variables $(I_u, I_\varphi)$ into
$(\sigma_1,\sigma_2)$ to obtain simple integration limits, and
integrate over $(\sigma_1, \sigma_2)$ by the ISPM.
In this way we obtain
\begin{eqnarray}
\delta g_{\rm EQ}^{(1)}(\eps)&=&
\sqrt{\frac{\pi}{2\hbar^3}}\Re\sum_\beta
e^{i(kL_\beta-\pi\nu_\beta/2)}
\frac{1}{\omega_v}\left|\pp{(I_u,I_\varphi)}{(\sigma_1,\sigma_2)}\right|
\sqrt{\frac{1}{J_\perp|\det J^{\rm EQ}|}} \nonumber\\
&& \times
\erf({\cal Z}_\perp^-,{\cal Z}_\perp^+;
 {\cal Z}_1^-,{\cal Z}_1^+;{\cal Z}_2^-,{\cal Z}_2^+)
\end{eqnarray}
where
\begin{eqnarray}
\erf\left(x^-,x^+;y^-,y^+;z^-,z^+\right)
=\left(\frac{2}{\sqrt{\pi}}\right)^{3}
\int_{x^-}^{x^+} dx \int_{y^-}^{y^+} dy
\int_{z^-}^{z^+} dz \,e^{-x^2-y^2-z^2}, \nonumber\\
\label{erf2}
\end{eqnarray}
Note that the integration limits for the internal integrals
over $y$ and $z$ in
$\erf\left(x^-,x^+; y^-,y^+; z^-,z^+\right)$
in general depend on the variable of the next
integrations, $y^\pm=y^\pm(x)$ and $z^\pm=z^\pm(x,y)$.
Here we define curvatures in the $(I_u, I_\varphi)$ variables as
\begin{displaymath}
J_u=\frac{\partial^2 S_\alpha}{\partial I_u^2}
  =2\pi Mn_v K_u, \quad
J_\varphi=\frac{\partial^2 S_\alpha}{\partial I_\varphi^2}
  =2\pi Mn_v K_\varphi,
\end{displaymath}
\begin{equation}
J_{u\varphi}=\frac{\partial^2 S_\alpha}{\partial I_u \partial I_\varphi}
  =2\pi Mn_v K_{u\varphi}.
\label{Jufi}
\end{equation}
Using (\ref{jacobperprelations}) and relations
\begin{equation}
\det J \equiv J_{11}J_{22}-{J_{12}}^2
=\left|\pp{(I_u,I_\varphi)}{(\sigma_1,\sigma_2)}\right|^2
(J_u J_\varphi-{J_{u\varphi}}^2),
\end{equation}
\begin{equation}
K_\varphi=\frac{1}{\pi pa\sin\phi},\qquad \omega_v=\frac{\pi
p}{ma\sin\phi},
\end{equation}
one finally obtains
\begin{equation}
\delta g_{\rm EQ}^{(1)}(\eps)=
\frac{1}{\eps_0}\Re\sum_{\rm EQ}~A_{\rm EQ}
\exp\left(ikL_{\rm EQ}-i\frac{\pi}{2}\nu_{\rm EQ}\right),
\label{pstrace3}
\end{equation}
\begin{equation}
A_{\rm EQ}=\frac12\sqrt{\frac{\sin^3\phi}{\pi M n_v kR \eta F_z}}
\erf\left({\cal Z}_\perp^-,{\cal Z}_\perp^+;
{\cal Z}_1^-,{\cal Z}_1^+;{\cal Z}_2^-,{\cal Z}_2^+\right),
\label{ampEQgen}
\end{equation}
where $L_{\rm EQ}$ represents length of the EQPO.
The ``triple'' error function in Eq.~(\ref{ampEQgen}) can be separated
into the product of three standard error functions,
\begin{eqnarray}
\erf\left({\cal Z}_\perp^-,{\cal Z}_\perp^+;
{\cal Z}_1^-,{\cal Z}_1^+;{\cal Z}_2^-,{\cal Z}_2^+\right)\approx
\erf\left({\cal Z}_\perp^-,{\cal Z}_\perp^+\right)
\erf\left({\cal Z}_1^-,{\cal Z}_1^+\right)
\erf\left({\cal Z}_2^-,{\cal Z}_2^+\right)
\nonumber \\
\label{errorproduct}
\end{eqnarray}
by taking the limits at the stationary points for all deformations,
except for a small region near the spherical shape. In this way we obtain
the simple results (\ref{ampEQ}).
The arguments of the error functions are given by (\ref{argerroru}) or
(\ref{argerroru2}) for ${\cal Z}_i^\pm$ ($i=1,2$) and
\begin{equation}
{\cal Z}_\perp^\pm = \pm
\frac{\pi}{2}\sqrt{-\frac{iJ_\perp^{\rm EQ}}{2\hbar}}
 = \pm\frac{\hbar(k\zeta)^2}{16}
\sqrt{\frac{i F_z^{\rm EQ}}{Mn_v ka\sin\phi\,\sigma^*(\sigma^*+1)
 \det K^{\rm EQ}}}.
\label{limitsEQ3}
\end{equation}

The spherical limit is easily obtained by using the spherical
action-angle variables
$\{\Theta_\theta,\Theta_r,\Theta_\varphi;I_\theta,I_r,I_\varphi\}$.
In these variables
\begin{equation}
A_{\rm EQ}=
\frac12\sqrt{\frac{\sin^3\phi}{\pi M n_r kR \eta F_z}}\,
\erf\left({\cal Z}_\perp^-,{\cal Z}_\perp^+;
{\cal Z}_{\theta}^-,{\cal Z}_{\theta}^+;
{\cal Z}_{\varphi}^-,{\cal Z}_{\varphi}^+\right),
\label{ampEQsph}
\end{equation}
where $n_r \equiv n_v$ for the equatorial-plane orbits with ($n_v,n_\varphi$),
the invariant stability factor $F_\theta \equiv F_z^{\rm EQ}$ given by
(\ref{gutzstabfactz}),
\begin{eqnarray}
{\cal Z}_\perp^\pm=
\sqrt{\frac{-i\pi F_\theta^{\rm EQ}}{16M n_r \hbar K_\theta^{\rm EQ}}}
(z^\pm - z^*),
\quad
{\cal Z}_{\brcase{\theta}{\varphi}}^\pm=
\sqrt{-i \pi M n_r K_{\brcase{\theta}{\varphi}}^{\rm EQ}/\hbar}
\left(I_{\brcase{\theta}{\varphi}}^\pm-
I_{\brcase{\theta}{\varphi}}^*\right).\hspace{-5em}
\nonumber\\
\label{boundsph}
\end{eqnarray}
The quantities $K_\theta^{\rm EQ}$ and $K_\varphi^{\rm EQ}$
\begin{equation}
K_{\brcase{\theta}{\varphi}}^{\rm EQ}=
\left(\frac{\partial^2 I_r}{\partial I_{\brcase{\theta}{\varphi}}^2}
 \right)_{\rm EQ}\,,\qquad
\label{curvaturesph}
\end{equation}
are the curvatures of the energy surface
$\eps=H(I_\theta,I_r,I_\varphi)$ in the spherical
coordinate system. In that system the maximum value
of $I_\varphi$ is equal to
the absolute value of the classical angular momentum $I_\theta$,
$I_\varphi^\pm=\pm I_\theta$,
$I_\theta^+$ being the maximum value of $|I_\theta|$,
and $I_\theta^-=0$. We note that for the diametric orbits the stationary points
$I_\theta^*$ and $I_\varphi^*$ are exactly zero and there are also
specific integration limits in Eq.~(\ref{ampEQsph}).
In this case the internal integral over $I_\varphi$
within a small region can be evaluated approximately as $2I_\theta$, and
one obtains for the ``triple'' error function
\begin{equation}
\erf\left({\cal Z}_\perp^-,{\cal Z}_\perp^+;
{\cal Z}_{\theta}^-,{\cal Z}_{\theta}^+;
{\cal Z}_{\varphi}^-,{\cal Z}_{\varphi}^+\right) \rightarrow
\sqrt{\frac{-4i F_z}{M\pi^2 n_r \hbar K_\theta^{\rm EQ}}}=
\sqrt{\frac{-4i F_z kR}{2\pi M}}.
\label{erfsph}
\end{equation}
We used here the fact that, in the spherical limit $F_z
\rightarrow 0$, the integral over ${\cal Z}_\theta$ can be
approximated by the upper limit ${\cal Z}_\theta^+$ given by
Eq.~(\ref{boundsph}). We omitted also the strong oscillating value
of $\int dz^2 e^{-z^2}$ at the upper limit since it
vanishes after any small averaging over $kR$ and equals 1 in this
approximation. We also accounted for the fact that $K_\theta^{\rm
EQ} \rightarrow {1/(\pi pR)}$ for the diameters; see
Eq.~(\ref{curvaturesph}) ($\phi=\pi/2$ for the diameters).
Finally, the stability factor $F_z$ is canceled and one obtains
the Balian-Bloch result (\ref{ampgeqbb}) for the contribution of
the diametric orbits in the spherical cavity.\cite{bablo}

For all other EQPO one has the stationary points
$I_\varphi^*=I_\theta^* \neq 0$ and $I_\varphi$ is identical to
its maximum value  $I_\theta$ in the spherical limit. This is the
reason why there is no next order ($1/\sqrt{kR}$) corrections to
the Balian-Bloch trace formula for the contribution of
the planar orbits with $n_r \geq 3$. The latter comes from the
spherical limit of the elliptic orbits in the meridian plane
(\ref{amp3d2d}), see Ref.~\citen{mfimbrk}.

\section{Separatrix}
\label{app:separatrix}

Like for the case of the turning points,
\cite{fed:jvmp,masl,fed:spm,masl:fed} we first expand
the exponent phase in Eq.~(\ref{pstraceactang1}) with respect to $I_u'$~:
\begin{eqnarray}
S_\alpha\left(\bI',\bI'',t_\alpha\right)-\left(\bI''-\bI'\right)
\cdot\bTheta'' &=&
c_0^\parallel+c_1^\parallel x +
c_2^\parallel x^2 +c_3^\parallel x^3+\ldots \nonumber\\
&\equiv&
\tau_0^\parallel+\tau_1^\parallel z +\frac13 z^3.
\label{expan3}
\end{eqnarray}
Here
\begin{equation}
x=\frac{1}{\hbar}\left(I_u'-{I_u'}^*\right),
\label{x}
\end{equation}
\begin{equation}
c_0^\parallel=\frac{1}{\hbar}
\left[S_\alpha^*\left(\bI',\bI'',\eps\right)-
\left(\bI'-\bI''\right)^*\cdot {\bTheta''}^*\right]=
\frac{1}{\hbar}
S_\alpha^*\left(\bTheta',\bTheta'',\eps\right),
\label{c0par}
\end{equation}
\begin{equation}
c_1^\parallel=
\left(\pp{S_\alpha}{I_u'}-\Theta_u''\right)^*=
\Theta_u'-\Theta_u'' \rightarrow 0, \qquad \sigma_1 \rightarrow 1,
\label{c1par}
\end{equation}
\begin{equation}
c_2^\parallel=\frac{\hbar}{2}
\left(\frac{\partial^2 S_\alpha}{\partial {I_u'}^2}\right)^*=
2 p \zeta M \hbar \tK_u^\alpha
\rightarrow \infty, \qquad \sigma_1 \rightarrow 1,
\label{c2par}
\end{equation}
\begin{equation}
c_3^\parallel=\frac{\hbar^3}{6}
\left(\frac{\partial^3 S_\alpha}{\partial {I_u'}^3}\right)^*=
\frac{2\pi^3 \hbar^2 M}{3 (p \zeta^2)^2}
\left(\pp{\tK_u^\alpha}{\tI_u}\right)<0, \quad
\sigma_\parallel \rightarrow 1,
\label{c3par}
\end{equation}
where the superscript asterisk indicates the value at $I_u'=I_u''=I_u^*$.
The asymptotic behavior of the constants $c_i^\parallel$ near the separatrix
$\sigma_1 \approx 1$ is found from
\begin{equation}
\tK_u^\alpha
\rightarrow
\frac{\log\left[(1+\sin\theta)/(1-\sin\theta)\right]}{(\sigma_1-1)
  \log^3(\sigma_1-1)},
\qquad \sigma_1 \rightarrow 1,
\label{Ksepasymp}
\end{equation}
$\theta \rightarrow \theta_h(\eta)$~ formally, see (\ref{kappatheta}),
\begin{equation}
\pp{\tK_u^\alpha}{\tI_u} \rightarrow
-\frac{2 \log\left[(1+\sin\theta)/(1-\sin\theta)\right]}{\left(
  (\sigma_1-1) \log^2(\sigma_1-1)\right)^2},
\qquad \sigma_1 \rightarrow 1.
\label{dKsepasymp}
\end{equation}
The rightmost part of Eq.~(\ref{expan3}) is
obtained by a linear transformation with
some constants $\alpha$ and $\beta$,
\begin{equation}
x=\alpha z +\beta,\qquad
\alpha=\left(3 c_3^\parallel\right)^{-1/3},\qquad
\beta=-c_2^\parallel/(3 c_3^\parallel),
\label{alphabeta}
\end{equation}
\begin{equation}
\tau_0^\parallel=\left(c_0-c_1c_2/(3c_3)+2c_2^3/(27c_3^2)\right)^\parallel,
\qquad
\tau_1^\parallel=\alpha\left[c_1-c_2^2/(3c_3)\right]^\parallel.
\label{taupar}
\end{equation}
Near the stationary point for $\sigma_1 \rightarrow 1$,
one obtains $c_1^\parallel \rightarrow 0$ and
$\tau_1^\parallel \rightarrow -w^\parallel$ with the positive quantity
\begin{equation}
w^\parallel=\left(\frac{c_2^2}{(3 c_3)^{4/3}}\right)^\parallel \rightarrow
\left|\frac{M\log\left[(1+\sin\theta)/(1-\sin\theta)\right]
(\sigma_1-1)}{\hbar\log(\sigma_1-1)}\right|^{2/3}.
\label{wpar}
\end{equation}
Using expansion (\ref{expan3}) in Eq.~(\ref{pstraceactang1})
and taking the integral over $\Theta_v''$ exactly
(i.e., putting $2 \pi$ for this integral), we obtain
\begin{eqnarray}
\delta g_{\rm LD}^{(0)} &=&
-\frac{2}{2\pi\hbar^2}\Re\sum_\alpha
\int d\Theta_\varphi'' \int dI_\varphi'
\int d\Theta_u'' \frac{1}{|\omega_v^*|}
e^{i\left(\tau_0^\parallel-\frac{\pi}{2}\nu_\alpha\right)}
\qquad\nonumber\\
&&\times
\sqrt{\frac{\sqrt{w^\parallel}}{c_2^\parallel}}
\left[\Ai\left(-w^\parallel,{\cal Z}_\parallel^-,
{\cal Z}_\parallel^+\right)
+i\Gi\left(-w^\parallel,{\cal Z}_\parallel^-,
{\cal Z}_\parallel^+\right)\right]~
\nonumber\\
&& \hspace{-3em}
\approx -\frac{2}{\hbar}\Re\sum_\alpha\int d\Theta_u''
\frac{1}{|\omega_v^*|}
\sqrt{\frac{\sqrt{w^\parallel}}{c_2^\parallel}}
      \left[\Ai\left(-w^\parallel\right)+i
      \Gi\left(-w^\parallel\right)\right]
e^{i\left(\tau_0^\parallel-\frac{\pi}{2}\nu_\alpha\right)}\,,
\nonumber\\
\label{pstracelactang}
\end{eqnarray}
where
\begin{equation}
{\cal Z}_\parallel^-=\sqrt{w^\parallel},\qquad
{\cal Z}_\parallel^+=\sqrt{\frac{c_2^\parallel}{\sqrt{w^\parallel}}}
\frac{I_u^{\rm(cr)}}{\hbar}+
\sqrt{w^\parallel},
\label{zmimalpar}
\end{equation}
$\Ai(-w,z_1,z_2)$ and $\Gi(-w,z_1,z_2)$ are the incomplete Airy and Gairy
functions defined by
\begin{equation}
\brcase{\Ai}{\Gi}(-w,z_1,z_2)
=\frac{1}{\pi}\int_{z_1}^{z_2}dz\,
\brcase{\cos}{\sin}\left(-wz+\frac{z^3}{3}\right),
\label{airynoncomp}
\end{equation}
$\Ai(-w)$ and $\Gi(-w)$ are the corresponding standard complete functions,
$I_u^{\rm(cr)}=I_u(\sigma_1^{\rm(cr)},\sigma_1^{\rm(cr)})$
is the ``creeping'' elliptic 2DPO value defined in section 2.
In the second equation of (\ref{pstracelactang}), we used the fact
that for any finite deformation $\eta$~ and large $kR$~
near the separatrix ($\sigma_1 \rightarrow 1$)
\begin{eqnarray}
{\cal Z}_\parallel^- &\rightarrow& 0, \nonumber\\
{\cal Z}_\parallel^+ &\rightarrow&
4\left[\frac{M\log\left[(1+\sin\theta)/(1-\sin\theta)\right]p\zeta}
{2 (\sigma_1-1)^2 \log^4(\sigma_1-1)}\right]^{1/3}
\left[\frac{\eta}{\sqrt{\eta^2-1}}
\rE\left(\frac{\sqrt{\eta^2-1}}{\eta}\right)-1\right] \nonumber\\
&\rightarrow& \infty. \label{asymptlim}
\end{eqnarray}
Using an analogous expansion of the action $\tau_0^\parallel$
in (\ref{pstracelactang})
with respect to the angle $\Theta_u''$ to the third order and
a linear transformation like (\ref{alphabeta}) one arrives at
\begin{eqnarray}
\delta g_{\rm LD}^{(0)}(\eps) &=& \frac{b}{2 \eps_0 \pi^2 R
\hbar}\, \Re \sum_\alpha \int d\Theta_\varphi''\int dI_\varphi'
\frac{1}{kR}\,\frac{\left(w^\parallel
w^\perp\right)^{1/4}}{\sqrt{|c_2^\parallel
   c_2^\perp|}} \nonumber\\
&&\hspace{-4em}\times \left[\Ai\left(-w^\parallel\right)
+i \Gi\left(-w^\parallel\right)\right]
\left[\Ai\left(-w^\perp,{\cal Z}_\perp^-,{\cal Z}_\perp^+\right)
+i\Gi\left(-w^\perp,{\cal Z}_\perp^-,{\cal Z}_\perp^+\right)
\right]\nonumber\\
&&\hspace{-4em}\times \exp\left\{\frac{i}{\hbar}
\left[S_\alpha^*\left(\bI',\bI'',\eps\right)-
\left(\bI'-\bI''\right)^* \cdot {\bTheta''}^*\right]\right.\nonumber\\
&&\hspace{4em}\left.
 + \frac{2i}{3}\left[(w^\parallel)^{3/2}+(w^\perp)^{3/2}\right]
-i \frac{\pi}{2}\nu_\alpha\right\}, \label{pstracelactang1}
\end{eqnarray}
where
\begin{equation}
w^\perp=\left(\frac{c_2^2}{(3c_3)^{4/3}}\right)^\perp~>~0,
\label{wperp}
\end{equation}
\begin{equation}
{\cal Z}_\perp^-=\sqrt{w^\perp},
\qquad
{\cal Z}_\perp^+=\frac{\pi}{2}\left|3 c_3^\perp\right|^{1/3}+
\sqrt{w^\perp},
\label{zmimalperp}
\end{equation}
\begin{equation}
c_2^\perp=\frac{1}{2\hbar}\left(J_{u,\alpha}^\perp\right)^*
=\left(\pp{^2 S_\alpha}{{\Theta_u'}^2}
+2 \pp{^2 S_\alpha}{\Theta_u'\partial \Theta_u''}
+\pp{^2 S_\alpha}{{\Theta_u''}^2}\right)_{\rm LD}^*
=-\frac{F_{xy}^{\rm LD}}{2\pi M K_u^\alpha}.
\label{c2perp}
\end{equation}
$F_{xy}^{\rm LD}$ is the stability factor for the long diameters,
\begin{equation}
F_{xy}^{\rm LD}= -4\sinh^2\left[M\arccosh\left(2 \eta^2- 1\right)\right]~,
\label{gutzstabfactld}
\end{equation}
\begin{eqnarray}
c_3^\perp&=&\frac{1}{6 \hbar}\left[
\pp{^3 S_\alpha}{{\Theta_u'}^3}
+3\pp{^3 S_\alpha}{{\Theta_u'}^2 \partial \Theta_u''}
+3\pp{^3 S_\alpha}{\Theta_u' \partial {\Theta_u''}^2}
+\pp{^3 S_\alpha}{{\Theta_u''}^3}\right]^*
\nonumber\\
&=&\frac{1}{6 \hbar}
\left[\pp{J_{u,\alpha}^\perp}{\Theta_u'}
     +\pp{J_{u,\alpha}^\perp}{\Theta_u''}\right]^* ~<~ 0\,.
\label{c3perp}
\end{eqnarray}
Note that according to (\ref{c2perp}) the quantity $c_2^\perp$
approaches  zero near the separatrix ($\sigma_1 \rightarrow 1$)
like in the caustic case. This is the reason why we apply the
Maslov-Fedoryuk theory
\cite{fed:jvmp,masl,fed:spm,masl:fed} for the transformation of
the integral over angle $\Theta_u''$ from (\ref{pstracelactang})
to (\ref{pstracelactang1}). The remaining two integrals over the
azimuthal variables ($I_\varphi'$ and $\Theta_\varphi''$) can be
calculated in a similar way as explained in the text.

Divergence of the curvature $K_\varphi$, Eq.~(\ref{Jufi}), for
the long diameters ($\sigma_1 \rightarrow 1$~, $\sigma_2
\rightarrow 0$) can be easy seen from the following expression
valid for any polygon orbit having a vertex on the symmetry axis,
\begin{equation}
K_\varphi^\beta = \frac{L_0c}{\rho_0^2 n_v M \hbar}
\left[\frac{2\eta^2}{1+\eta^2\tan^2\psi}-1\right],
\label{kfipol}
\end{equation}
where $L_0$ denotes the length of the side having a vertex at the pole,
$\rho_0$ the cylindrical coordinate of another end of this side,
and $\psi $ the angle between this side and the symmetry axis.
For the long diameters, $L_0 \rightarrow 2 b M$, $\rho_0
\rightarrow 0$ and $\psi \rightarrow 0$,
so that $K_\varphi^\beta \rightarrow \infty$~.

\section{Derivation of the third-order term}
\label{app:3rd_order}

\subsection{Third-order curvatures}
\label{app:thirdordercurv}

For the curvature $K_1^{(3)}$ which appears in the third-order
terms in the expansion of the action $S/\hbar$ with respect to
$\sigma_1$, one obtains
\begin{equation}
\tK_1^{(3)}=\frac{\pi}{p\zeta}K_1^{(3)}=
\pp{^3\tI_v}{\sigma_1^3}+\frac{n_u}{n_v}\pp{^3\tI_u}{\sigma_1^3},
\label{k31}
\end{equation}
where
\begin{eqnarray}
\pp{^3 \tI_v}{\sigma_1^3} &=& -\frac{1}{4 z_+^3}
\left[\pp{B_v}{\sigma_1} + 6z_+ \pp{z_+^2}{\sigma_1}\,
     \pp{^2 \tI_v}{\sigma_1^2}\right]\,,  \nonumber\\
\pp{^3\tI_u}{\sigma_1^3} &=& \frac{1}{2 z_+^3}
\left[\pp{B_u}{\sigma_1} - 3z_+ \pp{z_+^2}{\sigma_1}\,
     \pp{^2 \tI_u}{\sigma_1^2}\right]\,,
\label{d3ivuds131}
\end{eqnarray}
\begin{equation}
B_v=\left[\rPi(\varphi,k^2,k)-\rF(\varphi,k)\right]\,\tilde{\partial}_k
-\pp{z_+^2}{\sigma_1}\,\rF(\varphi,k)+\frac{2 z_+^2}{\Delta_\varphi}\,
\pp{\varphi}{\sigma_1}\,,
\label{bv1}
\end{equation}
\begin{eqnarray}
B_u &=& \left[\rPi(k^2,k)-\rF(k)\right]\,\tilde{\partial}_k-
\pp{z_+^2}{\sigma_1}\rF(k)\,,\qquad
\label{bu1}
\nonumber\\
\tilde{\partial}_{k} &=&
\frac{z_+^2}{k^2}\,\pp{k^2}{\sigma_1}=
\frac{1}{k^2}\pp{z_-^2}{\sigma_1}-\pp{z_+^2}{\sigma_1}\,,
\label{budkc1}
\end{eqnarray}
with the derivatives
\begin{eqnarray}
\pp{B_u}{\sigma_1} &=&
k\left[\pp{\rPi(k^2,k)}{\partial k} - \pp{\rF(k)}{k}\right]\,
\frac{\tilde{\partial}_k^2}{2z_+^2}
+\left[\rPi(k^2,k)-\rF(k)\right]\,
 \left[-\frac{\tilde{\partial}_k}{z_-^2}\,
 \pp{z_-^2}{\sigma_1}\right.
\nonumber\\
&&
\left.
+\frac{1}{k^2}\,\pp{^2 z_-^2}{\sigma_1^2}
-\pp{^2 z_+^2}{\sigma_1^2}\right]
-k \pp{\rF(k)}{k}\,
\frac{\tilde{\partial}_k}{2z_+^2}\,
\pp{z_+^2}{\sigma_1}-\pp{^2 z_+^2}{\sigma_1^2} \rF(k)\,,
\label{dbuds11}
\nonumber\\
\pp{^2 z_\pm^2}{\sigma_1^2} &=&
\pm\frac{2\sigma_2}{\left[(\sigma_1-1)^2+4\sigma_2\right]^{3/2}}\,,
\label{dbuds1d2dzds121}
\end{eqnarray}
\label{d3ivds131}
\begin{eqnarray}
\pp{B_v}{\sigma_1}=
\tilde{\partial}_k\left[\pp{\rPi(\varphi,k^2,k)}{\sigma_1}
-\pp{\rF(\varphi,k)}{\sigma_1}+\left(1-\frac{1}{z_-^2}
 \pp{z_-^2}{\sigma_1}\right)\right]
-\pp{^2 z_+^2}{\sigma_1^2}\,\rF(\varphi,k) \nonumber\\
-\pp{z_-^2}{\sigma_1}\,\pp{\rF(\varphi,k)}{\sigma_1}
+\frac{1}{\Delta_\varphi}
\left[\left(2\pp{z_+^2}{\sigma_1}-\frac{z_+^2}{\Delta_\varphi^2}\,
\pp{\Delta_\varphi^2}{\sigma_1}\right)
\pp{\varphi}{\sigma_1}+2z_+^2\pp{^2\varphi}{\sigma_1^2}\right].
\label{dbvds11}
\end{eqnarray}
Here
\begin{equation}
\pp{\rPi(k^2,k)}{k}=
\frac{k^2 \tilde{\partial}_k}{k_1^2z_+^2}\,
\left[\rPi(k^2,k)+\frac{1}{2k^2}\,\left(\rE(k)-\rF(k)\right)\right],
\label{dpikkdk1}
\end{equation}
\begin{equation}
\pp{\rF(k)}{k}=\frac1k\left[\rPi(k^2,k)-\rF(k)\right],
\label{dfkdk1}
\end{equation}
\begin{eqnarray}
\pp{\rPi(\varphi,k^2,k)}{\sigma_1} &=& \frac{1}{k_1^2}
  \biggl[\frac{k^2 \tilde{\partial}_k}{z_+^2}
  \biggl\{\rPi(\varphi,k^2,k)+\frac{1}{2k^2}
  \left(\rE(\varphi,k)-\rF(\varphi,k)\right)
\nonumber\\
&& \hspace{-6.5em}
  -\frac{\sin(2\varphi)}{4\Delta_\varphi^3}
  \left(1+\Delta_\varphi^2\right)\biggr\}
  +\Delta_\varphi\,\pp{\varphi}{\sigma_1}\,
  \left\{1-\frac{k^2}{4\Delta_\varphi^4}\,
  \left[4\Delta_\varphi^2\cos(2\varphi)+
  k^2\sin^2(2\varphi)\right]\right\}\biggr]\,, \nonumber\\
\label{dpifkkds11}
\end{eqnarray}
\begin{equation}
\pp{\rF(\varphi,k)}{\sigma_1}=\frac{\tilde{\partial}_k}{2 z_+^2}
\left[\rPi(\varphi,k^2,k)-\rF(\varphi,k)\right]
+\frac{1}{\Delta_\varphi}\,\pp{\varphi}{\sigma_1},
\label{dffkds11}
\end{equation}
\begin{eqnarray}
\pp{^2\varphi}{\sigma_1^2}&=&\frac{1}{\sin(2\varphi)}
\left\{\frac{1}{\left(z_b^2-z_-^2\right)^3}
\left[\left(\pp{^2z_-^2}{\sigma_1^2}\,\left(z_b^2-z_+^2\right)
-\pp{^2z_+^2}{\sigma_1^2}\left(z_b^2-z_-^2\right)\right)
\right.\right.
\nonumber\\
&& +\left.\left. 2\pp{z_-^2}{\sigma_1}\left(\pp{z_-^2}{\sigma_1}
\left(z_b^2-z_+^2\right)-\pp{z_+^2}{\sigma_1}
\left(z_b^2-z_-^2\right)\right)\right]
 - 2\cos(2\varphi)
\left(\pp{\varphi}{\sigma_1}\right)^2\right\}, \nonumber\\
\label{d2fids121}
\end{eqnarray}
\begin{equation}
\pp{\Delta_\varphi^2}{\sigma_1}=
-k^2\left[\frac{\tilde{\partial}_k}{z_+^2}\,\sin^2\varphi+
\sin(2\varphi)\pp{\varphi}{\sigma_1}\right].
\label{ddfi2ds111}
\end{equation}

\subsection{Stationary phase method with third-order expansions}

After the expansion of the action in the Poisson-sum trace formula
(\ref{Poisvarsigma}) up to the second order with respect to $\sigma_2$
and up to the third order with respect to $\sigma_1$, one obtains
\begin{eqnarray}
\delta g^{(2)}(\eps)&=&\frac{k\zeta^2}{4\pi^2R\eps_0}\,
\Re\sum_{\beta}\frac{L_\beta}{M n_v R\sqrt{\sigma_2^*}}
\left(\pp{\tI_u}{\sigma_1}\right)^*
\exp{\left(ikL_\beta-i\frac{\pi}{2}\nu_\beta\right)}
\nonumber\\
&& \times
\int_0^{\sigma_2^+}d\sigma_2 \int_{x^-}^{x^+}dx\,
  \exp\Bigl[ik\zeta Mn_v\tK_{22}
  \left(\sigma_2-\sigma_2^*+\frac{K_{12}}{K_{22}}(\sigma_1-\sigma_1^*)
  \right)^2 \nonumber\\
&& \hspace{12em}
  +i\left(c_1x+c_2x^2+c_3x^3\right)\Bigr]
\label{poissexp31}
\end{eqnarray}
where
\begin{equation}
c_1 \to 0, \quad
c_2 = k\zeta M n_v\frac{\det \tK}{\tK_{22}}, \quad
c_3 = \frac13 k\zeta M n_v\tK_1^{(3)},
\label{ci1}
\end{equation}
\begin{equation}
x=\sigma_1-\sigma_1^*,\quad x^\pm=\sigma_1^\pm-\sigma_1^*.
\label{xpm1}
\end{equation}
After transformation from $\sigma_2$ to the new ${\cal Z}_2$ variable,
\begin{equation}
{\cal Z}_2 = \sqrt{-ik \zeta M n_v \tK_{22}}
 \left(\sigma_2-\sigma_2^*+\frac{K_{12}}{K_{22}}(\sigma_1-\sigma_1^*)
 \right),
\label{z21}
\end{equation}
and a linear transformation from $x$ to $z$,
\begin{equation}
x=q_1z+q_2,\quad\mbox{with}
\quad q_1=(3c_3)^{-1/3}, \quad q_2=-\frac{c_2}{3c_3},
\label{xz1}
\end{equation}
one obtains Eq.~(\ref{deltag3d2d}) with the ISPM3 amplitude
\begin{eqnarray}
A_{\rm 3D}^{(2)}(\eps)&=&
\frac{L_\beta}{8M n_v R\left(k\zeta M n_v \tK_1^{(3)}\right)^{1/3}}
\sqrt{\frac{ik\zeta^3}{\pi M n_v R^2 \tK_{22}\sigma_2^*}}\,
\left(\pp{\tI_u}{\sigma_1}\right)^*
\exp\left(\frac23 i\tau^{3/2}\right)
\nonumber\\
&&\times
\erf\left({\cal Z}_2^{-},{\cal Z}_2^{+}\right)
\left[\Ai\left(-\tau,z_-,z_+\right)+
i\Gi\left(-\tau,z_-,z_+\right)\right],
\label{ampispm31}
\end{eqnarray}
Here ${\cal Z}_2^\pm$ is defined by Eq.~(\ref{argerroru2}b) and
\begin{equation}
\tau=(3c_3)^{-1/3}\,\left(\frac{c_2^2}{3c_3}-c_1\right), \quad
z_\pm=\frac{x^\pm-q_2}{q_1}.
\label{tauccalz2pm1}
\end{equation}
In the limit $c_1\rightarrow 0$
\begin{equation}
\tau = \frac{c_2^2}{(3c_3)^{4/3}}
= \frac{(k\zeta M n_v)^{2/3}(\det\tK/\tK_{22})^2}{
  \left(\tK_1^{(3)}\right)^{4/3}}.
\label{tauc1to0lim1}
\end{equation}

For finite curvatures far from the bifurcations, one can extend
the limits of the Airy and Gairy functions ($z_-\rightarrow 0$ and
$z_+ \rightarrow \infty$) and obtains the complete Airy
$\Ai(-\tau)$ and Gairy $\Gi(-\tau)$ functions. Then, using the
asymptotics of these functions for large $\tau \propto (kR)^{2/3}$
(large $kR$)
\begin{eqnarray}
\brcase{\Ai}{\Gi}(-\tau)
\sim\frac{1}{\sqrt{\pi}\tau^{1/4}} \brcase{\sin}{\cos}
\left(\frac23\tau^{3/2}+\frac{\pi}{4}\right)
\end{eqnarray}
and of the $\erf$-function in Eq. (\ref{ampispm31}) one obtains
the SSPM limit (\ref{amp3d2das}).

\end{document}